\documentclass[a4paper]{JHEP3}
\usepackage[centertags]{amsmath}
\usepackage{amssymb}
\usepackage{fixltx2e}
\MakeRobust{\eqref} 
\usepackage{graphicx}
\usepackage{subfigure}
\preprint{TIFR/TH/13-02\\
ICTS-2012-14}

\newcommand{\Tr}{\text{Tr}}

\def\cN{{\cal N}}

\def \beal#1 {\begin{align}#1\end{align}}

\def\bZ{{\mathbf Z}}
\def\tr{\mathrm{tr}}
\def\Tr{\mathrm{Tr}}

\def\nn{\nonumber\\}

\def\[{\left[}
\def\]{\right]}
\def\({\left(}
\def\){\right)}

\def\={\stackrel{\bullet}{=}}
\def\nn{\notag\\}

\def\Tr{\mathrm{Tr}}

\def\[{\left[}
\def\]{\right]}
\def\({\left(}
\def\){\right)}

\def\cN{{\cal N}}

\def\bZ{{\mathbf Z}}

\def \be {\begin{equation}}
\def \ee {\end{equation}}
\def \bea {\begin{eqnarray}}
\def \eea {\end{eqnarray}}
\def \beal#1 {\begin{align}#1\end{align}}
\def \nn {\notag\\}

\newcommand{\lm}{\lambda}

\def \be {\begin{equation}}
\def \ee {\end{equation}}
\def \bea {\begin{eqnarray}}
\def \eea {\end{eqnarray}}
\def \beal#1 {\begin{align}#1\end{align}}
\def \nn {\notag\\}

\usepackage{enumerate}

\title{Phases of large $N$ vector Chern-Simons  
theories on $S^2 \times S^1$}

\author{
Sachin Jain$^{a),1}$, 
Shiraz Minwalla$^{a),2}$, 
Tarun Sharma$^{a),3}$, 
Tomohisa Takimi$^{a),4}$, 
Spenta R. Wadia$^{a),b),5}$, 
Shuichi Yokoyama$^{a),6}$
\\
$^{a)}$Department of Theoretical Physics, Tata Institute of Fundamental Research,
Homi Bhabha Road, Mumbai 400005, India\\
$^{b)}$International Centre for Theoretical Sciences, Tata Institute of Fundamental Research,
TIFR Centre Building, Indian Institute of Science, Bangalore 560004, India \\
{\small \tt E-mail: $^1$sachin, ${}^2$minwalla, ${}^3$tarun, ${}^4$takimi, ${}^5$wadia, ${}^6$yokoyama(at)theory.tifr.res.in}
}

\abstract{We study the thermal partition function of level $k$ $U(N)$ Chern-Simons theories on $S^2$ interacting with matter in the fundamental 
representation. We work  in the 't~Hooft limit, $N,k\to\infty$,  with $\lambda = N/k$ and $\frac{T^2 V_{2}}{N}$ held fixed
 where $T$ is the temperature and $V_{2}$ the volume of the sphere. 
An effective action proposed in arXiv:1211.4843 relates the partition 
function to the expectation value of a `potential' function of the 
$S^1$ holonomy in pure Chern-Simons theory; in several examples 
we compute the holonomy potential as a function of $\lambda$. 
We use level rank duality of 
pure Chern-Simons theory to demonstrate the equality of thermal partition
functions of previously conjectured dual pairs of theories as a function 
of the temperature. We reduce the partition function to 
a matrix integral over holonomies. The summation over flux sectors quantizes 
the eigenvalues of this matrix in units of  ${2\pi \over k}$ and the 
eigenvalue density of the holonomy matrix is bounded from above by 
$\frac{1}{2 \pi \lambda}$. The corresponding matrix integrals generically 
undergo two phase transitions as a function of temperature.
For several Chern-Simons matter theories 
we are able to exactly solve the relevant matrix models in the low temperature
phase, and determine the phase transition temperature as a function of 
$\lambda$. At low temperatures our partition function smoothly matches onto the 
$N$ and $\lambda$ independent free energy of a gas of non renormalized 
multi trace operators. We also find an  exact solution to a simple toy 
matrix model; the large $N$ Gross-Witten-Wadia matrix integral subject to an 
upper bound on eigenvalue density.}

\begin{document}
\section{Introduction}

The AdS/CFT correspondence maps deconfinement transitions of large $N$
gauge theories on spheres to gravitational phase transitions involving
black hole nucleation \cite{Witten:1998zw}. This observation has motivated
the intensive study of deconfinement phase transitions of large $N$
$p+1$ dimensional Yang Mills theories (coupled to adjoint and fundamental
matter) on $S^p$. The thermal partition function of a Yang-Mills theory on
$S^p$ is given by the Euclidean path integral of the theory on $S^p \times S^1$. Upon integrating out all massive modes this path integral reduces to an
integral over the single unitary matrix $U$
 \begin{equation} \label{mint}
 Z_{\text{YM}}=\int D U \exp[-V_{\text{YM}}(U)]=
\prod_{m=1}^N \int_{-\infty}^\infty d\alpha_m
\left[ \prod_{l \neq m} 2 \sin \left(\frac{\alpha_l-\alpha_m}{2}
\right)
e^{-V_{YM}(U)}\right]
 \end{equation}
where $U$ is the zero mode (on $S^p$) of the holonomy around the
thermal circle, $e^{i \alpha_i}$ ($i=1 \ldots N$) are the eigenvalues 
of $U$.  $V_{\text{YM}}(U)$ is a potential function whose precise form 
depends on the theory under study. At least in
perturbation theory \cite{Aharony:2003sx} and perhaps beyond
\cite{AlvarezGaume:2005fv,AlvarezGaume:2006jg}, the potential $V_{\text{YM}}(U)$
is an analytic function of $U$.
\eqref{mint} may be thought of as a Landau Ginzburg or Wilsonian
description of the holonomy $U$, the lightest degree of freedom of the
finite temperature field theory.

The effective potential $V_{\text{YM}} (U)$ was computed in free gauge theories
\cite{Sundborg:1999ue, Aharony:2003sx}; it has also been evaluated at
higher orders in perturbation theory in special examples
\cite{Aharony:2005bq, Aharony:2006rf,Papadodimas:2006jd, Mussel:2009uw}.
In all these
examples $V_{\text{YM}}(U)$ is an attractive potential for the eigenvalues of the
unitary matrix. t' Hooft counting and the requirement of extensivity
force $V_{\text{YM}} (U)$ to scale like $N^2 V_{p}$. In the special case of a
conformal theory the potential scales like $N^2 V_{p} T^p$ and the attraction
between eigenvalues grows arbitrarily large at high temperatures.%
\footnote{Most of the general discussion of this paragraph applies also to
non conformal theories in the appropriate high temperature and/or large
volume limit.} On the other hand the integration measure
$D U$ (which vanishes when any two eigenvalues coincide) supplies a temperature independent
repulsive potential for the eigenvalues. At large $N$ the leading piece of the partition function
is determined by a saddle point distribution of eigenvalues of $U$.
Repulsion from the measure dominates over attraction
from the potential at low temperatures and the eigenvalues of $U$ are
distributed all over the unit circle in the complex plane.%
\footnote{If the matter content of the theory consists only of
 fundamental plus adjoint fields, and if the matter content
is held fixed as $N$ which is taken to infinity and the low
temperature saddle point for $U$ is the clock matrix; a matrix
whose eigenvalues are uniformly distributed on the unit
circle in the complex plane.} The integral \eqref{mint} undergoes
Gross-Witten-Wadia type deconfinement transitions \cite{Gross:1980he,Wadia:2012fr,Wadia:1980cp} at $V_{p} T^p$ of order unity.%
\footnote{At weak coupling at least the system undergoes
either a single first order transition or a second order transition followed
by a third order phase transition depending on the details of a quartic
term in the potential $V(U)$ \cite{Aharony:2003sx}.}
At high enough temperatures the saddle point eigenvalue distribution of $U$
has support on only a small arc in the unit circle of the complex plane.
The size of this arc goes to zero as $V_{p}T^p \to \infty$. In this
`decompactification' limit, the holonomy matrix $U$ is localized around the
unit matrix.

In this note we study the finite temperature phase structure of renormalized level $k$ U($N$) Chern-Simons theories coupled to a finite number of fundamental fields on $S^2$ in the t' Hooft limit $N \to \infty$, $k \to \infty$ with 
$\lambda=\frac{N}{k}$ fixed.%
\footnote{
We use the dimensional reduction regulation scheme through this paper. 
In this case $|k|=|k_{YM}|+N$, where $k_{YM}$ is the level of the Chern-Simons theory regulated by including an infinitesimal Yang Mills term in the action,
and $k$ is the level of the theory regulated in the dimensional reduction scheme
\cite{Pisarski:1985yj,Chen:1992ee}. 
For this reason this class of Chern-Simons theories may be well-defined only in the range $|\lambda| \leq 1$.
Hereafter we assume the 't~Hooft coupling is always positive for simplicity. 
Our results are easily generalized to the case $\lambda$ is negative by taking the absolute value of it. 
}
We address the following question: Does the Landau
Ginzburg description \eqref{mint} continue to hold for Chern-Simons matter
theories? If not what replaces it? Our analysis of this structural question 
applies to {\it all} fundamental matter Chern-Simons 
theories in the high temperature and large $N$ limit; in addition we make 
significant progress towards an exact determination of the partition function,  
the large $N$ free energy, as a function of temperature and $\lambda$, in 
particular examples of these theories.

To start our discussion let us  
first review the behaviour of matter Chern-Simons theories in the free limit $ \lambda \to 0$.
In particular examples such a thermal partition function  was studied
in \cite{Shenker:2011zf,Giombi:2011kc,Chang:2012kt}. The free
partition function {\it continues} to be governed by the Landau Ginzburg form
\eqref{mint} even at finite $N$; in fact the only qualitative difference
 from the Yang Mills case is that
$V(U)$ scales like $N V_{2} T^2$ (rather than $N^2 V_{2} T^2$) at
high temperatures.%
\footnote{The potential scales like $N$ rather than $N^2$ because
the matter is in the fundamental representation and pure Chern-Simons
theory has no propagating degrees of freedom.}
As the repulsion from the
measure $D U$ in \eqref{mint} continues to scale like $N^2$, the partition
function displays interesting dynamics in the large $N$ limit  only if we set
$V_{2}T^2=\zeta N$ and take the limit $N \to \infty$ holding $\zeta$ fixed.
At small $\zeta$ the eigenvalue density of the holonomy matrix has support
everywhere on the unit circle in the complex plane; however at $\zeta$
of order unity the system undergoes a third order clumping or
deconfinement phase transition. At higher temperature the eigenvalue
distribution has support only on a finite arc (centered about unity) on
the complex plane. In the infinite $\zeta$ limit the size of this arc goes
to zero and the holonomy reduces to the identity matrix.

Large $N$ fundamental matter Chern-Simons theories have recently been studied
intensively at finite values of the t' Hooft coupling $\lambda$ 
(see e.g. \cite{Giombi:2011kc,Aharony:2011jz,Maldacena:2011jn,Maldacena:2012sf,Banerjee:2012gh,Chang:2012kt,Aharony:2012nh,Jain:2012qi,Yokoyama:2012fa,Banerjee:2012aj,GurAri:2012is,Aharony:2012ns}).
In particular it has been argued in the important recent paper
\cite{Aharony:2012ns} that the entire effect
of matter loops on gauge dynamics, at temperatures of order $\sqrt{N}$
and at leading order in $N$, is to generate a term of the form
$$T^2 \int d^2x ~ \sqrt g v(U(x)) $$
for the gauge theory effective action. Here $v(U)$ is an effective potential
of order $N$,\footnote{The trace of $U$ is counted as order $N$.
An example of a potential of order $N$ is 
$$v(U)=a \Tr U + b \frac{ (\Tr U)^2 \Tr U^\dagger}{N^2} + \ldots + c.c. $$
} whose detailed form depends on the matter content and couplings
of the Chern-Simons matter system.
It follows that in the large $N$ limit, the thermal Chern-Simons matter
path integral is given by 
\begin{equation}\label{purecsi}
Z_{\text{CS} }=\int D A \exp \left[i \frac{k}{4\pi} \Tr  \int
\left( AdA + \frac{2}{3} A^3 \right)
 -T^2 \int d^2x \sqrt g ~v(U(x)) \right].
\end{equation}
\eqref{purecsi} is simply the {\it pure} Chern-Simons theory with an added
potential  $ v(U(x)) $ the local value of the holonomy matrix $U(x)$.
As pure Chern-Simons theory is topological, the expectation values of
observables are independent of $x$ and \eqref{purecsi} may be rewritten
as
\begin{equation}\label{purecsci}
Z_{\text{CS}}=\langle e^{ -T^2 V_{2} v(U)} \rangle_{N,k}
\end{equation}
where $\langle O \rangle_{N, k}$ denotes the expectation value of $O$
in the pure Chern-Simons theory at rank $N$ and level $k$ and $V_{2}$ 
is the proper volume of the spatial manifold ($S^2$ in this case). 
In other words the partition function of the matter Chern-Simons theory
may be re-expressed as a linear combination of Wilson loops
(in various representations of $U(N)$) that wind the time circle. The only
memory of the fundamental matter lies in the form of the function $v(U)$,
whose structure is determined by the fundamental matter content and
interactions.

Before turning to the important question of how $v(U)$ may be determined
in any given matter Chern-Simons theory, we pause to note an important
property of the formula \eqref{purecsci}. As \eqref{purecsci} is the
expectation value of (a sum of) Wilson loops in rank $N$ and (renormalized)
level $k$ Chern-Simons theory, the well established level rank duality of
{\it pure} Chern-Simons theory relates the expectation value in \eqref{purecsci}
to the expectation value of a dual operator in rank $k-N$ and level $k$
pure Chern-Simons theory.%
\footnote{If we use the unrenormalized Chern-Simons level $k_{YM}$ 
then the level rank duality states the invariance of {\it pure} Chern-Simons theory under the exchange of $k_{YM}$ and $N$.
}  (see \cite{Kapustin:2010xq,Kapustin:2010mh} for a relatively 
recent discussion of level rank duality and references to earlier work).
The specific relationship turns out to be the
following. Any gauge invariant function $v(U)$ may be regarded as a function
of the  variables $\Tr U^n$. For any such function we define a
corresponding dual function ${\tilde v}(U)$ by the equation
\begin{equation}\label{dualpot}
{\tilde v} \left( \tr U^n \right)=v\left( (-1)^{n+1} \tr U^n \right).
\end{equation}
Level rank duality of pure Chern-Simons theory turns out to imply that
\begin{equation}\label{lrdi}
\langle e^{-T^2 V_{2} v(U)} \rangle_{N,k}=\langle e^{-T^2 V_{2} {\tilde v}(U)} \rangle_{k-N,k}.
\end{equation}
\eqref{lrdi} will have interesting implications for matter Chern-Simons theories
as we will see below.

Let us now return to computation of $v(U)$ for any given Chern-Simons
matter theory. It is a remarkable fact that the function $v(U)$ appears
to be exactly computable as a function of $\lambda$ 
in the large $N$ limit for arbitrary
matter Chern-Simons theories. As $v(U)$ is an ultra local expression (i.e $v(U)$ depends only 
on the value of $U$ at a point and not its derivatives; similarly $v(U(x))$ is independent of the 
derivatives of the metric - i.e. curvatures - at the point $x$) it may be
computed on $R^2 \times S^1$ rather than $S^2 \times S^1$. Using standard
large $N$ techniques and employing an unusual lightcone gauge,
the authors of  \cite{Giombi:2011kc} were able to compute
$v(U)$ for the special case that $U=I$ as an arbitrary function of $\lambda$
for the theory of fundamental fermions minimally coupled to the Chern-Simons
field. This computation was later generalized to other theories in
\cite{Jain:2012qi},\cite{Oferb:2012}. 
It was also generalized to
the computation of $v[U]$ for a matrix $U$ whose eigenvalue distribution
is given by \eqref{evd} below (more below on why this is important) in
\cite{Aharony:2012ns}. A straightforward generalization of these computations
permits the evaluation of $v[U]$ for arbitrary $U$; our results for
$v[U, \lambda]$ are presented in section \ref{cvu} below for several 
examples of matter Chern-Simons theories. Pairs of the matter Chern
Simons theories we have studied in section \ref{cvu} below have been
conjectured to be related via level rank type dualities. In section
\ref{cvu} below we demonstrate that, in each case, the potentials $v[U]$
for conjecturally dual Chern-Simons matter theories are related by
\eqref{dualpot}. It follows from \eqref{lrdi} above
that $S^2$ partition functions for conjectural dual matter Chern-Simons
theory pairs are equal at every value of $\lambda$ and temperature.
Such dual pairs include  Giveon-Kutasov duals for a  chiral theory
\cite{Benini:2011mf}, minimally coupled fermions
and gauged critical bosons, as well as minimally coupled bosons with
gauged critical fermions \cite{Maldacena:2012sf,Aharony:2012nh,Aharony:2012ns,GurAri:2012is}, see also \cite{Giombi:2011kc} for a preliminary suggestion
for duality. Our results may be viewed as additional evidence
in support of these conjectured dualities.

In order to find explicit formulas for the partition function of 
Chern-Simons matter theories on $S^2$ we need to evaluate the expectation 
value \eqref{purecsci}. This may be achieved using path integral techniques.
In fact the partition function of Chern-Simons theories on $\Sigma_g \times S^1$
was evaluated by using path integral techniques long ago in the 
beautiful older paper \cite{Blau:1993tv}, the analysis in \cite{Blau:1993tv} 
is easily generalized to include the effect of $v(U)$. Briefly 
(see section \ref{pathinteva} for details) we follow \cite{Blau:1993tv}
and work in the `temporal' gauge  $\partial_3 A_3=0,$ (in this paper 
the Euclideanized time direction is $x^3$ so $A_{3}$ denotes temporal 
component of the gauge field).  We then abelianize the residual two dimensional 
gauge invariance by an additional gauge fixing condition; the 
holonomy matrix $U(x)=e^{ \beta A_3(x)}$ is required to be diagonal.
The integral over the off diagonal and Kaluza Klein modes of the 
spatial part of the gauge field is quadratic and yields a determinant, which 
turns out to largely cancel the Fadeev-Popov determinant of gauge fixing. 
The integral of the diagonal (unfixed abelian part) of the two dimensional 
gauge field yields a delta function that fixes $U(x)$ to be constant on 
$S^2$. And the summation of $U(1)^N$ flux sectors discretizes the eigenvalues 
of $U$ in units of $\frac{1}{2 \pi k}$. These manipulations reduce the 
expectation value \eqref{purecsci} to a `discretized' version of 
integral over the holonomy matrix $U$ \eqref{mint}
\begin{equation}\label{csmint}
Z_{\text{CS} }=\prod_{m=1}^N \sum_{n_m=-\infty}^\infty
\left[ \prod_{l \neq m} 2 \sin \left(\frac{\alpha_l(\vec n)-\alpha_m(\vec n)}{2}
\right)
e^{-V(U)}\right]
\end{equation}
where
\beal{
V(U) = T^2 V_{2} ~v(U)
\label{effectivepot}
}
and $U(\vec n)$ is the unitary matrix whose eigenvalues are
$e^{i \alpha_m(\vec n)}$, where $\alpha_m(\vec n) = \frac{2 \pi n_m}{k} (n_m \in \bZ)$ with $m$ running from $1$ to $N$. Notice that the discretization interval
between two allowed values of eigenvalues in \eqref{csmint}, 
$\frac{2 \pi}{k}$,  tends to zero in the 't~Hooft limit 
$k \to \infty, N \to \infty$. It follows that the eigenvalue density
function $\rho(\alpha)$ in discretized one \eqref{csmint} and 
non-discretized one \eqref{mint} obey identical
large $N$ saddle point equations upon identifying the two potentials. 
Nonetheless the saddle points of 
\eqref{csmint} and \eqref{mint} in the 't~Hooft limit are not always identical.
Recall that the classical large $N$ saddle point descriptions of
\eqref{csmint} and \eqref{mint} are written in terms of the
eigenvalue density function $\rho(\alpha)$ defined by
\begin{equation} \label{den}
\rho(\alpha)=\frac{1}{N} \sum_{m=1}^N \delta(\alpha -\alpha_m).
\end{equation}
As the summation in \eqref{csmint} effectively excludes terms
with coincident $n_m$'s,\footnote{In the case that the base manifold is an $S^2$, 
as studied in our paper, the exclusion of such configurations follows from 
the fact that the measure factor in \eqref{mint} eliminates their 
contributions. The generalization of \eqref{csmint} to the partition function of 
Chern-Simons theory on $\Sigma_g \times S^1$ where $\Sigma_g$ is a 
genus $g$ manifold of arbitrary metric is given by the formula \eqref{csmintg}. In these cases 
the measure factors is either constant (in the case of $g=1$) or diverges 
when two eigenvalues are equal. Nonetheless the correct prescription (the 
one that agrees with Chern-Simons computations using other techniques) 
appears to be to omit the contribution of such sectors. The justification 
for this prescription does not appear to be clearly understood from
first principle path integral reasoning. We hope to clear up this 
point in the future. We thank O. Aharony, S. Giombi and J. Maldacena 
for extensive discussions on this point.}  
it follows that the maximum number of eigenvalues in
an interval $\Delta
\alpha$ in the summation in \eqref{csmint}
is given by $\frac{\Delta \alpha}{\frac{2 \pi}{ k}}$, so that the
eigenvalue density \eqref{den} is bounded from above by
$\frac{k}{2 \pi} \times \frac{1}{N}=\frac{1}{
2 \pi \lambda}$.
In other words the eigenvalue distribution in
\eqref{csmint} is constrained to obey the inequalities
\begin{equation} \label{ineqali}
0 \leq \rho(\alpha) \leq \frac{1}{2 \pi \lambda}.
\end{equation}
On the other hand $\rho(\alpha)$ for the integral \eqref{mint} obeys 
the lower bound listed in \eqref{ineqali} (this is because
a density is an intrinsically positive quantity) but no upper 
bound.%
\footnote{
The same upper bound for the density function appears in two dimensional Yang-Mills theory ($p=1$), in which case the situation becomes similar due to the fact that there is no propagating degrees of freedom for the gauge field. 
A phase transition relevant to this upper bound was studied in 2d (q-deformed) Yang-Mills theory on $S^2$ \cite{Douglas:1993iia,Arsiwalla:2005jb,Caporaso:2005ta,Jafferis:2005jd}, which we will see from Chern-Simons theory below. We noticed these relevant papers when we were completing this paper.
} Note that the upper bound in \eqref{ineqali} agrees perfectly with the 
conclusions of \cite{Aharony:2012ns}, obtained using Hamiltonian methods for  
Chern-Simons theory on $T^2$.

Any saddle point of \eqref{mint} that happens to
everywhere obey the inequality \eqref{ineqali} is also
a saddle point of \eqref{csmint}. However saddle points of
\eqref{mint} that anywhere violate the upper bound \eqref{ineqali}
are not large $N$ solutions of \eqref{csmint}. Instead
\eqref{csmint} admits new classes of solutions; those that
saturate the upper bound of the inequality \eqref{ineqali} over one
or more arcs along the unit circle. While saddle points of
\eqref{mint} may be classified in terms of the number of gaps in the
solution (i.e. the number of arcs over which the eigenvalue
density vanishes), saddle points  of \eqref{csmint} are classified by
the number of `lower gaps' (regions over which the eigenvalue density
vanishes) together with the number of `upper gaps'
(arcs over which the upper bound of
\eqref{ineqali} are saturated).

In order to understand the generic phase structure of matrix model \eqref{csmint}
we found it useful to first study a toy model in which $V(U)$ takes the 
simple Gross -Witten -Wadia form 
$V(U)=-{N \zeta \over 2} (\Tr U+ \Tr U^\dagger)$.
We have found the exact solution to the saddle point equations of this 
toy model. The phase structure of this solution 
is depicted in Fig.\ref{Phase} (see \S\ref{summary} for a a more 
detailed summary of this solution); we pause to elaborate on this phase 
diagram. In this toy model it turns out that for 
$\lambda < \lambda_c=\frac{1}{2}$ the eigenvalue distribution has no gaps for 
$\zeta<1$, a single lower gap for $1<\zeta<\frac{1}{4 \lambda^2}$ and 
a lower gap plus an  upper gap for $\zeta > \frac{1}{4 \lambda^2}$.
This sequence of eigenvalue distributions is depicted graphically in
Fig.~\ref{nocut},\ref{gwwfig}, \ref{twocut}, \ref{twocut2}.
For $\lambda>\frac{1}{2}$,
on the other hand, the eigenvalue distribution has no gaps for
$\zeta < \frac{1}{\lambda}-1$, a single upper gap
in the range $\frac{1}{\lambda}-1< \zeta < \frac{1}{4 \lambda(1-\lambda)}$, 
and an upper plus a lower gap for still larger $\zeta$. This phase sequence 
is depicted in Figure \ref{lbig1}, \ref{lbig2}, \ref{lbig3}, \ref{lbig4} 
in \S\ref{summary}. Moreover, in the large $\zeta$ limit the eigenvalue 
distribution tends to the universal configuration
\begin{equation}\begin{split} \label{evd}
\rho(\alpha)&=\frac{1}{2 \pi \lambda} ~~~(|\alpha|< \pi \lambda) \\
&= 0 ~~~~~~~(|\alpha|>\pi \lambda)
\end{split}
\end{equation}
at every value of $\lambda<1$. The distribution \eqref{evd} is 
the nearest thing to a $\delta$ function permitted by the effective 
Fermi statistics of the eigenvalues in perfect agreement with the results of 
the \cite{Aharony:2012ns} which was the inspiration for the current work.

For any given matter Chern-Simons theory $V(U)$ is more 
complicated than the toy model of the previous paragraph. 
Even for these more complicated functions $V(U)$, however, it turns 
out to be easy to solve exactly for the eigenvalue density function in the 
`no gap' phase (see section \ref{est} for details). The sequence of 
phases described above for the toy model is 
also true for every particular CS matter theory we have studied. 
Our exact solution allows us to determine the first phase transition 
temperature as a function of $\lambda$ as well as the value of $\lambda_c$
in the theories we have studied (the value of $\lambda_c$ varies from theory 
to theory). On general grounds we expect the eigenvalue distribution to 
tend to the universal function \eqref{evd} in the high temperature limit 
in all Chern-Simons matter theories, even though we have not (yet) found 
exact solutions to the relevant matrix models in the two gap phase.\footnote{
Note that the case $\lambda=1$ is special. In this case the number of distinct
allowed `slots' for the eigenvalues (number of distinct allowed values
of $n_m$) is exactly equal to $N$. As a consequence the eigenvalue
distribution at $\lambda=1$ is given by
\begin{equation}\label{ffs}
\rho(\alpha)=\frac{1}{2 \pi}
\end{equation}
at all values of the temperature. In this strong coupling limit
the holonomy distributes uniformly around the circle.
}

We pause to describe an important property of our solution to the matrix model obtained
from real Chern-Simons theories in the no gap phase at low temperature 
(for notational simplicity we 
work on a round sphere of volume $V_2= 4 \pi$ 
in the rest of this paragraph). 
In the low temperature phase the  partition function takes the form
$$ \ln Z= N^2 f\left(\frac{V_2 T^2}{N}, \lambda \right).$$ 
In every theory we have studied, the 
Taylor expansion of the function $f(\frac{T^2}{N})$ takes form 
\begin{equation}\label{ft}
f=a\frac{T^4}{N^2} + b(\lambda) \frac{T^8}{N^4} + \ldots
\end{equation}
where $a$ is a constant independent of $\lambda$. It follows that at very 
low temperatures 
$$ \ln Z= a T^4 \left ( 1+ {\cal O}(T^4/N^2) \right).$$ 
This is precisely the form of the partition function of a gas of 
multi `traces'; the value of the coefficient $a$ also works out to 
the value predicted by this expectation. The fact that the free energy 
is independent of $\lambda$ at temperatures $T^2 \ll N$ 
is in perfect agreement with the non renormalization theorem of the 
spectrum of single trace operators in these 
theories. In other words  the results for the
partition function obtained in this paper at small $\frac{T^2}{N}$ 
matches smoothly onto the high temperature expansion of the 
$\lambda$ independent ${\cal O} (N^0)$ free energy. 

We have explained above that the eigenvalue distributions of two
theories related by level rank type dualities are related by the map
$$ \tr U^n \leftrightarrow (-1)^{n+1} \tr U^n. $$
This relationship implies that the saddle point eigenvalue densities
of two dual theories are related by the formula
$${\tilde \rho}(\alpha)= \frac{1}{2 \pi (1-\lambda)} -
\frac{\lambda}{1-\lambda} \rho(\alpha+\pi)$$
where $\rho(\alpha)$ is the eigenvalue distribution of the theory with coupling
$\lambda$ while ${\tilde \rho}(\alpha)$ is the eigenvalue distribution of
the theory with coupling $1-\lambda$. An interesting feature of this
map is that it interchanges lower and upper gaps, consistent with the
fact that the first phase transition is always from the no gap to the lower gap
phase at small $\lambda$ while it is from the no gap to the upper gap
phase at $\lambda$ near unity.

We end this introduction on a cautionary note. All of the results of 
this paper have been obtained starting from the effective action 
\eqref{purecsi}. While the arguments for this effective action presented 
in \cite{Aharony:2012ns} and reviewed in subsection \ref{hta} below are 
persuasive, they are not completely beyond question. It would be useful 
to present a clear justification of \eqref{purecsi} 
and understand how to compute 
corrections to the results from 
this action (such corrections are presumably suppressed in $\frac{1}{N}$). 
It would also be useful to better understand the exclusion of equal 
eigenvalues from \eqref{csmint}.
We feel that it is important 
to clear up these two issues before we can regard the large number of 
exact results presented in this paper as beyond reproach. We hope to 
return to these issues in the future.

\section{Effective action and level rank duality}\label{pi}

In this section we review the effective description of large $N$ 
Chern-Simons theories at temperature of order $\sqrt{N}$ 
proposed in \cite{Aharony:2012ns}. This effective description allows us 
to relate the partition function of any fundamental 
matter Chern-Simons theory, in an appropriate high temperature limit, 
to the expectation value of an appropriate `Wilson loop' in the {\it pure}
Chern-Simons theory. Using this fact we then explore the implications of 
level rank duality of Wilson loops in pure Chern-Simons theories for the 
partitions under study.

\subsection{The high temperature effective action} \label{hta}

Consider a Chern-Simons-fundamental matter theory on $S^2\times S^1$ in the 
't~Hooft large $N$ limit. As described in the introduction, we study a theory 
at very high temperatures with 
\begin{equation}\label{tsc}
V_{2}T^2= N \zeta
\end{equation}
 where $\zeta$ is held fixed in the large $N$ limit. 

Our strategy for evaluating the partition function is, roughly speaking, 
first to find an effective action for the two dimensional holonomy 
field, $U(x)$, around the thermal $S^1$ (here $x$ is a point on the 
base $S^2$) by integrating out all other fields, and then 
to perform the path integral over two dimensional unitary matrix valued field 
$U(x)$. In practice we find it more convenient to break up the first 
step (computation of the effective action of $V(U)$) into two steps 
as we now explain.

The effective action for $U(x)$ is obtained by summing all vacuum graphs 
in which all non-holonomy fields appear in an unrestricted manner. 
We will find it useful to break up these graphs into those that involve 
at least one matter field, and those that do not. Let the result of 
summing over all graphs of the first sort be denoted by $e^{-S_{eff}[U(x)]}$. 
Once we have $S_{eff}(U)$, the remaining task is to sum over all graphs 
that do not involve a matter field; and then integrate over $U(x)$.\footnote{Our conventions of the gauge field are those of \cite{Witten:1988hf}.
That is, the gauge field $A_\mu$ takes values of antihermitian matrix, the gauge covariant derivative is $D_\mu=\partial_\mu+A_\mu$ and its field strength is 
$F_{\mu\nu}=[D_\mu, D_\nu]=\partial_\mu A_\nu-\partial_\nu A_\mu +[A_\mu,A_\nu]. $
The gauge transformation is $A_\mu  \rightarrow A_\mu -[D_\mu, \epsilon]$, where $\epsilon$ is a gauge parameter taking values of antihermitian matrix. 
The Chern-Simons action is 
\beal{
\frac{k}{4\pi} \int \Tr \left( AdA + \frac{2}{3} A^3 \right) = 
\frac{k}{4\pi} \int d^3 x \Tr \epsilon^{\mu\nu\rho} \left(A_\mu \partial_\nu A_\rho  +  \frac{2}{3} A_\mu A_\nu A_\rho \right)
} where $\epsilon^{123}=1.$
Under an infinitesimal change of $A$ the action above changes as $\frac{k}{2\pi} \Tr \int \delta A \wedge F$.
} In 
other words the full partition function is given in terms of $S_{eff}(U(x))$
by the formula
\begin{equation}\label{purecsa}
Z_{\text{CS}}=\int DA 
e^{i \frac{k}{4\pi} \Tr  \int \left( AdA + \frac{2}{3} A^3 \right)
 -S_{eff}(U) }.
\end{equation}

While the formula \eqref{purecsa} is always correct in principle, it is 
useful in practice only when $S_{eff}(U)$ is a local function. The key 
observation of  \cite{Aharony:2012ns} is that $S_{eff}(U)$ is in fact 
ultralocal in the leading large $N$ limit, when the temperature is scaled 
as in \eqref{tsc}. At leading order in large $N$, in other words 
the effective action 
\begin{equation}\label{uls}
S_{eff}(U)=T^2 \int d^2x \sqrt g  ~ v(U)
\end{equation}
(the power of temperature has been inserted on dimensional grounds, see 
below for more details) and the formula for the partition function are
simplified to 
\begin{equation}\label{purecs}
Z_{\text{CS}}=\int DA e^{i \frac{k}{4\pi} \Tr  \int \left( AdA + \frac{2}{3} A^3 \right)
 -T^2 \int d^2x \sqrt g  ~ v(U)  }.
\end{equation}
The path integral in \eqref{purecs} is evaluated in Euclidean space and the 
temporal circle 
is compact with circumference $\beta$. 

The argument for the locality of $S_{eff}$ is essentially 
that the graphs that we integrate out to find $S_{eff}$ always 
include a matter propagator, and all matter fields develop 
large thermal masses. It appears to follow that $S_{eff}(U)$
can be expanded in a series of local operators in the form 
\begin{equation}\label{ltdf}
S_{eff} = \int d^2x \left( T^2 v(U)   + v_1(U) \Tr D_i U D^i U        
+ \ldots \right) 
\end{equation}
where we have inserted powers of $T$ by dimensional analysis.
Now the scaling \eqref{tsc} \footnote{We are emphasizing that the 
$S^2$ partition function of 
fundamental matter -Chern-Simons theories are trivial (and $\lambda$ 
independent) at temperatures of order unity. These partition functions have
interesting (and generically $\lambda$ dependent) structure only when 
the temperature is scaled as in \eqref{tsc}. }
converts the high temperature expansion \eqref{ltdf} into an expansion 
in inverse powers of $N$. For instance the terms listed above are 
respectively of order $N^2$, and $N$.
\footnote{Every term in the effective 
action \eqref{ltdf}  is of order $N$ at fixed $T$ (powers of $N$ are counted as explained in the 
introduction), as the action is generated by integrating out fundamental 
fields.} Terms that we have ignored in this
expansion (the $\ldots$ in \eqref{ltdf}) are further suppressed at large
$N$. At leading order in the large $N$ expansion it seems to be justified to 
truncate \eqref{ltdf} to the first term, yielding \eqref{purecs}.
\footnote{\cite{Aharony:2012ns} considered a slightly different effective 
action, one in which they integrated out all graphs of 
matter fields, Kaluza Klein modes of the 
gauge fields, but not the massless 2d gauge field. They expanded their 
effective action in the form
\begin{equation}\label{ltdf-aharony}
S^W_{eff} = \int d^2x \left( T^2 v_W(U)   + v_1(U) \Tr D_i U D^i U     
+ v_2(U) \epsilon^{ij} \Tr F_{ij}   
+ \frac{1}{T^2} v_3(U) \Tr F_{ij}F^{ij}   
+ \ldots \right); 
\end{equation}
once again all terms but the first are sub leading in the $\frac{1}{N}$ 
expansion. It would be interesting to carefully investigate the 
relationship between $S_{eff}$ and $S^W_{eff}.$}

The arguments leading up to \eqref{purecs} could conceivably have loopholes; 
in order to gain more confidence in \eqref{purecs} it would be useful to 
estimate the contributions of the neglected terms in \eqref{ltdf} to the 
partition function and verify that their contribution is indeed sub leading 
in the $\frac{1}{N}$ expansion. We leave this exercise to future work.

In order to make \eqref{purecs} a concrete formula we must explain how the 
function $v(U)$ may be computed for any particular vector matter Chern 
Simons theory. We take this question up in the next section, and 
present exact and reasonably explicit  formulas - as a function of 
the 't~Hooft coupling $\lambda$  for $v(U)$ for sample matter Chern-Simons 
theories. In the rest of this section, however, we first explore a 
structural `duality' of the formula \eqref{purecs}.

\subsection{Relations from level rank duality of pure Chern-Simons}

The equation \eqref{purecs} may be rewritten as 
\begin{equation}\label{purecsb}
Z_{\text{CS}}=\langle e^{-T^2 \int d^2x \sqrt g  ~ v(U(x))  } \rangle_{N,k}
\end{equation}
where the symbol 
$$\langle \Psi \rangle_{N,k}$$
denotes the expectation value of $\Psi$ in {\it pure} $U(N)$ Chern-Simons 
theory at level $k$. As pure Chern-Simons theory is topological, 
an expectation value of the form
$$\langle e^{v(U(x))} \rangle_{N,k}$$
is independent of $x$. It follows that \eqref{purecsb} may be rewritten as 
\begin{equation}\label{purecsc}
Z_{\text{CS}}=\langle e^{-T^2 V_{2} v(U)} \rangle_{N,k}.
\end{equation}
Recall that $v(U)$ is a gauge invariant function of the holonomy matrix. 
Any such function may be expanded in a basis of characters of $U(N)$, 
i.e. in a basis of Polyakov or Wilson loops, in arbitrary representations 
of the gauge group $U(N)$, around the curve that winds the thermal circle. 
In other words the remarkable formula \eqref{purecsc} relates the 
thermal partition function of fundamental matter Chern-Simons theories 
(in an appropriate coordinated high temperature and large $N$ limit) 
to the expectation value of a complicated thermal 
Wilson loop expectation value in {\it pure} Chern-Simons theory.

Now pure Chern-Simons theories of rank $N$ and renormalized level $k$ are 
dual to pure Chern-Simons theories of rank $k-N$ and level $k$ 
via the well known level rank duality. Wilson loops transform under this 
duality as follows: a Wilson loop in the representation labeled by 
the Young Tableaux $Y$ maps to the Wilson loop in the representation 
labeled by Young Tableaux ${\tilde Y}$ where $Y$ and 
${\tilde Y}$ are related by `transposition' 
(the interchange of rows with columns). This duality 
map for Wilson loops may be restated in a slightly simpler fashion for
arbitrary gauge invariant functions (like $e^{V(U)}$) of the holonomy 
matrix $U$ as we now explain.

The Wilson loop in any representation of $U(N)$ is expressed 
as a polynomial of the variables $\Tr U^n$ ($n \leq N$) by 
the character polynomials of group theory. 
Let $\chi_Y(U)$ denote the character polynomial of the $U(N)$ 
representation with Young Tableaux $Y$. Let $n$ denote the 
number of boxes in the  Young Tableaux $Y$. The character 
polynomial of $\chi_Y(U)$ is given by the so called 
Schur Polynomial (see e.g. eq (2.15) 
of \cite{Corley:2001zk}) 
\begin{equation}\label{sch}
\chi_Y(U)=\frac{1}{n!} \sum_{\sigma \in S_n} \chi_Y(\sigma)
\left( \prod_{m=1}^n(\Tr U^m)^{k_m} \right)
\end{equation}
where the summation on the RHS runs over all elements $\sigma$ of the 
permutation group $S_n$, $\chi_Y(g)$ is the character of the permutation 
group element $g$ in the representation labeled by the Young Tableaux 
$Y$ and $k_1$, $k_2 \ldots k_n$ denote the number of cycles of length 
$1, 2, \ldots n$ in the conjugacy class of the permutation element $g$. 
We now recall that characters of the permutation group have a well 
known transformation property under transposition of the Young Tableaux
\begin{equation}\label{tyt}
\chi_{\tilde Y}(g)=sgn(g) \chi_{Y}(g)
\end{equation}
where $sgn(g)$ is the sign of the permutation $g$
(see e.g. \cite{Musili}). 
The sign of the permutation $g$ is given in terms of its number of cycles 
$k_n$ by 
\begin{equation}\label{sgnperm}
sgn(g)=\prod_{m=1}^n \left[(-1)^{m+1} \right]^{k_m} .
\end{equation}
Plugging \eqref{tyt} and \eqref{sgnperm} into \eqref{sch} we find 
\begin{equation}\label{schp}
\chi_{\tilde Y}(U)=\frac{1}{n!} \sum_{\sigma \in S_n} \chi_Y(\sigma)
\left( \prod_{m=1}^n((-1)^{m+1}\Tr U^m)^{k_m} \right).
\end{equation}
In other words transposition of Young Tableaux in {\it all} 
character polynomials (i.e. in arbitrary gauge invariant functions of 
$U$) is achieved by the uniform interchange%
\footnote{For the reader who dislikes appealing to mathematical authority 
for proofs, we present a `physics' check of \eqref{schp}. 
Let $\chi_k$ denote the character polynomial of the representation
with $k$ boxes in the first row of the Young Tableaux and no boxes in any 
other row. In a similar manner let $\psi_k$ denote the character polynomial 
of the representation with $k$ boxes in the first column of the Young Tableaux 
and no boxes in any other columns. Then it follows from the usual formulas 
of Bose and Fermi statistics that 
\begin{equation}\begin{split}\label{bfs}
&\sum_{k=0}^\infty \chi_k x^k =\exp \left[
\sum_{m=1}^\infty \frac{ \Tr U^m x^m}{m} \right]\\
&\sum_{k=0}^\infty \psi_k x^k =\exp \left[
\sum_{m=1}^\infty \frac{(-1)^{m+1} \Tr U^m x^m}{m} \right]\\
\end{split}
\end{equation}
where $x$ is any real number. Level rank duality interchanges
$\psi_k$ and $\chi_k$. At the level of the generating functions in \eqref{bfs}
this interchange is simply achieved by \eqref{interchange}.} 
\begin{equation}\label{interchange}
\Tr U^n \leftrightarrow (-1)^{n+1} \Tr U^{n}.
\end{equation}
It follows that
level rank duality of pure Chern-Simons theory asserts that 
\begin{equation}\label{lrdd}
\langle e^{V( \Tr U^n)} \rangle_{N,k}=\langle e^{V\left((-1)^{n+1} \Tr U^n \right)} \rangle_{k-N,k}
\end{equation}
for  any gauge invariant function $V(U)$. 

In the large $N$ limit, it is sometimes more convenient to regard $V(U)$ as 
a functional of the eigenvalue distribution of $\rho(\alpha)$ than as 
a function of $\Tr U^n$. Recall that $\Tr U^n = N \rho_n$ 
where $\rho_n$ is the $n^{th}$ Fourier mode of the eigenvalue density function 
$\rho(\alpha)$. It follows that the interchange \eqref{interchange} 
amounts to 
\begin{equation}\label{edint}
N \rho_n =(k-N)(-1)^{n+1} {\tilde \rho}_n.
\end{equation}
Where ${\tilde \rho}_n$ are the Fourier modes of the holonomy in the CS theory 
with rank $k$.
\eqref{edint} implies 
\begin{equation}\label{rhot}
{\tilde \rho}_n=(-1)^{n+1} \frac{N}{k-N}\rho_n=
(-1)^{n+1} \frac{\lambda}{1-\lambda} \rho_n
\end{equation}
(for $n\neq 0$). For $n=0$ we define ${\tilde \rho}_0=\rho_0=1$. 
It follows that 
\begin{equation}\label{evdex} 
{\tilde \rho}(\alpha)=
\sum_n \frac{{\tilde \rho}_n}{2 \pi} e^{i n \alpha}
=\frac{1}{2 \pi (1-\lambda)} - \frac{\lambda}{1-\lambda} 
\rho(\alpha+\pi)
\end{equation}
or 
\begin{equation}\label{evdexc} 
(1-\lambda){\tilde \rho}(\alpha)+\lambda \rho(\alpha+\pi)
=\frac{1}{2 \pi}.
\end{equation}
\eqref{lrdd} is thus equivalent to the assertion that  
\begin{equation}\label{lrdde}
\langle e^{V[\rho]} \rangle_{N,\lambda}=
\langle e^{{\tilde V}[\rho]} \rangle_{k-N,1-\lambda}
\end{equation}
where the functional ${\tilde V}$ is defined via the relationship 
\begin{equation}\label{tvd}
{\tilde V}[\rho]= V[{\tilde \rho}]
\end{equation}
for arbitrary functions $\rho(\alpha)$. 

\subsection{Implications for Chern-Simons matter theories}

Let us label the class of fundamental matter Chern-Simons theories by an 
abstract label $a$. As we have emphasized above, the partition function of 
such theories at rank $N$ and level $k$ is given by an expression of 
the form 
$$\langle e^{V^a_{N, \lambda}[\rho]} \rangle_{N, \lambda}$$
where the index $a$ emphasizes that  $V^a_{N, \lambda}[\rho]$ depends on the 
particular Chern-Simons matter theory under study. 

Now suppose it happens to be true that for two different 
CS matter theories, theory $a$ and theory 
$b$
\begin{equation}\label{Veq}
V^a_{N, \lambda}[\rho]={\tilde V}^b_{k-N, 1-\lambda}[\rho].
\end{equation} 
It then follows from \eqref{lrdde} that the partition function of theory 
$a$ at rank $N$ and level $k$ is identical to the partition function 
of theory $b$ at rank $k-N$ and level $k$. The identity of partition 
functions holds at all temperatures (of order $\sqrt{N}$) on $S^2$
and also on arbitrary genus Riemann surfaces. 

In the next section we will 
find examples of simple Chern-Simons matter theories that obey \eqref{Veq},
strongly suggesting (previously conjectured) level rank type dualities 
between the relevant non-topological matter Chern-Simons theories. 

\cite{Kapustin:2010xq,Kapustin:2010mh} had previously supplied evidence 
for several supersymmetric Giveon-Kutasov theories by demonstrating that 
the $S^3$ partition functions of these theories - as computed by supersymmetric 
localization -  are in fact equal. In fact the equality of $S^3$ partition functions 
was demonstrated in \cite{Kapustin:2010xq,Kapustin:2010mh} by relating these partition
functions to the expectation value of unknotted Wilson loops
in pure Chern-Simons theory, and then demonstrating the equality of the resulting 
expressions using level rank duality of pure Chern-Simons theory. It is interesting 
that a similar mechanism appears to ensure the equality of {\it thermal} partition
functions of the dual theories (at least at high temperature), even though the thermal 
partition function (unlike the $S^3$ partition function) is far from a topological quantity.
This suggests that it may be possible to prove duality between matter Chern-Simons theories 
starting from the level rank duality of pure Chern-Simons theory. We leave further development
of this idea to future work.

\section{Computation of $v(U)$}\label{cvu}

In the previous section we explained how the high temperature large $N$ 
limit of the partition function for {\it any} fundamental matter Chern 
Simons theory takes the form \eqref{purecsc} for some function $v(U)$. 
The form of the function $v(U)$, of course, depends on the specific 
Chern-Simons fundamental matter theory under study. In this section
we present exact results for the potential function $v(U)$, as a function 
of the 't~Hooft coupling $\lambda$, for several simple Chern-Simons matter 
theories. We then go onto observe that these potential functions happen 
to be related by \eqref{Veq} for three pairs of matter Chern-Simons 
theories that had previously been conjectured to be related via level 
rank type dualities. Our results may be regarded as further evidence 
for both these dualities as well as the correctness of the effective 
description \eqref{purecs}.

We have obtained our results for $v(U)$ by making only minor modifications 
to several existing computations in the literature. In order to explain this 
we pause to present a brief review of the recent relevant literature on 
fundamental matter Chern-Simons theories. Such theories at nonzero $\lambda$ 
have recently received serious attention, starting with the papers 
\cite{Giombi:2011kc, Aharony:2011jz} and subsequently \cite{Banerjee:2012gh,Jain:2012qi,Yokoyama:2012fa}. In particular  the authors of 
\cite{Giombi:2011kc} worked in a lightcone gauge to present an exact, 
all orders computation of the infinite volume finite temperature 
partition function of the fundamental fermion Chern-Simons theory
in the background of the identity holonomy matrix.\footnote{The 
authors of \cite{Giombi:2011kc} (and extensions to other theories 
\cite{Giombi:2011kc,Jain:2012qi,Oferb:2012}) proposed 
that the partition function evaluated in these works should be identified 
with the thermal partition function of the relevant theories. This 
identification was later realized to be inconsistent  
\cite{Jain:2012qi},\cite{Oferb:2012} with Giveon-Kutasov duality for a  
chiral theory \cite{Benini:2011mf}. The very beautiful recent paper 
\cite{Aharony:2012ns} pointed out that this identification 
\cite{Giombi:2011kc,Jain:2012qi,Oferb:2012} is incorrect. They   
used the Hamiltonian methods of \cite{Douglas:1994ex} to explain  
that the $R^2$ thermal partition function of matter Chern-Simons theories is 
given by the partition function of the relevant theories 
in the holonomy \eqref{evd}. This proposal yields results consistent 
with Giveon Kutasov duality, as well as a new non supersymmetric version 
of this duality conjectured and tested in 
\cite{Maldacena:2012sf,Aharony:2012nh,GurAri:2012is}.  }
The recent paper \cite{Aharony:2012ns} pointed out that the computation 
of \cite{Giombi:2011kc} is easily generalized to another holonomy 
background; \cite{Aharony:2012ns} repeated the computation of 
\cite{Giombi:2011kc} (and generalizations) in the holonomy background 
\eqref{evd}. In fact the partition function computation of \cite{Giombi:2011kc}
may easily be repeated in an arbitrary holonomy background. In this 
paper we propose that the result of this computation may be identified with 
$T^2 V_{2} v(U)$. This proposal allows us to explicitly compute $v(U)$. 
The computations that yield $v(U)$ are a straightforward generalization
of the work of \cite{Giombi:2011kc,Jain:2012qi,Oferb:2012,Aharony:2012ns}.
For this reason we do not describe the method employed to evaluate 
$v(U)$ but simply present our final results below.

\subsection{CS theory coupled to fundamental bosons with classically
marginal interactions}\label{marginalbos}

In this subsection we study the Chern-Simons matter theory coupled to massless
fundamental bosons with a $ \phi^6$ interaction. The Lagrangian 
of the theory we study is presented in equation (4.2) of \cite{Aharony:2012ns}.

We find 
\begin{equation}\label{rbosfree}
\begin{split}
v[\rho]&=-\frac{N}{6\pi}\sigma^3 (1+\frac{2}{\hat\lambda (\lambda, \lambda_6)})
+ \frac{N}{2\pi}
\int_{-\pi}^{\pi} d\alpha \rho(\alpha) 
\int_\sigma^\infty dy~y \left(\ln(1-e^{-y+i\alpha})+\ln(1-e^{-y-i\alpha})\right),
\end{split}
\end{equation}
where 
\begin{equation}\label{dljh}
\hat\lambda(\lambda, \lambda_6)=\sqrt{\frac{\lambda_6}{8\pi^2}+\lambda^2}.
\end{equation}
The potential $v[\rho]$ above is determined in terms of the constant 
$\sigma$ that has a simple interpretation; it determines the thermal 
mass of the bosonic field via the equation 
\begin{equation}
\Sigma_{B}=\sigma^2 T^2.
\end{equation} 
The value of $\sigma$ is determined by the requirement that it extremizes 
$v[\rho]$ at constant $\rho$,  i.e. $\sigma$ is required to obey the equation 
\begin{equation}\label{rboself}
\begin{split}
  \sigma &=-\frac{1}{2}\sqrt{\frac{\lambda_6}{8\pi^2}+\lambda^2} 
\int_{-\pi}^{\pi} d\alpha \rho(\alpha)     \left(\ln 2\sinh(\frac{\sigma- i \alpha}{2})+\ln 2\sinh(\frac{\sigma+ i \alpha}{2})\right)\\
&=-\frac{1}{2}\frac{\hat\lambda(\lambda, \lambda_6)}
{\lambda}\int_{-\pi}^{\pi} d\alpha \lambda \rho(\alpha)     \left(\ln 2\sinh(\frac{\sigma- i \alpha}{2})+\ln 2\sinh(\frac{\sigma+ i \alpha}{2})\right).     
\end{split}
\end{equation}
The expressions presented above agree with previous computations in the 
literature for special choices of $\rho(\alpha)$. Note in particular 
that, \eqref{rboself} reduces to Eq.(4.22) of \cite{Aharony:2012ns} 
upon replacing the arbitrary density function in 
 \eqref{rboself} with \eqref{evd}. Similarly \eqref{rbosfree} reduces to 
Eq.(4.33) of \cite{Aharony:2012ns} upon making the same replacement.

\subsection{CS theory minimally coupled to Fermions}\label{minimalfer}

In this subsection we study the Chern-Simons matter theory coupled to massless
fundamental fermions. The Lagrangian of the theory we study is presented in 
equation (2.1) of \cite{Giombi:2011kc}. We find
\begin{equation}\label{freefer}
 \begin{split}
 v[\rho]&=-\frac{N}{6\pi}\left(\frac{\tilde c^3}{\lambda}-\tilde c^3
+3 \int_{-\pi}^{\pi} d \alpha \rho(\alpha) \int_{\tilde c}^{\infty}dy ~ y
(\ln(1+e^{-y-i\alpha})+\ln(1+e^{-y+i\alpha}))\right), 
 \end{split}
\end{equation}
where ${\tilde c}$ once again determines the thermal mass of the fermions.%
\footnote{As in the bosonic case 
$\Sigma_T = {\tilde c}^2 T^2$ is the thermal mass of the 
fundamental fermions. More precisely, the fermionic self energy is given by 
$\Sigma_{T}(p)=f(\beta p_s)p_s I+i p^{-}g(\beta p_s)\gamma^{-}$, where 
\begin{equation}\label{selffer}\begin{split}
f(y)&=\frac{\lambda}{y}\int_{-\pi}^{\pi} d\alpha~\rho(\alpha)\left(\ln 2\cosh(\frac{\sqrt{y^2+{\tilde c}^2}+i \alpha}{2})+\ln 2\cosh(\frac{\sqrt{y^2+{\tilde c}^2 }-i \alpha}{2})\right)\\
g(y)&=\frac{{\tilde c}^2}{y^2}-f(y)^2.\\
\end{split}
\end{equation} }
The value of ${\tilde c}$ is obtained by extremizing $v[\rho]$ w.r.t 
${\tilde c}$ at constant $\rho$, i.e. ${\tilde c}$ obeys the equation
\begin{equation}\label{tc}
\tilde c=\lambda \int_{-\pi}^{\pi} d\alpha~\rho(\alpha)\left(\ln 2\cosh(\frac{\tilde c+i \alpha}{2})+\ln 2\cosh(\frac{\tilde c-i \alpha}{2})\right).
\end{equation}
The expressions presented above agree with previous computations in the 
literature for special choices of $\rho(\alpha)$. 
Upon setting $\rho(\alpha)=\delta(\alpha),$ \eqref{selffer} and \eqref{tc} 
reduce to Eq.(2.80), Eq.(2.74) of \cite{Giombi:2011kc} while \eqref{freefer}
reduces to Eq.(2.92) of \cite{Giombi:2011kc}. On the other hand, on setting $\rho(\alpha)$ to 
the expression given in \eqref{evd},  
\eqref{tc} and \eqref{selffer} reduces to Eq.(5.13), Eq.(5.14) and Eq.(5.15) of 
\cite{Aharony:2012ns} (upon setting $\tilde \sigma=0$ where ${\tilde \sigma}$ 
is defined in that paper ) while \eqref{freefer} reduces to Eq.(5.28) of 
\cite{Aharony:2012ns} in appropriate limit as discussed above.

\subsection{Chern-Simons coupled to critical boson}\label{cccb}

In this subsection we study the Chern-Simons matter theory coupled to massless
critical bosons in the fundamental representation, i.e. a Chern-Simons 
gauged version of the $U(N)$ Wilson Fisher theory. The Lagrangian of the UV
theory is simply that of massless minimally coupled fundamental bosons 
deformed by the interaction 
\begin{equation}
 \delta S=\int d^{3}x A \bar\phi\phi,
\end{equation}
where $A$ is a Lagrange multiplier field (see 
subsection 4.3 of \cite{Aharony:2012ns} for details, and in particular 
eq.(4.35) for the Lagrangian; note \cite{Aharony:2012ns} employs the symbol 
$\sigma$ for our field $A$). We find 
\begin{equation}\label{vcrit}
 v[\rho]=-\frac{N }{6\pi}\left(\sigma^3 +\frac{2 (\sigma^2-A \beta^2)^{\frac{3}{2}}}{\lambda}\right) + \frac{N}{2\pi}\int_{\sigma}^{\infty} y dy \int_{-\pi}^{\pi}\rho(\alpha)d\alpha \left(\ln(1-e^{-y+i\alpha})+\ln(1-e^{-y-i\alpha})\right). 
\end{equation}
The squared thermal mass of our system is $\sigma^2 T^2$. The constants 
$A$ and $\sigma$ above are determined by the requirement that they 
extremize $v[\rho]$ at constant $\rho$.
This requirement yields the equations 
\begin{equation}\label{saddle1}
 A=\sigma^2 T^2,
\end{equation}
and
\begin{equation}\label{selfboscri}
  \int_{-\pi}^{\pi} \rho(\alpha)     \left(\ln 2\sinh(\frac{\sigma- i \alpha}{2})+\ln 2\sinh(\frac{\sigma+ i \alpha}{2})\right)=0.
\end{equation}
The expression for $v[\rho]$ simplifies somewhat upon inserting 
\eqref{saddle1} into \eqref{vcrit}; we find 
\begin{equation}\label{vcritb}
 v[\rho]=-\frac{N }{6\pi}\sigma^3  + \frac{N}{2\pi}\int_{\sigma}^{\infty} dy\int_{-\pi}^{\pi} d\alpha~y \rho(\alpha) \left(\ln(1-e^{-y+i\alpha})+\ln(1-e^{-y-i\alpha})\right), 
\end{equation} 
where $\sigma$ is obtained by extremizing \eqref{vcritb}, i.e. 
it is determined by the equation \eqref{selfboscri}.
Notice that neither \eqref{vcritb} nor \eqref{selfboscri} depend on the 
't~Hooft coupling $\lambda$. It follows that $v[\rho]$ is independent of 
$\lambda$ as well as $\lambda_6$ (this was not the case in the previous two subsections).

The expressions presented above reduce to formulas that have previously 
appeared in the literature for special values of $\rho$. In particular 
for $\rho$ given by \eqref{evd}, \eqref{vcritb} reduces to 
Eq.(4.41) of \cite{Aharony:2012ns}, while \eqref{selfboscri} reduces to 
Eq.(4.42) of \cite{Aharony:2012ns}.

\subsection{Chern-Simons theory coupled to critical fermions} \label{crifer}

In this subsection we study the Chern-Simons matter theory coupled to massless
critical fermions in the fundamental representation. The Lagrangian of our 
system is simply that of massless fundamental fermions minimally coupled 
to Chern-Simons theory deformed by the terms
\begin{equation}
 \delta {\cal L} = B \bar\psi\psi+\frac{N}{6}\lambda_6^{f} B^3
\end{equation}
where $B$ is a Lagrange multiplier field and $\lambda_6^f$ is a coupling 
that is marginal in the critical fermion theory as $B$ turns out to have
unit dimension.
We find  
\begin{equation}\label{cferfree}\begin{split}
  v[\rho]&=-\frac{N }{6\pi}\left(\tilde c^3(\frac{1}{\lambda}-1)-\frac{3 \beta B \tilde c^2}{2\lambda}+\frac{(\beta B)^3}{2\lambda}-\pi\lambda_{6}^{f}(\beta B)^3\right)\\
&-\frac{N }{2\pi} \int_{\tilde c}^{\infty}dy\int_{-\pi}^{\pi}d\alpha~ y ~\rho(\alpha)(\ln(1+e^{-y-i\alpha})+\ln(1+e^{-y+i\alpha})) ,
\end{split}
\end{equation}
where ${\tilde c}~T$ is the thermal mass of the fermions.\footnote{ The fermion 
self energy (the correction to the quadratic term in the fermion effective 
action) is given by $\Sigma_{T}(p)+B I=f(\beta p_s)p_s I+i p^{-}g(\beta p_s)\gamma^{-},$ where 
\begin{equation}\label{selfferc}\begin{split}
 f(y)&=\frac{\beta}{y}B+\frac{\lambda}{y}\int_{-\pi}^{\pi} d\alpha~\rho(\alpha)\left(\ln 2\cosh(\frac{\sqrt{y^2+\tilde c^2}+i \alpha}{2})+\ln 2\cosh(\frac{\sqrt{y^2+\tilde c^2}-i \alpha}{2})\right),\\
g(y)&=\frac{\tilde c^2}{y^2}-f(y)^2,\\
\tilde c-\beta B &=\lambda \int_{-\pi}^{\pi} d\alpha~\rho(\alpha)\left(\ln 2\cosh(\frac{\tilde c+i \alpha}{2})+\ln 2\cosh(\frac{\tilde c-i \alpha}{2})\right).
\end{split}
\end{equation}}
The constants 
$B$ and $\tilde c$ above are determined by the requirement that they 
extremize $v[\rho]$. This requirement yields the equations   
\begin{equation}\begin{split} \label{cfs}
\tilde c\left(1- \frac{\beta B}{\tilde c} \right)&=\lambda \int_{-\pi}^{\pi}d\alpha~\rho(\alpha)\left(\ln 2\cosh(\frac{\tilde c+i \alpha}{2})+\ln 2\cosh(\frac{\tilde c-i \alpha}{2})\right),\\
                 \beta B&=\pm \frac{1}{\sqrt{1-2\pi\lambda\lambda_{6}^{f}}} \tilde c .
\end{split}
\end{equation}
Assuming that ${\tilde c}$ is positive, we are forced to choose the negative  
root in the second of \eqref{cfs}.\footnote{This may be seen as follows. 
Inserting the second equation in \eqref{cfs} into the first equation of 
\eqref{cfs} we  obtain
\begin{equation}
 \tilde c\left(1\mp \frac{1}{\sqrt{1-2\pi\lambda\lambda_{6}^{f}}} \right)=\lambda \int_{-\pi}^{\pi}d\alpha~\rho(\alpha)\left(\ln 2\cosh(\frac{\tilde c+i \alpha}{2})+\ln 2\cosh(\frac{\tilde c-i \alpha}{2})\right).
\end{equation}
As the R.H.S. of this equation is positive, we must choose the negative root, 
i.e. $\beta B=-\frac{1}{\sqrt{1-2\pi\lambda\lambda_{6}^{f}}} \tilde c.$}
Inserting the second equation in \eqref{cfs} into the first we find
\begin{equation}\label{cfers1}
 \tilde c\left(1+ \frac{1}{\sqrt{1-2\pi\lambda\lambda_{6}^{f}}} \right)=\lambda \int_{-\pi}^{\pi}d\alpha~\rho(\alpha)\left(\ln 2\cosh(\frac{\tilde c+i \alpha}{2})+\ln 2\cosh(\frac{\tilde c-i \alpha}{2})\right).
\end{equation} 
Inserting the same equation into \eqref{cferfree} yields
\begin{equation}\label{cferfree1}
  v[\rho]=-\frac{N }{6\pi\lambda}\left(\tilde c^3(1-\lambda+
\hat g(\lambda, \lambda_6^f) )+3 \lambda\int_{\tilde c}^{\infty}dy \int_{-\pi}^{\pi}\rho(\alpha)d\alpha~ y(\ln(1+e^{-y-i\alpha})+\ln(1+e^{-y+i\alpha}))\right), 
\end{equation}
where 
\begin{equation} \label{hgd}
\hat g(\lambda, \lambda_6^f) = 
\frac{1}{\sqrt{1-2\pi\lambda\lambda_{6}^{f}}}.
\end{equation}
The expressions presented above reduce to formulas that have previously 
appeared in the literature for special values of $\rho$. In particular 
for $\rho$ given by \eqref{evd}, \eqref{cfers1} reduces to 
Eq.(5.35) of \cite{Aharony:2012ns}, while \eqref{cferfree1} reduces to 
Eq.(5.34) of \cite{Aharony:2012ns} with appropriate factors. 

\subsection{${\cal N}=2$ theory with a single fundamental chiral multiplet} \label{susycase}
The Lagrangian of the theory we study is presented in 
equation (2.1), (3.58) of \cite{Jain:2012qi} (see also (6.1) and (6.20) of \cite{Aharony:2012ns}). We find
 \beal{\label{vsusy}
  { v[\rho] } 
=& -\frac{N }{6 \pi |\lambda|}\left(\tilde c^3 - 6 |\lambda|  \int_{-\pi}^{\pi} d\alpha \rho(\alpha) \text{Re} \int_{\tilde c} ^\infty dy y \log \tanh {y+i\alpha \over 2}\right),
}
where for  the thermal mass (for both the boson as well as the fermion in the 
supermultiplet) is denoted by $\tilde c T$. The constant 
 $\tilde c$ above is determined by the requirement that it extremizes $v[\rho]$;
i.e. 
\be \label{susyct}
\tilde c=2  \biggl | \text{Re} \int_{-\pi}^{\pi} d\alpha \lambda \rho(\alpha) \log \coth {\tilde c + i \alpha \over 2} \biggl |. 
\ee
The expressions presented above reduce to formulas that have previously 
appeared in the literature for special values of $\rho$. In particular 
for $\rho$ given by \eqref{evd}, \eqref{susyct} reduces to 
Eq.(6.22) of \cite{Aharony:2012ns}, while \eqref{vsusy} reduces to 
Eq.(6.23) of \cite{Aharony:2012ns} with appropriate factors. For the case when $\rho(\alpha)=\delta(\alpha),$ 
\eqref{susyct},  \eqref{vsusy} are straight-forward generalization of  (3.61), (3.62) of \cite{Jain:2012qi}.

\subsection{Level-Rank duality}

The Chern-Simons matter theories we have studied, earlier in this section, 
have been conjectured to be related to one another by strong weak coupling 
dualities~\cite{Maldacena:2012sf,Aharony:2012nh,Aharony:2012ns,GurAri:2012is}, see also \cite{Giombi:2011kc} for a preliminary suggestion. 
The best established of these conjectures is the conjectured
strong weak coupling self duality of the supersymmetric theory. 
According to this conjecture the supersymmetric theory at rank $N$ and 
renormalized level $k$ may be identified with the {\it same} theory 
at rank $k-N$ and level $k$. The non supersymmetric theories we have 
studied above have also been conjectured to be related to one another 
by strong weak coupling type dualities. In particular the theory of minimally 
coupled fermions at rank $N$ and renormalized level $k$ has been conjectured 
to be identical to the theory of Chern-Simons gauged critical bosons at 
rank $k-N$ and level $k$. 

Finally, the minimally coupled scalar theory and the critical fermion theory 
are conjectured to obey the following duality. The minimally coupled 
scalar with rank $N$, 't~Hooft coupling $\lambda$ and ${\hat \lambda}(\lambda, 
\lambda_6)$ is conjectured to be dual to the fermionic theory with rank 
$k-N$, t' Hooft coupling $1-\lambda$ and ${\hat g}(1-\lambda, \lambda_6^f)
=\frac{2 \lambda}{{\hat \lambda}(\lambda, \lambda_6)}$ 
(the last relationship effectively 
determined the relationship between the two marginal parameters) where $\hat\lambda,~\hat g$ are defined in \eqref{dljh},\eqref{hgd} respectively.

In this subsection we will demonstrate that the potentials, $v[\rho]$, 
for (conjectured) dual pairs all obey \eqref{Veq}. It follows that 
all the partition functions computed in this paper are consistent 
with all three conjectured dualities listed above. This agreement may be 
regarded as evidence in favor of all three dualities, as well as 
the effective action framework of section \ref{pi} on which the current paper 
is based.

\subsubsection{Self-duality of $\cN=2$}

$v[\rho]$ for the supersymmetric theory was reported in subsection 
\ref{susycase}. This potential depends on the values of 
$N$ and $\lambda$; in the rest of this subsubsection we use notation 
for this potential that makes its dependence on parameters explicit; 
we denote the potential by $v^{\text{susy}}_{N, \lambda}[\rho]$. We also denote 
the thermal mass of the supersymmetric theory by 
${\tilde c}^{\text{susy}}_{\lambda}[\rho]$. In the rest of this subsubsection we 
will demonstrate that 
\begin{equation}\label{dualsusy} \begin{split}
{\tilde c}^{\text{susy}}_{\lambda}[\rho]&={\tilde c}^{\text{susy}}_{1-\lambda}[{\tilde \rho}],\\
v^{\text{susy}}_{N, \lambda}[\rho]&=v^{\text{susy}}_{k-N, 1-\lambda}[{\tilde \rho}]
\end{split}
\end{equation}
where $\tilde \rho$ is the transformed eigenvalue distribution function
defined by the equation 
\begin{equation}\label{rhosus} 
(1-\lambda)\tilde \rho(\alpha) 
= \frac{1}{2\pi}-\lambda \rho(\alpha+\pi).\end{equation}

Let us first demonstrate the first equation in \eqref{dualsusy}.
According to \eqref{susyct} 
\be
{\tilde c}^{\text{susy}}_{1-\lambda}[{\tilde \rho}] = 2 \biggl | \text{Re}  \int_{-\pi}^{\pi} d\alpha (1-\lambda)\tilde \rho(\alpha) \log \coth {{\tilde c}^{\text{susy}}_{1-\lambda}[{\tilde \rho}] + i \alpha \over 2} \biggl |. 
\ee
Using \eqref{rhosus} the RHS of \eqref{susyct} may be rewritten as
\be
2 \biggl | \text{Re}  \int_{-\pi}^{\pi} d\alpha  \( { 1 \over 2\pi} - \lambda \rho(\alpha+\pi) \)\log \coth {{\tilde c}^{\text{susy}}_{1-\lambda}[{\tilde \rho}] + i \alpha \over 2} \biggl | = 
2 \biggl | \text{Re}  \int_{0}^{2\pi} d\alpha  \lambda \rho(\alpha) \log \coth {{\tilde c}^{\text{susy}}_{1-\lambda}[{\tilde \rho}] + i \alpha \over 2} \biggl |,
\ee
where we have used the integral identities
\be
 \int_{-\pi}^{\pi} d\alpha \log \coth {x + i \alpha \over 2} =0, \quad
\coth ({x + i\pi \over 2}) = \tanh ({x \over 2}) \quad  \text{for all} \; x. 
\label{formula1}
\ee
It follows that
\begin{equation}\begin{split}
{\tilde c}^{\text{susy}}_{1-\lambda}[{\tilde \rho}] &=2  \biggl | \text{Re}  \int_{0}^{2\pi} d\alpha  \lambda \rho(\alpha) \log \coth {{\tilde c}^{\text{susy}}_{1-\lambda}[{\tilde \rho}] + i \alpha \over 2} \biggl | \\
&=2  \biggl | \text{Re}  \int_{-\pi}^{\pi} d\alpha  \lambda \rho(\alpha) \log \coth {{\tilde c}^{\text{susy}}_{1-\lambda}[{\tilde \rho}] + i \alpha \over 2} \biggl |. 
\end{split}
\end{equation}
Comparing with \eqref{susyct} we see that 
${\tilde c}^{\text{susy}}_{1-\lambda}[{\tilde \rho}]$ obeys the same equation as 
${\tilde c}^{\text{susy}}_{\lambda}[ \rho]$. 
Hence we conclude
\begin{equation}\label{fts}
 {\tilde c}^{\text{susy}}_{1-\lambda}[{\tilde \rho}]={\tilde c}^{\text{susy}}_{\lambda}[ \rho],
\end{equation}
demonstrating the first equation in \eqref{dualsusy}.

We now establish the second equation in \eqref{dualsusy}. According 
to \eqref{vsusy} 
\beal{& v^{\text{susy}}_{k-N, 1-\lambda}[{\tilde \rho}]\nonumber\\
 &=-\frac{k-N }{6 |\lambda|\pi |1-\lambda|} \left(({\tilde c}^{\text{susy}}_{1-\lambda}[{\tilde \rho}])^3 - 6 |1-\lambda|  \int_{-\pi}^{\pi} d\alpha \tilde \rho(\alpha) \text{Re} \int_{{\tilde c}^{\text{susy}}_{1-\lambda}[{\tilde \rho}]} ^\infty dy y \log \tanh {y+i\alpha \over 2}\right)\nn
&= -\frac{N }{6 |\lambda|\pi }\left(({\tilde c}^{\text{susy}}_{1-\lambda}[{\tilde \rho}])^3 - 6   \int_{-\pi}^{\pi} d\alpha  \( { 1 \over 2\pi} - |\lambda| \rho(\alpha+\pi) \) \text{Re} \int_{{\tilde c}^{\text{susy}}_{1-\lambda}[{\tilde \rho}]} ^\infty dy y \log \tanh {y+i\alpha \over 2}\right)\nn
&= -\frac{N }{6 |\lambda|\pi }\left(({\tilde c}^{\text{susy}}_{1-\lambda}[{\tilde \rho}])^3 - 6   \int_{0}^{2\pi} d\alpha  \(  - |\lambda| \rho(\alpha) \) \text{Re} \int_{{\tilde c}^{\text{susy}}_{1-\lambda}[{\tilde \rho}]} ^\infty dy y \log \coth {y+i\alpha \over 2}\right)\nn
&=-\frac{N }{6 |\lambda|\pi }\left( ({\tilde c}^{\text{susy}}_{1-\lambda}[{\tilde \rho}])^3 - 6   \int_{0}^{2\pi} d\alpha   |\lambda| \rho(\alpha) \text{Re} \int_{{\tilde c}^{\text{susy}}_{1-\lambda}[{\tilde \rho}]} ^\infty dy y \log \tanh {y+i\alpha \over 2}\right)\nn
&= -\frac{N }{6 |\lambda|\pi }\left(({\tilde c}^{\text{susy}}_{\lambda}[\rho])^3 - 6   \int_{-\pi}^{\pi} d\alpha   |\lambda| \rho(\alpha) \text{Re} \int_{{\tilde c}^{\text{susy}}_{\lambda}[\rho]} ^\infty dy y \log \tanh {y+i\alpha \over 2}\right)\nn
&= v^{\text{susy}}_{N, \lambda}[\rho],
}
where we used \eqref{formula1} in going from the second to the third line, 
and \eqref{fts} in going from the fourth to the fifth line. This completes 
our demonstration of the self duality of $v[\rho]$ for the supersymmetric 
theory.

\subsubsection{Critical boson v.s. Regular fermion}
The potentials $v[\rho]$ for the critical boson and the minimally 
coupled fermion theory were reported in subsections \ref{cccb} and \ref{minimalfer} respectively. $v[\rho]$ depends on the values of 
$N$, $\lambda$; in the rest of this subsubsection we use notation 
for this potential that makes its dependence on parameters explicit; 
we denote the potential for the critical boson theory and the 
regular fermion theory respectively by $v^{\text{Cri.B.}}_{N, \lambda}[\rho],
~v^{\text{Reg.F.}}_{N, \lambda}[\rho]$. We also denote 
the thermal mass of the critical bosonic theory and the 
regular fermionic theory by $\sigma^{\text{Cri.B.}}_{\lambda}[\rho]$, 
${\tilde c}^{\text{Reg.F.}}_{\lambda}[\rho]$ respectively. 
In the rest of this subsubsection we 
will demonstrate that 
\begin{equation}\label{dualbosc} \begin{split}
v^{\text{Cri.B.}}_{k-N, 1-\lambda}[\tilde \rho]&=v^{\text{Reg.F.}}_{N, \lambda}[{ \rho}],\\
{\sigma }^{\text{Cri.B.}}_{1-\lambda}[\tilde \rho]&={\tilde c}^{\text{Reg.F.}}_{\lambda}[{ \rho}],
\end{split}
\end{equation}
where $\tilde \rho$ is the transformed eigenvalue distribution function
defined by the equation 
\eqref{rhosus}.

To start with, we first demonstrate the second equation in \eqref{dualbosc}. 
According to \eqref{selfboscri},
\begin{equation}
\begin{split}
  \int_{-\pi}^{\pi} d\alpha \tilde \rho(\alpha)     \left(\ln 2\sinh(\frac{{\sigma }^{\text{Cri.B.}}_{1-\lambda}[\tilde \rho]- i \alpha}{2})+\ln 2\sinh(\frac{{\sigma }^{\text{Cri.B.}}_{1-\lambda}[\tilde \rho]+ i \alpha}{2})\right)&=0\\
 \int_{-\pi}^{\pi} d\alpha (\frac{1}{2\pi}-\lambda \rho(\alpha+\pi))  \left(\ln 2\sinh(\frac{{\sigma }^{\text{Cri.B.}}_{1-\lambda}[\tilde \rho]- i \alpha}{2})+\ln 2\sinh(\frac{{\sigma }^{\text{Cri.B.}}_{1-\lambda}[\tilde \rho]+ i \alpha}{2})\right)&=0.\\
\end{split}
\end{equation}
Rearranging terms in this equation we find
\begin{equation}\label{selfboscri1}
\int_{-\pi}^{\pi} d\alpha  \left(\ln 2\sinh(\frac{{\sigma }^{\text{Cri.B.}}_{1-\lambda}[\tilde \rho]- i \alpha}{2})+\rm{c.c}\right)=2\pi\lambda\int_{-\pi}^{\pi} d\alpha \rho(\alpha+\pi)     \left(\ln 2\sinh(\frac{{\sigma }^{\text{Cri.B.}}_{1-\lambda}[\tilde \rho]- i \alpha}{2})+\rm{c.c}\right).
\end{equation}
Using the integral identity
\begin{equation}\label{selfboscri2}
\int_{-\pi}^{\pi} d\alpha  \left(\ln 2\sinh(\frac{{\sigma }^{\text{Cri.B.}}_{1-\lambda}[\tilde \rho]- i \alpha}{2})+\rm{c.c}\right)=2\pi {\sigma }^{\text{Cri.B.}}_{1-\lambda}[\tilde \rho],
\end{equation}
\eqref{selfboscri1} may be rewritten as 
\begin{equation}\label{ppp}\begin{split}
{\sigma }^{\text{Cri.B.}}_{1-\lambda}[\tilde \rho] &=
\lambda \int_{0}^{2\pi} d\alpha \rho(\alpha)  \left(\ln 2\sinh(\frac{{\sigma }^{\text{Cri.B.}}_{1-\lambda}[\tilde \rho]- i \alpha + i\pi }{2})+\rm{c.c}\right)\\
&=\lambda \int_{-\pi}^{\pi} d\alpha \rho(\alpha)     \left(\ln 2\cosh(\frac{{\sigma }^{\text{Cri.B.}}_{1-\lambda}[\tilde \rho]- i \alpha}{2})+\rm{c.c}\right).
\end{split}\end{equation}
 Comparing \eqref{ppp} with \eqref{tc} we conclude that 
${\sigma }^{\text{Cri.B.}}_{1-\lambda}[\tilde \rho]$ obeys the same equation as 
${\tilde c}^{\text{Reg.F.}}_{\lambda}[{ \rho}]$, establishing that
\begin{equation}\label{cribosfers}
 {\sigma }^{\text{Cri.B.}}_{1-\lambda}[\tilde \rho]={\tilde c}^{\text{Reg.F.}}_{\lambda}[{ \rho}].
\end{equation}

Now we proceed to demonstrate first equation of \eqref{selfboscri}.
According to \eqref{vcritb}
\begin{equation}\label{freenboscri1}\begin{split}
 & v^{\text{Cri.B.}}_{k-N, 1-\lambda}[\tilde \rho]\\&=-\frac{N }{6\pi}\frac{1-\lambda}{\lambda}({\sigma }^{\text{Cri.B.}}_{1-\lambda}[\tilde \rho])^3 + \frac{N }{2\pi}\frac{1-\lambda}{\lambda}\int_{\sigma}^{\infty} y dy\int_{-\pi}^{\pi} d\alpha\tilde\rho(\alpha) \left(\ln(1-e^{-y+i\alpha})+\ln(1-e^{-y-i\alpha})\right)\\
&=-\frac{N }{6\pi}\frac{1-\lambda}{\lambda}
({\tilde c}^{\text{Reg.F.}}_{\lambda}[{ \rho}])^3 - \frac{N }{2\pi}\int_{\sigma}^{\infty} dy y\int_{-\pi}^{\pi} \rho(\alpha)d\alpha \left(\ln(1+e^{-y+i\alpha})+\ln(1+e^{-y-i\alpha})\right),
\end{split}
\end{equation}
where we have used \eqref{cribosfers} in going from the first to the 
second line. Comparing with  \eqref{freefer} we conclude 
\begin{equation}
 v^{\text{Cri.B.}}_{k-N, 1-\lambda}[\tilde \rho]=v^{\text{Reg.F.}}_{N, \lambda}[{ \rho}].
\end{equation}
This completes our demonstration of \eqref{dualbosc}.

\subsubsection{Regular boson v.s. Critical fermion}
The potential $v[\rho]$ for the marginally boson and the 
critically coupled fermion theory were reported in subsections 
\ref{marginalbos} and \ref{crifer} respectively. 
This potential depends on the values of 
$N$, $\lambda,$ $\lambda_{6}$ for the boson, and on $N$, $\lambda$ and 
$\lambda_6^f$ for the fermion. 
In the rest of this subsubsection we use notation 
for this potential that makes its dependence on parameters explicit; 
we denote the potential of the 
regular bosonic theory and critical fermionic theory 
by $v^{\text{Reg.B.}}_{N, \lambda, \lambda_{6}}[\rho],
~v^{\text{Cri.F.}}_{N, \lambda,\lambda_{6}^{f}}[\rho]$ respectively. We also denote 
the thermal mass of the regular bosonic theory and critical fermionic theory 
by $\sigma^{\text{Reg.B.}}_{\lambda, \lambda_{6}}[\rho],$
${\tilde c}^{\text{Cri.F.}}_{\lambda, \lambda_{6}^{f}}[\rho]$ respectively. In the rest of 
this subsubsection we will demonstrate that 
\begin{equation}\label{dualbosr} \begin{split}
v^{\text{Reg.B.}}_{k-N, 1-\lambda, \lambda_{6} }[\tilde \rho]&=
v^{\text{Cri.F.}}_{N, \lambda, \lambda_{6}^{f}}[{ \rho}],\\
{\sigma }^{\text{Reg.B.}}_{1-\lambda, \lambda_{6} }[\tilde \rho]&=
{\tilde c}^{\text{Cri.F.}}_{\lambda, \lambda_{6}^{f}}[{ \rho}].
\end{split}
\end{equation}
 Consequently 
the relations \eqref{dualbosr} assert the duality between bosonic and 
fermionic theories with marginal deformation parameters related in a known 
but complicated manner. In the equation above 
$\tilde \rho$ is the transformed eigenvalue distribution function
defined by the equation 
\eqref{rhosus} and $\lambda_6^f$ is given as a function of $\lambda_6$ 
by the complicated relationship
\begin{equation}\label{rellm6}
 \frac{\sqrt{\frac{\lambda_{6}}{8\pi^2}+(1-\lm)^2}}{1-\lambda}=2 \sqrt{1-2\pi\lambda\lambda_{6}^{f}}.
\end{equation}

Let us first demonstrate the second equation in \eqref{dualbosr}. 
According to \eqref{rboself} 
\begin{equation}\label{mancf}
\begin{split}
 &{\sigma }^{\text{Reg.B.}}_{1-\lambda,\lambda_{6}}[\tilde \rho]\\
& =-\frac{\sqrt{\frac{\lambda_{6}}{8\pi^2}+(1-\lm)^2}}{2 (1-\lambda)}\int_{-\pi}^{\pi} d\alpha (1-\lambda) \tilde\rho(\alpha)     \left(\ln 2\sinh(\frac{{\sigma }^{\text{Reg.B.}}_{1-\lambda,\lambda_{6}}[\tilde \rho]- i \alpha}{2})+c.c.\right)\\     
&=-\frac{\sqrt{\frac{\lambda_{6}}{8\pi^2}+(1-\lm)^2}}{2 (1-\lambda)} 
\int_{-\pi}^{\pi} d\alpha (\frac{1}{2\pi}-\lambda \rho(\alpha+\pi))    \left(\ln 2\sinh(\frac{{\sigma }^{\text{Reg.B.}}_{1-\lambda,\lambda_{6}}[\tilde \rho]- i \alpha}{2})+c.c.\right)\\     
&=-\frac{\sqrt{\frac{\lambda_{6}}{8\pi^2}+(1-\lm)^2}}{2 (1-\lambda)}
\left({\sigma }^{\text{Reg.B.}}_{1-\lambda,\lambda_{6}}[\tilde \rho]-\int_{-\pi}^{\pi} d\alpha \lambda \rho(\alpha+\pi)    \left(\ln 2\sinh(\frac{{\sigma }^{\text{Reg.B.}}_{1-\lambda,\lambda_{6}}[\tilde \rho]- i \alpha}{2})+c.c.\right)\right)\\     
&=-\frac{\sqrt{\frac{\lambda_{6}}{8\pi^2}+(1-\lm)^2}}{2 (1-\lambda)}\left({\sigma }^{\text{Reg.B.}}_{1-\lambda,\lambda_{6}}[\tilde \rho]-\int_{0}^{2\pi} d\alpha \lambda \rho(\alpha)    \left(\ln 2\cosh(\frac{{\sigma }^{\text{Reg.B.}}_{1-\lambda,\lambda_{6}}[\tilde \rho]- i \alpha}{2})+c.c.\right)\right).     
\end{split}
\end{equation}
In going
 from the third to the fourth line of \eqref{mancf} we have used 
the integral identity \eqref{selfboscri2}. Rearranging terms, we obtain
\begin{equation}
\left(1+ \frac{2 (1-\lambda)}{\sqrt{\frac{\lambda_{6}}{8\pi^2}+(1-\lm)^2}}
\right){\sigma }^{\text{Reg.B.}}_{1-\lambda,\lambda_{6}}[\tilde \rho] =\int_{-\pi}^{\pi} d\alpha \lambda \rho(\alpha)    \left(\ln 2\cosh(\frac{{\sigma }^{\text{Reg.B.}}_{1-\lambda,\lambda_{6}}[\tilde \rho]- i \alpha}{2})+c.c.\right).
\end{equation}
Using \eqref{rellm6} we obtain\footnote{\eqref{rellm6} is just a slight rewriting of 
Eq.(5.37) of \cite{Aharony:2012ns} which reads
$
 \frac{1}{\lambda_{b}}{\sqrt{\frac{\lambda_{6}}{8\pi^2}+\lm_b^2}}
=2 \sqrt{1-2\pi(1-\lambda_{b})\lambda_{6}^{f}}.
$} 
\begin{equation}\label{criticalfer1}\begin{split}
\left(1+ \frac{1}{\sqrt{1-2\pi\lambda\lambda_{6}^{f}}}
\right){\sigma }^{\text{Reg.B.}}_{1-\lambda,\lambda_{6}}[\tilde \rho] &=\int_{-\pi}^{\pi} d\alpha \lambda \rho(\alpha)    \left(\ln 2\cosh(\frac{{\sigma }^{\text{Reg.B.}}_{1-\lambda,\lambda_{6}}[\tilde \rho]- i \alpha}{2})+c.c.\right). 
\end{split}\end{equation}
Comparing \eqref{criticalfer1} with \eqref{cfers1} we conclude
\begin{equation}\label{relasc}
 {\sigma }^{\text{Reg.B.}}_{1-\lambda, \lambda_{6}}[\tilde \rho]={\tilde c}^{\text{Cri.F.}}_
{\lambda, \lambda_{6}^{f}}[{ \rho}].
\end{equation}

The first of the \eqref{dualbosr} can be proved as follows. Under the duality one obtains
\begin{equation}\label{criticalferfree1}
\begin{split}
 &v^{\text{Reg.B.}}_{k-N, 1-\lambda,\lambda_{6}}[\tilde \rho] \\
&= -N\frac{1-\lambda}{\lambda} \frac{ ({\sigma }^{\text{Reg.B.}}_{1-\lambda,\lambda_{6}}[\tilde \rho])^3}{6\pi}\left(1+\frac{2}{\sqrt{\frac{\lambda_{6}}{8\pi^2}+(1-\lm)^2}}\right)
\\
&~~~+\frac{N }{2\pi}\frac{1-\lambda}{\lambda}\int_{-\pi}^{\pi}d\alpha\int_{\sigma}^{\infty} y \tilde\rho(\alpha) dy\left(\ln(1-e^{-y+i\alpha})+c.c.\right)\\
&= -N\frac{ ({\sigma }^{\text{Reg.B.}}_{1-\lambda,\lambda_{6}}[\tilde \rho])^3}{6\pi \lambda}
\left(1-\lambda+\frac{2 (1-\lm)}{\sqrt{\frac{\lambda_{6}}{8\pi^2}+(1-\lm)^2}}
\right)\\
&+\frac{N }{2\pi}\frac{1-\lambda}{\lambda}\int_{-\pi}^{\pi}d\alpha\int_{\sigma}^{\infty} dy y \left(\frac{1}{2\pi (1-\lambda)}-\frac{\lambda}{1-\lambda}\rho(\alpha+\pi)\right) \left(\ln(1-e^{-y+i\alpha})+\ln(1-e^{-y-i\alpha})\right).\\
\end{split}
\end{equation}
Now using \eqref{rellm6} and the fact that
$
 \int_{-\pi}^{\pi}  d\alpha \left(\ln(1-e^{-y+i\alpha})+\ln(1-e^{-y-i\alpha})\right)=0,
$
one obtains
\begin{equation}\label{criticalferfree2}
\begin{split}
 & v^{\text{Reg.B.}}_{k-N, 1-\lambda,\lambda_{6}}[\tilde \rho]\\
 &= -N\frac{ ({\sigma }^{\text{Reg.B.}}_{1-\lambda,\lambda_{6}}[\tilde \rho])^3}{6\pi \lambda}\left(1-\lambda+\frac{1}{\sqrt{1-2\pi\lambda\lambda_{6}^{f}}}\right)
-\frac{N }{2\pi}\int_{-\pi}^{\pi}d\alpha\int_{\sigma}^{\infty} dy \rho(\alpha+\pi) y\left(\ln(1-e^{-y+i\alpha})+c.c.\right)\\
&=-N\frac{ ({\sigma }^{\text{Reg.B.}}_{1-\lambda,\lambda_{6}}[\tilde \rho])^3}{6\pi \lambda}\left(1-\lambda+\frac{1}{\sqrt{1-2\pi\lambda\lambda_{6}^{f}}}\right)
-\frac{N }{2\pi}\int_{0}^{2\pi}d\alpha\int_{\sigma}^{\infty} y dy\rho(\alpha) \left(\ln(1+e^{-y+i\alpha})+c.c.\right)\\
&=-N\frac{ ({\sigma }^{\text{Reg.B.}}_{1-\lambda,\lambda_{6}}[\tilde \rho])^3}{6\pi \lambda}\left(1-\lambda+\frac{1}{\sqrt{1-2\pi\lambda\lambda_{6}^{f}}}\right)
-\frac{N }{2\pi}\int_{-\pi}^{\pi}d\alpha \int_{\sigma}^{\infty}dy y \rho(\alpha)\left(\ln(1+e^{-y+i\alpha})+c.c.\right).
\end{split}
\end{equation}
Using \eqref{hgd},\eqref{cfers1} and \eqref{relasc} we  conclude that above equation is same as \eqref{cferfree1}. So we have shown 
\begin{equation}
 v^{\text{Reg.B.}}_{k-N, 1-\lambda,\lambda_{6}}[\tilde \rho]=v^{\text{Cri.F.}}_{N, \lambda,\lambda_{6}^{f}}[{ \rho}].
\end{equation}

\section{Evaluation of the path integral \eqref{purecs}}\label{pathinteva}
At first sight \eqref{purecs} is the path 
integral over a nontrivial interacting field theory in 
three dimensions, and its evaluation appears to be a daunting task. 
However the rewritten form \eqref{purecsb} emphasizes that 
\eqref{purecs} evaluates a topological observable in the 
topological Chern-Simons theory. This special features of 
Chern-Simons theory make the evaluation of the path integral almost 
trivial. The path integral we need was in fact largely 
evaluated in \cite{Blau:1993tv}. 
In order to make our paper self contained, however, we reevaluate it below, 
very closely following \cite{Blau:1993tv}. Our focus in this section, and 
the rest of this paper, will largely be on Chern-Simons theories on 
$S^2 \times S^1$. However we will also briefly consider the generalization 
to Chern-Simons theories on $\Sigma_g \times S^1$, where $\Sigma_g$ is 
a genus $g$ Riemann surface.

\subsection{Gauge fixing} 

We impose the following gauge fixing conditions:
\begin{enumerate}[{(i)}]
\item  $\partial_3U \equiv \partial_3 e^{\beta A_3}=0.$
\item We simultaneously diagonalize $U(x)$ for all points $x$ on 
$S^2$. 
\item We impose the Coulomb gauge on the time independent 
diagonal elements of $A_1, A_2$.
\end{enumerate}
Here $A_3$ is the gauge field along the $S^1$ direction and 
$A_{1,2}$ are the ones along the $S^2$ directions.

The choice (i) is the nearest 
one that can come to the temporal gauge $A_3=0$ when the thermal time is compact. 
The gauge condition (i) leaves two dimensional gauge transformations (those 
that depend only on the coordinates of the sphere) unfixed. The gauge 
condition (ii) fixes some of this freedom, leaving only 
a two dimensional $U(1)^N$ gauge symmetry unfixed. The gauge condition 
(iii) fixes the residual 2d  Abelian gauge invariance by imposing (Abelian) 
Coulomb gauge for the time independent parts of the diagonal elements of 
$A_i$, where $i,j, \ldots=1,2$ are the sphere spatial indices.%

The gauge condition (ii) that we have employed to abelianize the residual  
two dimensional gauge invariance is very similar to the gauge employed 
by 't~Hooft in his classic study of confinement \cite{'tHooft:1981ht} 
due to monopole condensation. As noted by 't~Hooft the gauge condition 
(ii) is ambiguous (i.e. fails to completely fix gauge) when two or more 
eigenvalues of the matrix $U=e^{ \beta A_3}$ coincide. Two eigenvalues of a 
unitary matrix coincide on a surface of codimension 3 in the space of all 
unitary matrices. In 't~Hooft's study of  confinement of 4 dimensional gauge 
theories, such coincidences typically occurred on a $4-3=1$ dimensional 
surface in spacetime, i.e. on a line. The world line of such coincidences 
were monopoles for the gauge fixed abelian theory; 't~Hooft argued that 
the condensation of such monopoles is responsible for confinement.

In the current paper the `abelianizing' gauge condition (ii) is adopted
on a two dimensional manifold (i.e. the base manifold $S^2$). As 
a consequence we would generically expect eigenvalues to coincide 
on a $2-3=-1$ dimensional surface of our two dimensional base manifold 
($S^2$); in other words 
in the two dimensional context of this paper two eigenvalues of $U(x)$
do not generically coincide. Consequently configurations with coincident 
eigenvalues do not represent `objects' or `defects' in our path integral. 
Nonetheless such configurations play an important role in our path integral 
as we now explain.  

Over the next two paragraphs we characterize the space of smooth functions 
$U(x)$ before we impose gauge condition (ii). Smooth connections
 on $S^2 \times S^1$ are all 
topologically connected to $A_\mu=0$ 
\footnote{$U(N)$ connections have distinct topological sectors, 
labeled by the overall $U(1)$ flux.}; in particular 
the set of all smooth functions $U(x)$ form a single connected space.
 However our particular choice of 
gauge (ii) is singular when two eigenvalues of the matrix $U$ coincide. 
Such a singularity occurs on a subspace of codimension 1
in the space of all smooth functions $U(x)$ \footnote{This may be understood 
as follows. Consider a one parameter set of functions $U(c,x)$ where $c$ 
is a real valued parameter. Requiring that two of the eigenvalues 
of $U(c,x)$ coincide give a set of three equations, that will generically 
yield discrete solutions for the 3 variables $c$, $x_1$, $x_2$. As  
this condition is satisfied only at discrete values of the parameter $c$, 
it follows that the subspace of $U(x)$ with two equal eigenvalues is of 
codimension unity in the space of all smooth functions $U(x)$.}. 
These singular configurations (as far 
as our choice of gauge is concerned) constitute `domain walls' that chop up 
the space of smooth $SU(N)$ connections into a set of domains or 
unit cells. By definition, within a unit cell in the space of smooth 
functions $U(x)$, the eigenvalues of $U$ coincide nowhere 
on $S^2$. However on the domain 
walls two eigenvalues of $U$ coincide somewhere on $S^2$ (more than 
two eigenvalues coincide at a point - or two eigenvalues coincide  
at multiple points on $S^2$ where domain walls intersect). As our gauge
fails when this happens, we are forced to perform our path integral in 
each unit cell separately. Provided the path integral over every unit cell 
has no divergence when we approach the domain wall surfaces we can simply 
ignore the contribution of the domain walls to the path integral (as they 
are of codimension unity in field space); the full path 
integral is given by the sum of the path integrals in every unit cell.%
\footnote{We thank Xi Yin for a 
very useful discussion on this topic.}

In order to see how this works in more detail, we now focus on the special  
case $N=2$, the generalization to the $SU(N)$ theory is straightforward. 
Within any unit cell $U(x)$ is nowhere equal to $\pm I$ where $I$   
is the identity matrix (this is true by definition of a unit cell). Within any  
unit cell, therefore, the function $U(x)$ is a map from $S^2$ to  
${\tilde S}^3$ where ${\tilde S}^3$ is the $SU(2)$ group manifold 
 $S^3$ with two  points $U=\pm I$ removed  (the two points may be thought of as the  `north pole' and `south pole' of the $S^3$).  
While maps from $S^2$ to $S^3$ are trivial, maps from $S^2$ to  
${\tilde S}^3$ are characterized by an integer valued winding number. 
\footnote{This is analogous to the easily visualized fact that 
 maps from $S^1$ to  
${\tilde S}^2$ are characterized by an integer valued winding number,  
where ${\tilde S^2}$ is the sphere with north and south pole removed. 
Note also that for configurations for which $A_3$ is small the winding number  
may be characterized in a slightly more elementary manner. Consider a smooth  
configuration $A_3(x)$ before imposing the gauge choice  
(ii). As long as $A_3(x)$ is everywhere nonzero (i.e. provided its eigenvalues 
nowhere coincide), the unit vector field  
${\hat A}_3^a(x)=\frac{A_3^a}{\sqrt{A_3^b A_3^b}}$ is well  
defined on $S^2$. This unit vector field is characterized by an  
integer valued topological invariant; it's wrapping number over the $S^2$.} 
We have reached an important conclusion; unit cells in the space of smooth  
connections are labeled by an integer valued winding number. Distinct   
unit cells have distinct winding numbers. The winding number is 
ill defined on the domain walls that constitute the boundaries of  
unit cells; i.e.  at the points at which the gauge fixing (ii) will fail.

We now turn to the implementation of the gauge condition (ii) 
(we continue to work with the special case of $SU(2)$). 
We wish to construct gauge transformation $g(x)$ that implements the 
gauge fixing condition (ii); we proceed as follows.  We first perform a global 
gauge transformation that diagonalizes $U$ at the south pole of 
the base manifold $S^2$ (note that the north and south pole of the 
spatial $S^2$ are completely distinct from the north and south pole 
of the group manifold $S^3$ discussed in the previous paragraph). 
We then 
diagonalize $U(x)$ everywhere else on the $S^2$ as follows. Starting from the 
north pole, we move along lines of constant longitude 
(i.e. constant $\phi$ using the usual conventions for polar coordinates on 
$S^2$) and determine a $g(\theta, \phi)$ that 
diagonalizes $U(x)$ along that longitude; it is easy to check that this is 
always possible. In the neighborhood of the south pole this construction 
gives us a function $ H(\phi) =\lim_{\epsilon \to 0} g(\pi-\epsilon, \phi)$, i.e. 
a map from a circle to the gauge group. Recall, however, that $U$ was already 
diagonal at the south pole. In order that $U$ continue to be diagonal 
after the gauge transformation it must be that $H(\phi)$ lies in the 
$U(1)$ subgroup of $SU(2)$ that leaves any diagonal element invariant. 
In other words $H(\phi)$ defines a map from $S^1$ to $U(1)$; such maps 
are characterized by an integer valued winding number. It is a famous 
result in the study of non abelian solitons that the winding number of the map 
$H(\phi)$ is the same as the wrapping number of the map from $S^2$ 
to ${\tilde S}^3$ defined by $U(x)$. \footnote{ The function 
$g(\theta, \phi)$ is not unique; if $g(\theta, \phi)$ diagonalizes $U(x)$
then the same is true of ${\tilde h}(\theta, \phi) g(\theta, \phi)$ where 
${\tilde h}$ is an arbitrary $U(1)$ diagonal gauge transformation. This 
ambiguity maps $H(\phi)$  to ${\tilde h}(\pi, \phi) H(\phi)$. However 
as ${\tilde h}(0, \phi)$ is independent of $\phi$ (as $g$ is chosen to 
be well defined at the north pole) it follows that ${\tilde h}(\pi, \phi)$
is smoothly connected to a constant (by varying $\theta$) and so has 
no winding number. It follows that  the winding number of $H(\phi)$ is 
unaffected by this ambiguity and is also the only convention independent
information in the function $H(\phi)$.}

Clearly the function $H(\phi)$ can be chosen to be a constant (so that 
the gauge transformation $g(\theta, \phi)$ is well defined on the south 
pole) only if its winding number vanishes. It follows that the gauge 
choice (ii) can be implemented by a smooth gauge function $g(x)$ if and only if 
the wrapping number of $U(x)$ on $S^2$ vanishes. 

When the wrapping number of $U(x)$ is non vanishing on the $S^2$, 
the singularity of the gauge transformation function at the south pole 
has a straightforward physical interpretation. The winding number of the 
function $H(\phi)$ is simply equal, by Stokes theorem, to the $U(1)$ flux
that runs through the south pole. In other words once the gauge condition
(ii) has been implemented starting with a function $U(x)$ that had wrapping
number $n$ on the $S^2$, the background $U(1)$ connection has $n$ units
of flux over the $S^2$ (and a compensating Dirac string threading the 
south pole). Consequently when we implement our gauge condition (ii)
within a unit cell characterized by winding number $n$, we are finally 
left with the path integral over the abelian theory on a sphere with 
$n$ units of $U(1)$ flux for the abelian gauge group. 

Let us summarize. Before fixing the gauge condition (ii) we were 
required to perform a path integral over each unit cell, labeled by 
a wrapping number of $U(x)$ on $S^2$. Upon fixing the gauge condition 
(ii) sum of the path integrals over each unit cell turns into a sum 
over path integrals in the abelian gauge fixed theory on an $S^2$, 
where the abelian path integral is performed over connections 
with a nonzero abelian flux on the $S^2$, equal to the wrapping 
number of $U(x)$ on $S^2$ before we implemented the gauge condition (ii). 
In each flux sector we must now integrate unrestrictedly over the diagonal 
elements of $U(x)$ and the abelian gauge fields $A_\mu(x)$. 

In our presentation it might appear that the path integral 
in nonzero flux sectors has to be performed about a singular background, 
but of course the singularity of the gauge field at the south pole of $S^2$
is of the Dirac string type and is a gauge artifact. Within the residual 
$U(1)$ theory we can desingularize this configuration a la Wu and Yang 
using coordinate patches. The path integral to be performed is 
perfectly smooth in every sector; it merely has to be performed in 
the background of smooth fluxes for the residual abelian 
group. We will implement this procedure in what follows.

\subsection{Fadeev-Popov Determinants}

The Fadeev-Popov determinants for the gauge fixing conditions (i) and (ii) is 
\begin{equation}\label{fpdet}
\Delta_{FP}(A_3)=Det_S'(D_3)\equiv Det_S'(\partial_3 +[A_3,])
\end{equation}
where $Det_S'$ is the determinant of the operator acting on all (monopole) 
scalar functions on the $S^2$, in the given flux sector, 
 exempting diagonal time independent 
modes. The gauge fixing condition (iii) fixes the residual {\it abelian}
gauge invariance. As a consequence the associated Fadeev-Popov determinant
is (divergent) constant, which will be regularized in a suitable way. 
In summary the full Fadeev-Popov determinant for our problem
is given by \eqref{fpdet}. In Appendix \ref{fp} we heuristically evaluate
and interpret this determinant.

\subsection{Evaluation of the path integral over vectors}

In our gauge the path integral \eqref{purecs} reduces to 
\begin{equation}\label{pcp}
Z=\int DA_3 ~\Delta_{FP}(A_3) 
~\exp\left(-T^2 \int d^2x  \sqrt g ~v(U) \right) ~ Z'[A_3]
\end{equation}
where 
\begin{equation}\label{pcpp} \begin{split}
Z'[A_3]& =\int DA_i e^{-(S_1+S_2)}\\
S_1&=\frac{ki }{2 \pi} 
\int d^3x  \Tr \left( A_3(\partial_1 A_2-\partial_2 A_1)\right)\\
S_2&=\frac{k i}{2 \pi} \int d^3x \Tr \left( D_3 A_1 A_2 \right).\\
\end{split}
\end{equation}
It is easily verified that the action $S_1$ receives contributions 
only from time independent diagonal modes of $A_i$. It may also be verified
that these modes (time independent diagonal modes of $A_i$) do not 
contribute to $S_2$. As a consequence the path integral in \eqref{pcpp}
splits into the product of two path integrals, the first over time 
independent diagonal modes of $A_i$, and the second over all other modes. 
The second path integral is quadratic; its evaluation yields 
the determinant $$\frac{1}{\sqrt{Det_V'(\partial_3+[A_3,])}}$$
where by $Det_V'$ we mean the determinant of the operator $D_3$ acting 
in wedge product over (monopole) vector functions (i.e. the determinant of the 
operator in the quadratic form $\int d^3x \epsilon^{ij} A_i D_3 A_j$) on the 
space of all (monopole) vector fields baring time independent diagonal 
modes. Note that this determinant generically depends on the flux 
sector we work in.  

We use the notation $i\frac{\alpha_m(x)}{\beta}$ ($m=1, \ldots, N$) 
for the $N$ eigenvalues of $A_3$ and $a^i_m(x)$
for the time independent diagonal modes of $A_i$. We work in the 
flux sector in which we have $M_m$ units of flux for the $m^{th}$ 
$U(1)$ factor. The path integral over $a^i_m(x)$ may be evaluated 
as follows. In each $U(1)_m$ factor, the most general gauge field 
configuration is given by a constant flux whose integral is given by 
$2 \pi M_m$ plus a fluctuating component of $a^i_m$ that is everywhere 
well defined on the sphere. The most general well defined vector field 
on the sphere is given in terms of two scalars by the formula 
\begin{equation}\label{vexp}
a^i_m(x)=\partial^i \psi_m(x) + \epsilon^{ij} \partial_j \chi_m(x).
\end{equation}
The Coulomb gauge condition sets $\psi_m(x)$ to zero. 
Substituting this solution into $S_1$ in \eqref{pcpp} yields a term proportional to 
$$ \sum_m \int d^3x~ \chi_m \nabla^2 \alpha_m$$
where $\nabla^2$ is the Laplacian defined on the two sphere. 
The integral over $\chi_m$ yields  the delta function 
$$\prod_{m=1}^N \delta(\nabla^2 \alpha_m).$$
This $\delta$ function reduces the functional integral over  
$\alpha_m(x)$ to the ordinary integral over the constant pieces 
$\alpha_m$. Consequently, the path integral over diagonal time 
independent modes in the relevant flux sector is simply given by  
$$\int d \alpha_m e^{ik M_m \alpha_m}.$$
In order to evaluate the full path integral we must sum over all 
flux sectors. Before we can do that, however, we need to evaluate 
the dependence of the ratio of determinants on the flux sectors. 
We turn to that evaluation now. 

\subsection{Evaluation of the ratio of determinants}

The ratio of the ghost determinant (our Fadeev-Popov determinant) 
and the determinant obtained from integrating over vector fields 
was very carefully performed in the paper of \cite{Blau:1993tv}, and 
we will simply quote their result below. In order to give the 
reader a flavor of the result, however, we present a hand waving 
`evaluation' of the ratio of determinant \eqref{purecs}
 in the zero flux sector.  

Let us consider the vector quadratic form in more detail. Let us 
first work in the zero flux sector. We have 
$$\int d^{2}x\epsilon^{ij} A_i D_3 A_j.$$
In the zero flux sector an arbitrary vector field can be written as 
$\partial_i \phi +
\epsilon_{ij}\partial^j \chi$. Plugging in, we find that the quadratic form 
above is proportional to 
$$ \int d^2 x \left( \phi D_3 \nabla^2 \chi \right) $$ 
where we have integrated by parts and used the fact that $A_3$ is a constant
matrix. Ignoring possible subtleties involving regulation of these 
infinite products, it follows that the quadratic path integral over vectors 
is given by  
$$\frac{1}{\sqrt{Det_V'(D_3)}}=\frac{1}{Det''_S(D_3)}$$
where $Det''_S(D_3)$ is the determinant of the operator $D_3$ acting on scalar 
functions \eqref{fpdet} {\it excluding} the zero mode of the scalar field on $S^2$. 
It follows that the ratio of the ghost determinant \eqref{fpdet} and the determinant
that arises from integrating over vector fields is  
\begin{equation}\label{pd}
\frac{Det'_S(D_3)}{Det''_S(D_3)}=\prod_{m\neq l} 
2\sin \left(\frac{\alpha_m(n_m)-\alpha_l(n_l)}{2} \right)
\end{equation}
where the nonzero contribution to \eqref{pd} arises purely from the zero modes on $S^2$
(see Appendix \ref{fp} for the explicit formula). 

The result \eqref{pd} was obtained much more carefully in the paper 
\cite{Blau:1993tv}. In that paper it was also demonstrated that \eqref{pd}
is valid on a space with the topology of the sphere independent of both 
the background metric of the space as well as the background fluxes 
(the Chern classes). We will not attempt to re-derive this result here, but 
will use it in what follows.\footnote{It is very important below that 
the ratio of determinants is independent of the fluxes. The following 
comments may point towards the intuitive explanation of this rigorously 
derived fact. 
 In a background with $q$ units of flux, an arbitrary 
scalar field may be expanded in scalar spherical harmonics which 
transform in the spin $q, q+1,  \ldots $ representations of $SU(2)$. 
An arbitrary vector field, however, can be expanded in vector spherical 
harmonics which transform in the spin $(q-1, q, \ldots)$ + $(q+1, q+2 \ldots)$ 
representations of $SU(2)$. The number of vector spherical harmonics with 
angular momentum $q+1$ or greater is exactly twice the number of 
scalar spherical harmonics with the same quantum numbers; as a consequence
the contributions of these modes cancel in the ratio of determinants.
However the number components of in a scalar spherical harmonics of angular 
momentum $q$ is precisely one more than half the number of modes 
in vector spherical harmonics of quantum numbers  $q-1$ and $q$ respectively.
These observations may underlie the reason that \eqref{pd} independent of the 
flux sector. }

The evaluation of the ratio of determinants presented here was heuristic 
because we cancelled infinite ratios of determinants without regulating 
them. It turns out that our heuristic derivation is actually accurate 
in the dimensional reduction regulation scheme which we use throughout this 
paper. In other regulation schemes, like regulation with a Yang Mills 
term - the determinant ratio has a phase which shifts the Chern-Simons level. 
Throughout this paper $k$ refers to the renormalized Chern-Simons level; the 
effective level of the theory after accounting for this shift.

\subsection{Summation over flux sectors}

As the ratio of determinants is independent of the flux sector, 
the summation over fluxes operates entirely in the zero mode sector 
and is very easily carried out. We must evaluate
$$\prod_{m=1}^N \int d \alpha_m \sum_{M_m=-\infty}^\infty e^{ik M_m \alpha_m}$$
(recall $M_m$ are the constant units of flux in the $U(1)_m$ factors). 
The summation over $M_m$ constrain $\alpha_m$ 
to take the discrete values\footnote{We thank O. Aharony and S. Giombi for 
emphasizing the importance of the summation over flux sectors for our 
problem.}  
$$\alpha_m= \frac{2 \pi n_m}{k}.$$
In other words the integral over $\alpha_i$ is replaced by a sum over 
the uniformly spaced discrete values.\footnote{In \cite{Grignani:2007xz} the thermal partition function
for $N=8$ SYM on $S^2\times S^1$ was computed in the presence of monopole vacua.} 

Putting the various pieces together, it follows that 
the path integral \eqref{purecs} 
is given by  \eqref{csmint}.

\subsection{Squashed spheres and genus $g$ surfaces}

Our derivation of the formula \eqref{csmint} actually used no feature 
of the $S^2$ spatial manifold apart from its topology (the topology was 
used in the assertion that every regular vector field can be written 
in the form \eqref{vexp}; this is not true when the manifold in question
has nontrivial one cycles). It follows that \eqref{csmint} applies for 
an arbitrary metric on the $S^2$ with the volume $V_{2}$ that appears in 
\eqref{effectivepot} interpreted as the volume of the manifold.

Although we will not consider it in the rest of this paper, there is a 
natural generalization of \eqref{csmint} to the partition function of 
Chern-Simons theory on $\Sigma_g \times S^1$ where $\Sigma_g$ is a 
genus $g$ manifold of arbitrary metric\footnote{The natural conformal 
coupling of a minimally coupled scalar to curvature appears to endow these 
scalars with a negative mass for $g \geq 2$. This suggests that a theory
with a regular scalar is unstable on a Riemann surface for $g \geq 2$. 
However the thermal mass appears to stabilize the theory at high 
enough temperatures.}
\begin{equation}\label{csmintg}
Z^g_{\text{CS} }=\prod_{m=1}^N \sum_{n_m=-\infty}^\infty 
\left[ \left(
\prod_{l \neq m}2 \sin \left(\frac{\alpha_l(\vec n)-\alpha_m(\vec n)}{2}
\right) 
\right)^{1-g}
e^{-V(U)}\right]
\end{equation} 
where the summation over $n_m$ is restricted so that no two $n_m$ are 
allowed to be equal. A straightforward generalization of the 
derivation of \eqref{csmint} yields \eqref{csmintg} \cite{Blau:1993tv}. 
The restriction 
that no two $n_m$ are allowed to be equal is presumably because 
the path integral actually vanishes when two eigenvalues of the holonomy 
matrix are exactly equal.  Note that such configurations have 
to be treated specially because our gauge fixing fails precisely 
where two eigenvalues of the holonomy matrix are equal on the sphere.%
\footnote{We were not forced to face up to this issue in the special case 
$g=0$ because the measure factor in \eqref{csmintg} automatically kills
the contribution of these sectors when $g=0$.} We leave a fuller exposition
of this point to future work.

As we have explained in the introduction and will see in detail below, 
the saddle point holonomy distribution of the $S^2$ partition function 
is an interesting and nontrivial function (with a couple of sharp 
phase transitions) of $\frac{V_{2} T^2}{N}$. This interesting behaviour 
results from a competition between potential $V(U)$ that tends to clump 
eigenvalues, and the measure factor in \eqref{csmint} which tends to 
repel them. At higher genus, on the other hand, the measure factor 
in \eqref{csmintg} is either absent (for $g=1$) or attractive 
(for $g>1$). As a consequence the eigenvalue presumably clump at all 
values of $\frac{V_{2} T^2}{N}$, and the partition function does not appear 
to display any interesting structure as a function of this parameter.
In particular the high temperature partition function on such 
manifolds appears always 
to be given by the $R^2$ partition function regulated by the volume of 
the manifold under study. 

Unlike the special case of $g=0$, the partition function of Chern-Simons 
matter theories on $\Sigma_g\times S^1$ for $g \geq 1$ 
has nontrivial dependence on $\lambda$ \cite{Banerjee:2012gh}
 (and potentially nontrivial 
behaviour) at $V_{2}T^2$ of order unity. It would 
be fascinating to find an exact formula that generalizes \eqref{csmintg}
down to temperatures of order unity  (of course \eqref{csmintg} is only 
conjectured to apply in the high temperature limit, in particular 
when $V_{2}T^2 = {\cal O}(N)$) however we do not consider that problem 
in this paper.

\section{Large $N$ solutions and level rank duality}\label{formal}

The `capped matrix model' \eqref{csmint} is dominated by a saddle point 
configuration in the 't~Hooft large $N$ limit ($N \to \infty$ with 
$\frac{V(U)}{N^2}$ and $\frac{N}{k}$ held fixed). In section \ref{cvu}
above we have determined the explicit form of the potential $V(U)$
for several fundamental matter Chern-Simons theories. In the next section 
we will turn to a study of the large $N$ limit of \eqref{csmint}
with $V(U)$ given by the specific expressions listed in section \ref{cvu}. 
In this section, we develop  a general method to obtain the saddle point 
eigenvalue distributions for `integrals' of the form \eqref{csmint}, 
for an arbitrary single trace potential $V(U)$ (recall the potentials
listed in section \ref{cvu} are all of this single trace form when viewed as 
a function of the thermal mass and the eigenvalue density $\rho$; the dependence on thermal mass 
may be eliminated by extremizing the final answer w.r.t. this variable).%
\footnote{In principle, multi-trace potentials can also be 
dealt with using the methods of this section, by treating the traces 
that appear on the LHS of the equivalent of \eqref{vefe} below as free 
parameters, solving the problem for arbitrary values of these parameters, 
and then determining these parameters from the requirement of self consistency.
We leave a fuller treatment of such potentials to future work.}

\subsection{Large $N$ Solution}

As we have explained in the previous section, the summation over allowed 
values of the eigenvalues $\alpha_i$ in \eqref{csmint} is dominated by a 
saddle point configuration of eigenvalues at large $N$. This saddle point 
minimizes the following effective potential 
\begin{equation}\begin{split}\label{vef}
V(U)-\sum_{m \neq l} \ln 2 \sin \frac{\alpha_m-\alpha_l}{2} {\rm ~~where~~} 
V(U)= \sum_m V(\alpha_m). \\
\end{split}
\end{equation}
Of course $\alpha_m$ can actually run over only a discrete set of 
possibilities. As explained in the introduction, however, the only effect 
of this discreteness, in the large $N$ limit, is an upper bound on the 
eigenvalue densities. The eigenvalues within an upper gap saturate a filled 
Fermi sea. Effective Pauli statistics prevents us from infinitesimally 
varying the positions of these eigenvalues. However those eigenvalues that 
are not located within an upper gap have unoccupied slots within which 
they can move around - effectively continuously in the large $N$ limit. 
The fact that the saddle point extremizes the effective potential \eqref{vef}
then imposes the equation
\begin{equation}\label{vefe}
V'(\alpha_m)=\sum_{m\neq l}\cot \frac{\alpha_m-\alpha_l}{2} 
\end{equation}
for those $\alpha_m$ that do not lie in a filled Fermi sea. 
In terms of the eigenvalue density function $\rho(\alpha)$ 
(see \eqref{den}) we have 
\begin{equation}\label{vefec}
V'(\alpha_0)= N {\cal P}  \int d \alpha \cot \frac{\alpha_0-\alpha}{2} 
\rho(\alpha) \end{equation}
where ${\cal P}$ represents the principal value (this has its origin 
in the $m \neq l$ in \eqref{vefe}). \eqref{vefe} applies only for 
$\alpha_0$ in the complement of the upper and lower gaps of the eigenvalue 
distributions.

In the absence of upper gaps, \eqref{vefec} is a generalization of the 
Gross-Witten-Wadia problem. This problem has been very well studied 
and admits a general solution in terms of an integral involving the 
potential $V(x)$ (see Appendix \ref{rgww} for a review). We are interested 
in eigenvalue density functions with upper as well as lower gaps. A simple 
change of variables in \eqref{vefec}, however, suffices to convert every 
upper gap into an effective lower gap, as we now explain.  
Let $\rho_0(\alpha)$ represent any trial 
eigenvalue distribution that has the right boundary conditions (i.e. is a continuous function
that vanishes on all lower gaps and equals $\frac{1}{2 \pi \lambda}$ on all upper
gaps). The detailed form of $\rho_0$ is arbitrary; we 
are free to choose it to suit our convenience. 
 Under the substitution $\rho(\alpha)= \rho_0(\alpha) 
+ \psi(\alpha)$, \eqref{vefec} reduces to 
\begin{equation}\label{vvec} \begin{split}
&N {\cal P} \int d \alpha ~\psi(\alpha) 
 \cot \left( \frac{\alpha_0-\alpha}{2} \right)
=U(\alpha_0)\\
&\int d \alpha~ \psi(\alpha)=A[\rho_o]\\
&U(\alpha)=V'(\alpha)-N {\cal P} \int d \theta ~\rho_0(\theta)
\cot \left( \frac{\alpha-\theta}{2} \right) 
\\
&A[\rho_0]=1-  \int d \alpha ~\rho_0(\alpha).
\end{split}
\end{equation}
The important point is that the new variable $\psi(\alpha)$ is 
nonzero only in the complement of {\it both} upper gaps as well 
as the lower gaps (if any). In other words the formulas reviewed
in Appendix \ref{rgww} - with $U$ and $A$ listed in \eqref{vvec} - 
apply to the determination of $\psi(\alpha)$. In Appendix \ref{cumm}
we have explicitly plugged \eqref{vvec} into the method of Appendix 
\ref{rgww} to find a formal solution for the eigenvalue distribution 
of the capped unitary matrix model with upper as well as lower gaps. Our final 
solution is independent of the arbitrary trial density $\rho_0$, as of 
course had to be the case. When $V(U)$ is an arbitrary potential our 
slightly implicit final answer is listed in the appendices. 
Specifically, the eigenvalue distribution 
is given by \eqref{rhofin} with $\psi_V$ defined 
in \eqref{psiv} in terms of the function $\Phi(u)_V$ defined in 
\eqref{hgp}. The contour of integration $L_{ugs}$ runs over all upper 
gap arcs on the unit circle. The locations of these arcs is determined by 
the normalization condition \eqref{bcpv}. 

\subsection{Level Rank duality of the large $N$ solution}\label{lrdln}

In this subsection we will demonstrate the following. Let $\rho(\alpha)$ 
be a solution to the saddle point equation
\begin{equation}\label{vefec2}
V'(\alpha_0)= N {\cal P}  \int d \alpha \cot \frac{\alpha_0-\alpha}{2} 
\rho(\alpha). \end{equation}
Then 
\begin{equation}
{\tilde \rho}(\alpha)=\frac{\lambda}{1-\lambda}
\left( \frac{1}{2 \pi \lambda}-\rho(\alpha +\pi) \right)
\end{equation}
solves the equation 
\begin{equation}\label{vefecb}
{\tilde V}'(\alpha_0)= (k-N) {\cal P}  \int d \alpha \cot \frac{\alpha_0-\alpha}{2} 
{\tilde \rho}(\alpha) 
\end{equation}
where the potential ${\tilde V}$ is defined in \eqref{tvd}.
In other words the large $N$ 
saddle point equations are consistent with level rank 
duality. As in the previous section, in this section we 
assume that the potential $V(\alpha)$ takes the 
single trace form $V(\alpha)=\sum_i V(\alpha_i)$. 
However the result that our saddle point equations are consistent 
with level rank duality applies also to multi-trace potentials 
$V(U)$, as we demonstrate in Appendix \ref{multitrace}.

It follows almost immediately from definitions that\footnote{\eqref{vtd} may be demonstrated as follows. Let
\begin{equation}
V(U)= 
\sum_{n} A_{n} \tr U^n + c.c. 
\end{equation}
so that 
$$V(\alpha)=\sum_n A_n e^{i n \alpha} +c.c$$
where $A_{n}$ are arbitrary constants.
Now the dual theory potential is given by making the replacement  
$\tr U^n \rightarrow (-1)^{n+1}\tr U^n$  
in $V(U)$, i.e. 
\begin{equation}
\tilde{V}(U) =
\sum_{n} A_{n} (-1)^{n+1}\tr  U^n + c.c
\end{equation}
so that 
$${\tilde V}(\alpha)=\sum_n A_n (-1)^{n+1} e^{i n \alpha}
=-\sum_n A_n e^{i n (\alpha + \pi)} =-V(\alpha+\pi)$$
\eqref{vtd} follows immediately by differentiating both sides of this equation.} 
\begin{equation}\label{vtd}
{\tilde V}'(\alpha)=-V'(\alpha +\pi).
\end{equation}
Using \eqref{vtd}, \eqref{vefecb} (which we wish to prove) may be rewritten as 
\begin{equation}\label{vtpp}
-{V}'(\alpha_0+\pi)= 
N \frac{1-\lambda}{\lambda} 
{\cal P}  \int d \alpha \cot \frac{\alpha_0-\alpha}{2} 
 \frac{\lambda}{1-\lambda}
\left( \frac{1}{2 \pi \lambda}-\rho(\alpha +\pi) \right).
\end{equation}
Using the integral identity 
\begin{equation}
\int_{-\pi}^\pi d\alpha  \cot \frac{\alpha_0-\alpha}{2}=0,
\end{equation}
\eqref{vtpp} reduces to 
\begin{equation}\label{vtppp}
-{V}'(\alpha_0+\pi)= 
-N  {\cal P}  \int d \alpha \cot \frac{\alpha_0-\alpha}{2} 
 \rho(\alpha +\pi).
\end{equation}
However \eqref{vtppp} is simply \eqref{vefec2} evaluated at argument 
$\alpha+\pi$. As we have assumed that \eqref{vefec2} holds at all values 
of the argument, it follows that \eqref{vtppp} and hence \eqref{vefecb} also 
holds.

On the saddle point, the partition function evaluates to
\begin{equation} \label{acm}
-\ln Z= V[\rho]- N^2{\cal P}\int d \alpha d \beta 
\rho(\alpha) \rho(\beta) \ln 
\left( 2 \sin \frac{\alpha-\beta}{2} \right).
\end{equation}
Using the integral identity 
\begin{equation}\label{idente}
\int_{-\pi}^\pi  d \alpha  \ln \left( 2 \sin \frac{\alpha-\beta}{2} \right) =0
\end{equation}
and the defining relation $V[\rho] = {\tilde V}[{\tilde \rho}]$
it is easy to verify that $\ln Z$ may also be written as 
\begin{equation} \label{acm-dual}
-\ln Z= {\tilde V}[{\tilde \rho}]- (k-N)^2 {\cal P}\int d \alpha d \beta
{\tilde \rho}(\alpha) {\tilde \rho}(\beta) \ln 
\left( 2 \sin \frac{\alpha-\beta}{2} \right).
\end{equation}
In other words the saddle point equations and the value of the partition 
function evaluated on the saddle point both respect level rank duality.

\section{Exact solution of the $S^2$ partition function of CS matter theories in the low temperature phase}\label{est}

In the previous section, section \ref{formal}, we have recast the problem of
computing the $S^2$ partition function of fundamental matter Chern-Simons 
theories into the evaluation of a `capped' large $N$ matrix model. 
We have also reduced the problem of evaluating this large $N$ matrix 
model to the determination of solutions of a saddle point equation, and 
presented a general - if slightly implicit- method to determine the 
solutions of these equations. In order to find explicit results for the 
$S^2$ free energy of any given matter Chern-Simons theory, we need to 
implement this general procedure for the particular functions $v[\rho]$ 
determined in section \ref{cvu}. In this section we partially 
implement this programme, in a manner we now explain.

As we have explained in the introduction (and in much greater detail 
in the context of an example in the next section), capped matrix models 
of the sort that arise from fundamental matter Chern-Simons theories 
typically have four qualitatively different phases. These consist of 
a low temperature `no gap' phase, an intermediate temperature `upper gap'
or `lower gap' phase, and a high temperature `two gap' phase. In this 
section, we present the exact evaluation of the $S^2$ partition function of 
all the Chern-Simons matter theories studied in section \ref{cvu} in 
the low temperature `no gap' phase. 
\footnote{In \cite{Takimi:2013zca}, which is a sequel of this paper, 
higher temperature phases with gaps are studied.}
This exercise allows us, in particular, 
to determine the first phase transition temperature of all these theories 
as a function of $\lambda$. It also allows us to determine the `critical' 
value of $\lambda,$ $\lambda_c$ for each of these theories (the value of 
$\lambda$ above which the intermediate temperature phase is the upper gap 
rather than the lower gap phase). Finally we are able to verify that 
the free energy of this phase matches smoothly onto the ($\lambda$ 
independent, non renormalized) spectrum of multi traces at low enough 
temperatures.

 Let the eigenvalue distribution in the no gap phase  be given by
\begin{equation}\label{ron}
 \rho(\alpha)=\frac{1}{2\pi}+\frac{1}{2\pi}\sum_{n=1}^{\infty}(\rho_{n}e^{-i n \alpha}+\rho_{-n}e^{i n \alpha}).
\end{equation}
The repulsive measure factor in the potential \eqref{vef} may be 
rewritten in terms of $\rho_n$ as 
\begin{equation}\begin{split}
\sum_{i,j}\ln 2\sin\frac{\alpha_{i}-\alpha_{j}}{2}&=N^2\int_{-\pi}^{\pi} d\alpha\int_{-\pi}^{\pi} d\beta \rho(\alpha)\rho(\beta)\ln 2\sin\frac{\alpha-\beta}{2}\\
&=-N^2\int_{-\pi}^{\pi} d\alpha\int_{-\pi}^{\pi} d\beta \rho(\alpha)\rho(\beta)\sum_{n=1}^{\infty}\frac{1}{2n}\left(e^{in(\alpha-\beta)}+e^{-in(\alpha-\beta)}\right)\\
&=-N^2\sum_{n=1}^{\infty}\frac{1}{n}\rho_{n}\rho_{-n},\\
\end{split}\end{equation}
where in the last line we have used 
\begin{equation}\label{fron}
 \int_{-\pi}^{\pi}d\alpha\rho(\alpha)e^{-in\alpha}=\rho_{n}.
\end{equation}
It follows that the potential \eqref{vef} can be written in terms of 
$\rho$ as  
\begin{equation}\label{extrmizerho}
V_S= N T^2 v[\rho]+N^2\sum_{n=1}^{\infty}\frac{1}{n}\rho_{n}\rho_{-n}.
\end{equation}
As we have seen in section \ref{cvu}, $v[\rho]$ for matter Chern-Simons 
theories always takes the form 
$$ v[\rho]= -\sum_{n=1}^\infty  \frac{a_n(c, \lambda)}{n} (\rho_n +\rho_{-n}), $$
where $a_n(c, \lambda)$ are coefficients that depend on a `thermal mass' 
constant $c$ which itself is determined (by a complicated 
constitutive equation) as a 
function of $\rho_n$. 
Throughout this section we will treat $c$ as a free 
parameter, solve for $\rho_n$ in terms of $c$ by extremizing the potential
 \eqref{extrmizerho} w.r.t. $\rho_n$ to obtain 
$$\rho_n= \frac{T^2}{N} a_n(c, \lambda).$$
(This result is correct only in the low temperature lower gap phase ). 
We then self consistently determine $c$ on this solution, using its 
constitutive equation for $c$, completing the determination of $\rho(\alpha)$. 
We now implement this procedure for the various theories we study. 
\footnote{As we have explained in section \ref{cvu}, the constitutive 
equation for $c$ may be obtained by extremizing $v[\rho]$ w.r.t $c$ 
at arbitrary constant $\rho$. Once $\rho_n$ have been obtained by 
extremization, however, it follows immediately that this value of 
$c$ also extremizes the potential $V_S$ in \eqref{extrmizerho}. This follows 
immediately from the observation that 
$$\frac{d V_S}{d c}=\partial_c V_S|_{\rho_n} + \partial_{\rho_n} V_S|_{\rho_m, c} 
\frac{d \rho_n}{dc},$$
together with the fact that  $\partial_{\rho_n} V_S|_{\rho_m, c}$ vanishes on 
shell. In other words the value of $c$ may be extremizing $V$, simultaneously
$V$ is extremized w.r.t. $\rho_n$ and $c$.}

\subsection{Supersymmetric theory}\label{rosusysub}
We first consider the supersymmetric theory. From \eqref{vsusy} we obtain
\begin{equation}\label{Vrosusy}\begin{split}
 &V(\rho)=V_{2}T^{2}v[\rho]\\
&=-\frac{N V_{2} T^2}{6 \pi \lambda}\left(\tilde c^3 - 6 \lambda  \int_{-\pi}^{\pi} d\alpha \rho(\alpha) \text{Re} \int_{\tilde c} ^\infty dy y \log \tanh {y+i\alpha \over 2}\right)\\
&=-\frac{N V_{2} T^2}{6 \pi \lambda}\left(\tilde c^3 - 3 \lambda  \int_{-\pi}^{\pi} d\alpha \rho(\alpha)  \int_{\tilde c} ^\infty dy y \left(-\sum_{m=0}^{\infty}\frac{2}{2m+1}e^{-(2m+1)y}(e^{i(2m+1)\alpha}+e^{-i(2m+1)\alpha})\right)\right)\\
&=-\frac{N V_{2} T^2}{6 \pi \lambda}\left(\tilde c^3 - 3 \lambda   \int_{\tilde c} ^\infty dy y \left(-\sum_{m=0}^{\infty}\frac{2}{2m+1}e^{-(2m+1)y}(\rho_{(2m+1)}+\rho_{-(2m+1)})\right)\right)\\
&=-\frac{N V_{2} T^2}{6 \pi \lambda}\left(\tilde c^3 - 3 \lambda    \left(-\sum_{m=0}^{\infty}\frac{2}{(2m+1)^3}(1+ (2m+1) {\tilde c})e^{-(2m+1) {\tilde c}}(\rho_{(2m+1)}+\rho_{-(2m+1)})\right)\right),
\end{split}
\end{equation}
where in the last equation, going from second line to third we have Taylor series expanded log terms in $y,$ going from third to
fourth line we have used \eqref{fron} and while going from 
 fourth to fifth line we have used 
\begin{equation}\label{integral}
 \int_{A}^{\infty}dy~ ye^{-ny} = \frac{1}{n^2}e^{-nA}(1+nA).
\end{equation}
The saddle point for $\rho_{n}, \rho_{-n}$ can be obtained by using 
\eqref{Vrosusy} and extremizing \eqref{extrmizerho}  with respect to $\rho_{-n}, \rho_{n}.$
This gives
\begin{equation}\label{romns}\begin{split}
 \rho_{-2n}&= \rho_{2n}=0,\\
\rho_{-(2n+1)} &= \rho_{2n+1}= \frac{V_{2} T^2}{ N\pi}\frac{1}{(2n+1)^2}e^{-(2n+1)\tilde c}(1+(2n+1)\tilde c).
\end{split}\end{equation}
Plugging \eqref{romns} into \eqref{ron}, we obtain
\begin{equation}\label{rhocri1}
 \rho(\alpha)=\frac{1}{2\pi}+\frac{V_2 T^2}{\pi^2 N}\sum_{m=0}^{\infty}\cos(2m+1)\alpha \frac{1+(2m+1)\tilde c}{(2m+1)^2}e^{-(2m+1)\tilde c}.
\end{equation}
Extremizing $v[\rho]$ w.r.t ${\tilde c}$ at constant $\rho$ yields 
\eqref{susyct}
(as explained in the previous subsection, this is the same as 
extremizing  \eqref{extrmizerho} with respect to $\tilde c,$ on shell) 
which takes the form
\begin{equation}
\begin{split}\label{critilc}
\tilde c &=\lambda\sum_{n=0}^{\infty}\frac{2}{2n+1}e^{-(2n+1)\tilde c}(\rho_{2n+1}+\rho_{-(2n+1)})\\
&=\frac{4 V_2 T^2 \lambda}{N \pi}\sum_{n=0}^{\infty}\frac{1}{(2n+1)^3}e^{-2(2n+1)\tilde c}(1+(2n+1)\tilde c),\\
\end{split}
\end{equation}
where going from first to second line we have used \eqref{romns}.

\eqref{critilc} 
completely determine $\rho(\alpha)$ in the 
no gap phase. These equations define a legal no gap solution provided that 
everywhere $\rho(\alpha) \geq 0$ and $\rho(\alpha) 
\leq \frac{1}{2 \pi \lambda}$. It is easy to convince oneself 
that $\rho(\alpha)$ is maximum at $\alpha=0$ and minimum at $\alpha=\pi$. 
\footnote{For any $\tilde{c}\ge 0$, 
actually, we can show that
\begin{equation}
0 \le \sum_{n=0}^{\infty} \frac{1}{(2n+1)^2}(1+(2n+1)\tilde{c})
e^{-(2n+1)\tilde{c}} \le 
\frac{\pi^2}{8}
\end{equation}
because the function is a monotonically decreasing function with respect to 
$\tilde{c}$.
Hence \eqref{rhocri1} is bounded as
\begin{equation}
\frac{1}{2\pi} + \frac{V_2T^2}{8N}\ge
\rho(\alpha) \ge
\frac{1}{2\pi} - \frac{V_2T^2}{8N}.
\end{equation}
So if $V_2T^2 \ll N$, it is legitimate to argue
$0 \le \rho(\alpha) \le \frac{1}{2\pi \lambda}$.
}
It follows that we have a legal no gap solution provided that 
\begin{equation}\label{ugc}
\rho(0) \leq \frac{1}{2 \pi \lambda},
\end{equation} and 
\begin{equation}\label{lgc}
\rho(\pi) \geq 0.
\end{equation}
For $\lambda <\lambda_{c}$ it turns out that \eqref{lgc} is violated 
at a lower temperature than the one which breaks \eqref{ugc}; 
the reverse is true for 
$\lambda>\lambda_{c}$. 
One can determine $\lambda_{c}$ for this theory by simultaneously solving 
\begin{equation}
\begin{split}
\rho(\pi)=0,~~\rho(0)&=\frac{1}{2\pi\lm_{c}},\\
              \end{split}
\end{equation} which respectively gives
\begin{equation}\label{lcd}
\begin{split}
\frac{1}{2\pi}-\frac{V_2 T^2}{\pi^2 N}\sum_{m=0}^{\infty} \frac{1+(2m+1)\tilde c}{(2m+1)^2}e^{-(2m+1)\tilde c}&=0, \\
\frac{1}{2\pi}+\frac{V_2 T^2}{\pi^2 N}\sum_{m=0}^{\infty} \frac{1+(2m+1)\tilde c}{(2m+1)^2}e^{-(2m+1)\tilde c}&=\frac{1}{2\pi\lm_{c}}.
\end{split}
\end{equation}
Clearly, $\lambda_{c}=\frac{1}{2}$ solves \eqref{lcd} 
(we could have anticipated
this result from the self duality of the susy solution). 

\begin{figure}[tbp]
  \begin{center}
  \subfigure[]{\includegraphics[scale=.6]{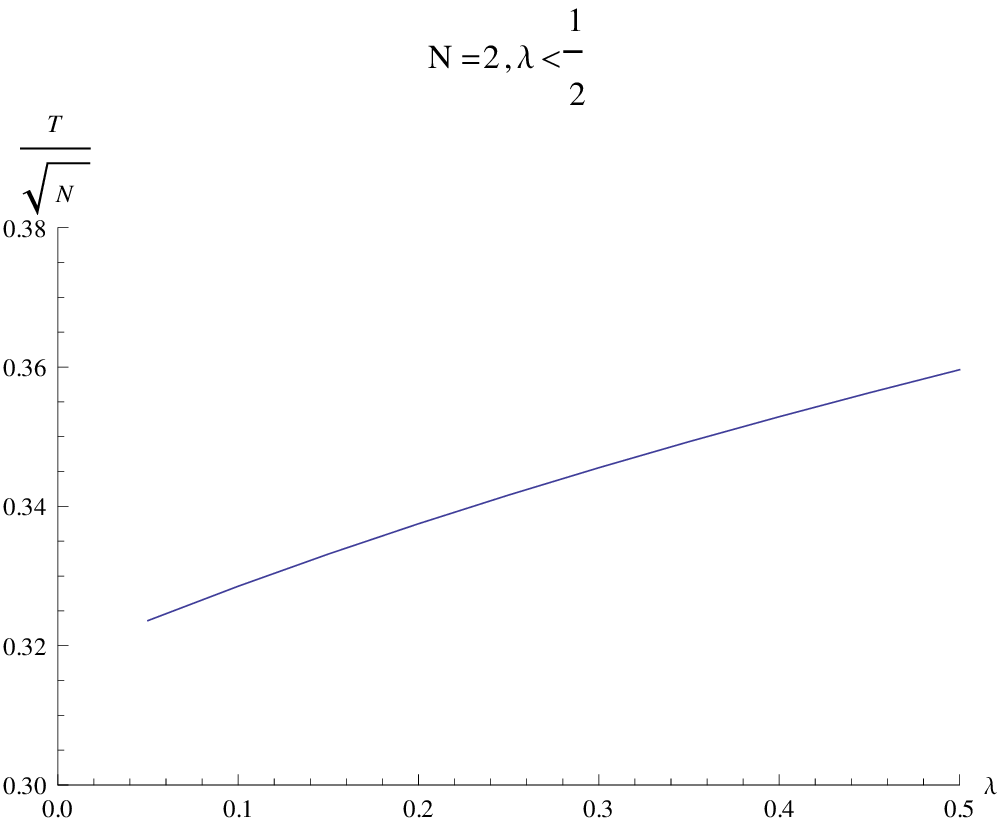}
\label{GWWsusy}
  }
  \qquad\qquad
  \subfigure[]{\includegraphics[scale=.6]{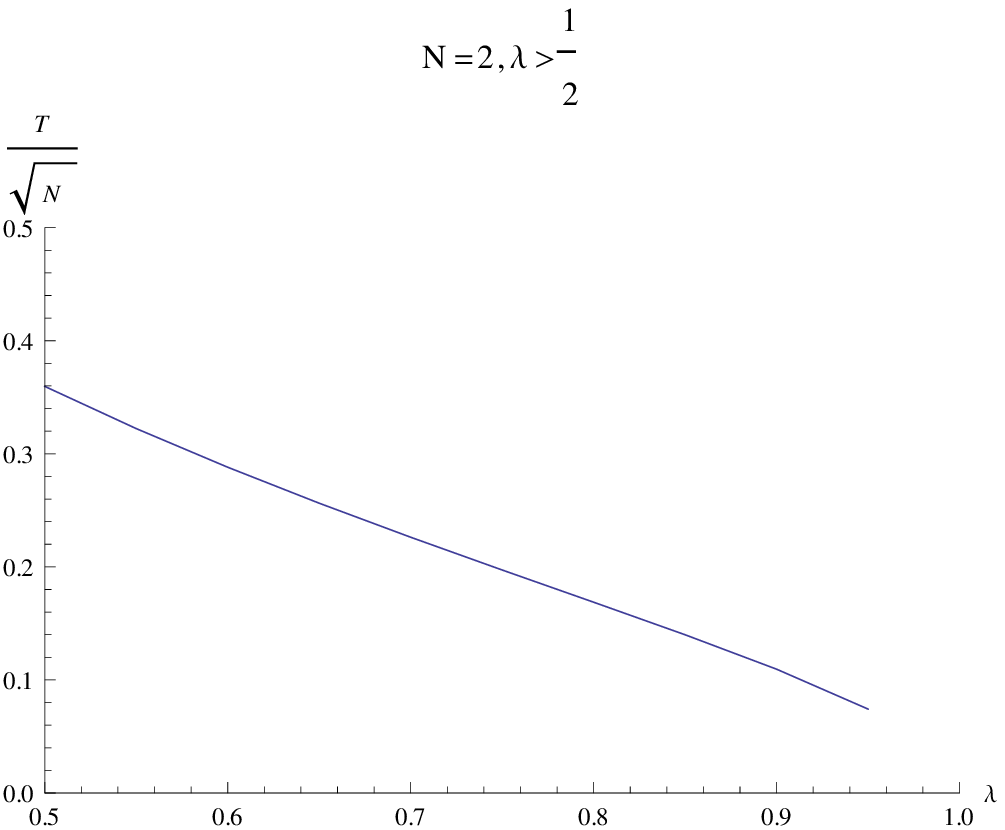}
\label{lmTsusy1}
  }
  \qquad\qquad
  \subfigure[]{\includegraphics[scale=.6]{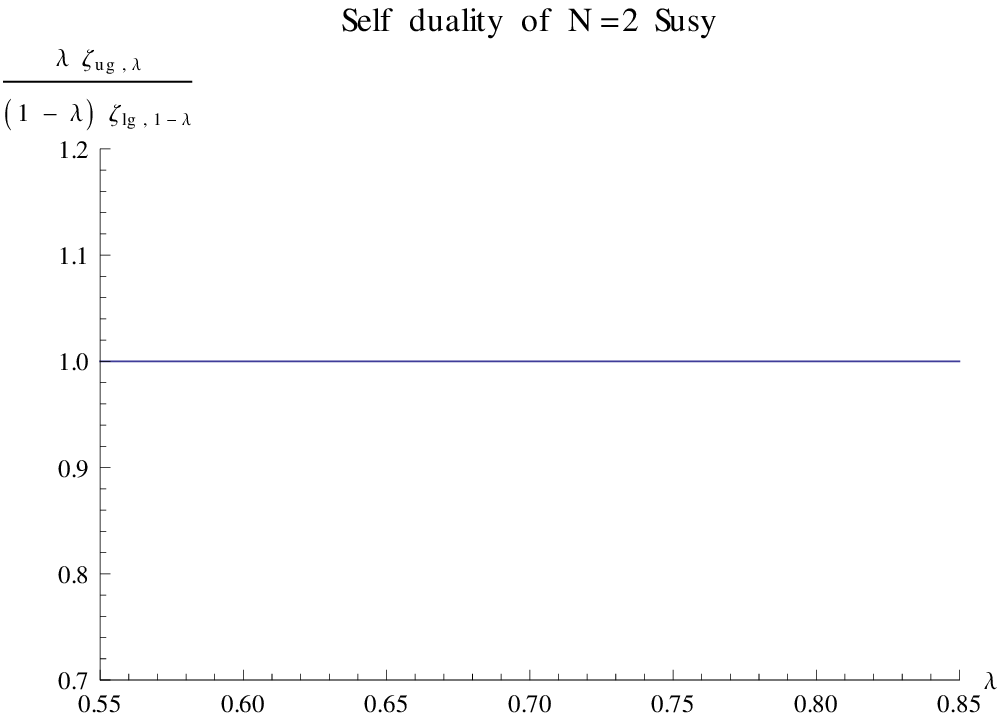}
\label{selfdualfig}
  }
  \end{center}
  \vspace{-0.5cm}
  \caption{We plot  $\frac{T}{\sqrt{N}}$ as a function of $\lm$ for ${\cal N}=2$ supersymmetric case for $\lambda<\frac{1}{2}$ in Fig.(a),
$\lambda >\frac{1}{2}$ in Fig.(b). In Fig.(a), $T$ denotes the phase transition temperature from no gap phase to lower gap phase. In
Fig.(b), $T$ denotes the phase transition temperature from no gap phase to upper gap phase.
We demonstrate \eqref{buglg} by Fig.(c). }
  \end{figure}

In the rest of this section we work on a round sphere of unit radius, so that 
$V_2=4 \pi$. 
For the values $\lambda < \frac{1}{2},$ the no gap phase transits into 
a lower gap phase via a Gross Witten Wadia type phase transition. We 
have determined this phase transition temperature 
$$ T_c= \sqrt{N} \sqrt{\frac{\zeta_{lg}(\lambda)}{V_{2}}},$$
by numerically solving the equations \eqref{critilc} and $\rho(\pi)=0$.
Our result is plotted in Fig \ref{GWWsusy}. Note that the phase transition 
temperature increases as a function of $\lambda$.

In a similar manner, for values $\lambda > \frac{1}{2},$ the no gap
 phase transits into an upper gap phase via a Gross Witten Wadia type phase 
at a temperature 
$$ T_c= \sqrt{N} \sqrt{\frac{\zeta_{ug}(\lambda)}{V_{2}}},$$
which we have determined by numerically solving the equations \eqref{critilc} 
together $\rho(0)= \frac{1}{2 \pi \lambda}$.
Our result is plotted in Fig \ref{lmTsusy1}. Note that the phase transition 
temperature decreases as a function of $\lambda$. The phase transition 
temperature at the critical value of $\lambda$ is $T_c=0.403133 \sqrt{N}$.

It follows from duality that 
$$N \zeta_{ug}(\lambda)=(k-N) \zeta_{lg}(1-\lambda),$$
i.e. that 
\begin{equation}\label{buglg}
\zeta_{ug}(\lambda)=\frac{1-\lambda}{\lambda}\zeta_{lg}(1-\lambda) .
\end{equation}
This relationship is graphically verified in Fig.\ref{selfdualfig}.

\subsubsection{Low temperature expansion}

On general grounds the free energy of the sphere partition function takes
the form \begin{equation}\label{sphereparti-f}
-\ln Z= N^2 f\left(\frac{T^2}{N}\right).
\end{equation}
A Taylor expansion of the function $f$ may be obtained as follows. 
To start with, the self energy equation \eqref{susyct} may easily be solved in 
a power series expansion in $\rho_n$. At ${\cal O}(\rho_n^0)$, 
$\tilde{c}$ becomes $\tilde c=0$ 
(for all $\lambda$); at higher orders we find 
\begin{equation}\label{selfroexp-susy}
\tilde c= 2\lm \sum_{n=0}^{\infty}\frac{1}{2n+1}\left(\rho_{(2n+1)}+\rho_{-(2n+1)}\right)+{\cal O}(\rho_n^2).
\end{equation}
Plugging this solution into \eqref{Vrosusy} we find 
\begin{equation}\label{Vroexp-susy}
\begin{split}
 V(\rho) &= -\frac{N V_{2}T^2}{\pi} \sum_{n=0}^{\infty}\frac{1}{(2n+1)^3}\left(\rho_{(2n+1)}+\rho_{-(2n+1)}\right)\\
&+\frac{2 N V_{2}T^2\lm^2}{3\pi} \left(\sum_{n=0}^{\infty}\frac{1}{(2n+1)}\left(\rho_{(2n+1)}+\rho_{-(2n+1)}\right)\right)^3+{\cal O}(\rho_n^4).
\end{split}
\end{equation}
The important point in \eqref{Vroexp-susy} is that the ${\cal O}(\rho_n^1)$ term is independent of $\lm$ and the ${\cal O}(\rho_n^2)$ term vanishes. Now plugging \eqref{Vroexp-susy} in \eqref{extrmizerho} and completing 
square and using the fact that (which follow from \eqref{romns})
\begin{equation}\label{ropnstexsusy}\begin{split}
 \rho_{-2n}=\rho_{2n}&=0,\\
\rho_{-(2n+1)}=\rho_{(2n+1)} &= \frac{V_{2} T^2}{ N\pi}\frac{1}{(2n+1)^2}+{\cal O}\left(\frac{T^6}{N^3}\right),
\end{split}\end{equation} we obtain
\begin{equation}\label{ftsusy1}
 f\left(\frac{T^2}{N}\right)
=-\frac{31 }{8}\frac{V_2}{\pi}\left(\frac{T^2}{N}\right)^{2}\zeta(5)+\frac{2}{3}\left(\frac{V_{2}}{\pi}\right)^{4}\left(\frac{7}{8} \zeta(3)\right)^3 \left(\lm^{\frac{1}{2}}\frac{T^2}{N}\right)^{4}
+{\cal O}\left(\left(\frac{T^2}{N}\right)^{5}\right).
\end{equation}
Note that, ${\cal O}\left(\frac{T^4}{N^2}\right)$ term 
is independent of $\lm$. Also note that, dependence of $\lm$ of ${\cal O}\left(\frac{T^8}{N^4}\right)$ in $f\left(\frac{T^2}{N}\right)$
is such that under duality, $\ln Z$ as defined in \eqref{sphereparti-f} is invariant.
Also note that at low temperature, upon setting $V_2=4\pi,$ ${\cal O}\left(\frac{T^4}{N^2}\right)$ term in $f$ (see \eqref{ftsusy1})
gives $$\ln Z=\frac{31}{2} T^4 \zeta(5),$$ where we have used \eqref{sphereparti-f}. This result is in precise agreement with the high temperature limit 
of the partition function over a gas of (non renormalized) multitrace 
operators (see \eqref{freelogzb}).

\subsection{Critical boson}\label{rocribossub}
For the critical boson, using \eqref{vcritb} we obtain
\begin{align}\label{Vrocb}
 V(\rho)=V_{2}T^{2}v[\rho]&=-\frac{N V_{2} T^2}{6\pi}\sigma^3  + \frac{N V_{2} T^2}{2\pi}\int_{\sigma}^{\infty} dy\int_{-\pi}^{\pi} d\alpha~y \rho(\alpha) \left(\ln(1-e^{-y+i\alpha})+\ln(1-e^{-y-i\alpha})\right)\nonumber\\
&=-\frac{N V_{2} T^2}{6\pi}\sigma^3  - \frac{N V_{2} T^2}{2\pi}\int_{\sigma}^{\infty} dy\int_{-\pi}^{\pi} d\alpha~y \rho(\alpha)\sum_{n=1}^{\infty}\frac{1}{n}
 \left(e^{-ny+in\alpha}+e^{-ny-in\alpha}\right)\nonumber\\
&=-\frac{N V_{2} T^2}{6\pi}\sigma^3  - \frac{N V_{2} T^2}{2\pi}\int_{\sigma}^{\infty} dy~y \sum_{n=1}^{\infty}\frac{1}{n}
 e^{-ny}\left(\rho_{n}+\rho_{-n}\right)\nonumber\\
&= -\frac{V_2 N T^2}{6\pi}\sigma^3  -\frac{V_2 N T^2}{2\pi}\sum_{n=1}^{\infty}\frac{1}{n^3}(1+n\sigma)(\rho_{n}+\rho_{-n})e^{-n\sigma},
\end{align} where 
going from first line to second, 
we have used Taylor series expanded log terms in $y,$ going from second to
third line we have used \eqref{fron} and while going from 
 third to fourth line we have used \eqref{integral}.

\begin{figure}[tbp]
  \begin{center}
  \subfigure[]{\includegraphics[scale=.6]{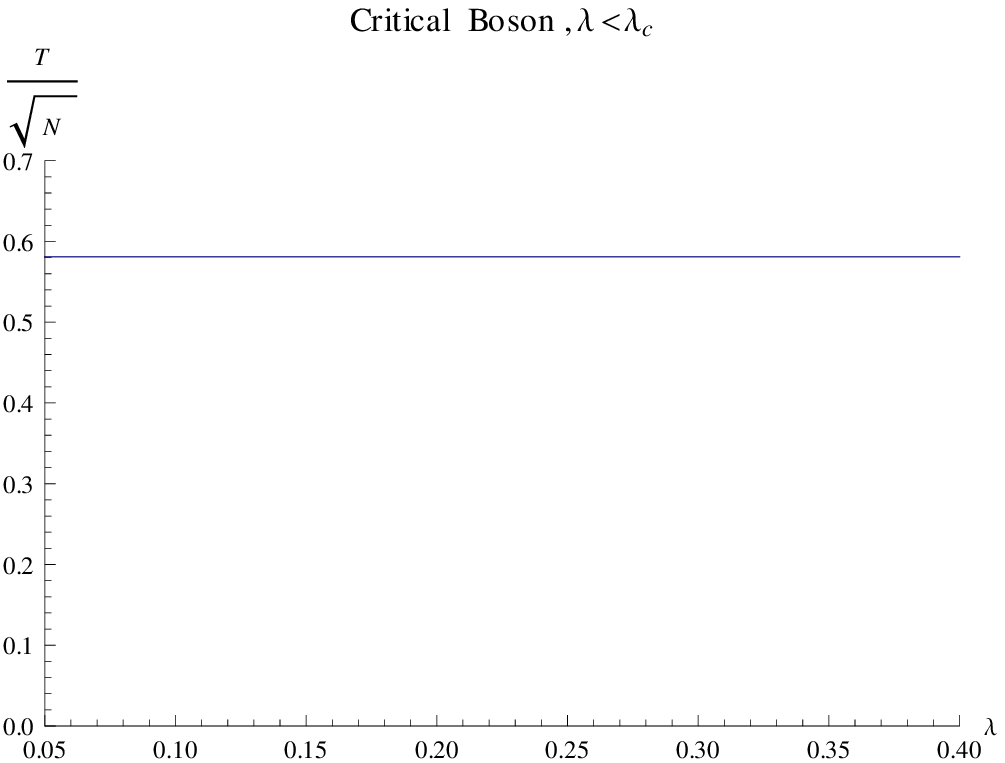}
\label{GWWcribos}
  }
  \qquad\qquad
  \subfigure[]{\includegraphics[scale=.6]{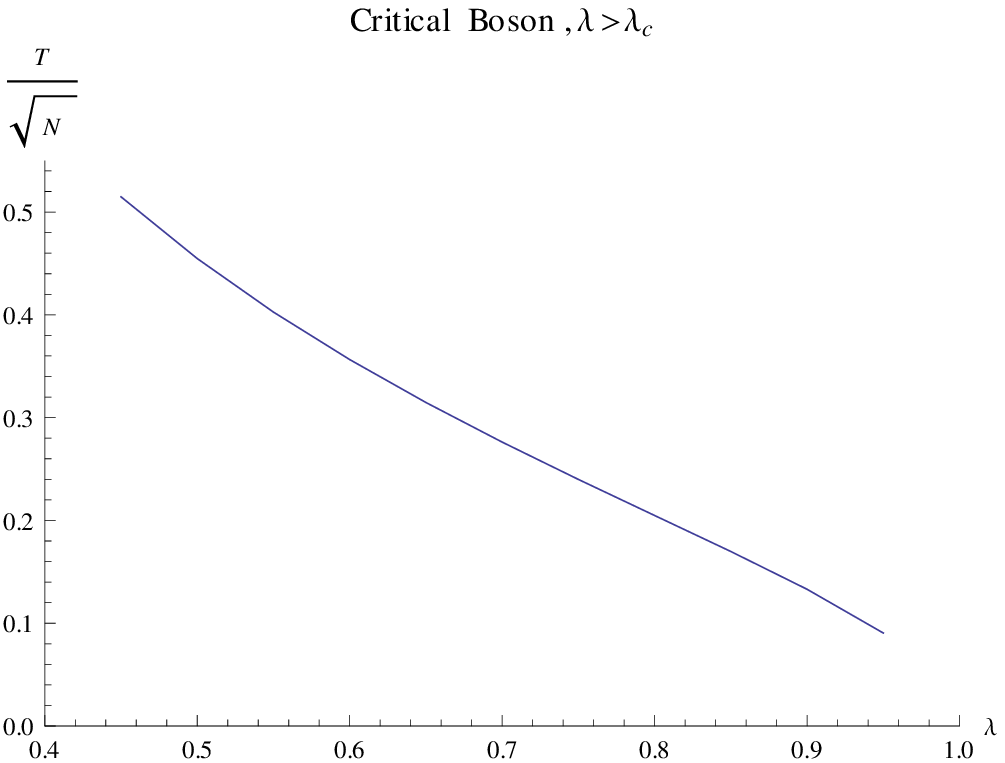}
\label{lmTcribos1}
  }
  \end{center}
  \vspace{-0.5cm}
  \caption{We plot phase transition temperature 
$\frac{T}{\sqrt{N}}$ as a function of $\lm$ for critical bosonic theory for $\lambda<\lambda_{c}^{\text{Cri.B.}}(=0.403033)$ in Fig.(a) and for $\lambda >\lambda_{c}^{\text{Cri.B.}}(=0.403033)$ in Fig.(b).
In Fig.(a), $T$ denotes the phase transition temperature from no gap phase to lower gap phase. In
Fig.(b), $T$ denotes the phase transition temperature from no gap phase to upper gap phase.}
  \end{figure}

The saddle point for $\rho_{n},\rho_{-n}$ can be obtained by  using \eqref{Vrocb} and extremizing \eqref{extrmizerho}  with respect to $\rho_{n},\rho_{-n}$
respectively.
This gives
\begin{equation}\label{ropn}
 \rho_{n}=\rho_{-n}=\frac{V_{2} T^2}{2 N\pi}\frac{1}{n^2}e^{-n\sigma}(1+n\sigma).
\end{equation}
Plugging \eqref{ropn} into \eqref{ron}, we obtain
\begin{equation}\label{rhocri}
 \rho(\alpha)=\frac{1}{2\pi}+\frac{T^2 V_2}{2 N\pi^2}\sum_{n=1}^{\infty}\frac{1}{n^2}\cos(n\alpha)e^{-n\sigma}(1+n\sigma).
\end{equation}
Extremizing $v[\rho]$ with respect to $\sigma$ at constant $\rho_n$
yields the equation \eqref{selfboscri} we obtain
\begin{equation}
\begin{split}\label{crisig}
\sigma &=\sum_{n=1}^{\infty}\frac{1}{n}e^{-n\sigma}(\rho_{n}+\rho_{-n})\\
&=\frac{ V_2 T^2}{N \pi}\sum_{n=1}^{\infty}\frac{1}{n^3}e^{-2n\sigma}(1+n\sigma)\\
&=  \frac{ V_2 T^2}{N \pi}(\sigma \text{Li}_{2}(e^{-2\sigma})+\text{Li}_{3}(e^{-2\sigma})),               
\end{split}
\end{equation}
where going from first to second line we have used \eqref{ropn}. The equations \eqref{Vrocb},\eqref{rhocri},\eqref{crisig} are independent of 
$\lambda.$ It follows that the partition function in the no gap phase 
(and also in the lower gap phase) is completely independent of $\lambda$ 
in this case (see Fig. (\ref{GWWcribos})).

 Note that
\eqref{rhocri} and \eqref{crisig} are in perfect agreement with Eq.(50) and Eq.(48) of \cite{Shenker:2011zf}  provided we choose $V_2=4\pi.$
As in the supersymmetric case, at low values of $\lambda$ the first 
phase transition in this system occurs when $\rho(\pi)=0.$ This occurs 
at a temperature
$$T_{c}=0.581067\sqrt{N}.$$ 
As both the no gap phase quantities 
as well as the condition for the phase transition 
are independent of $\lambda$, it follows that this phase transition temperature
is independent of $\lambda$.

Above a critical value of $\lambda$ (denoted by $\lambda_c$), 
the first phase transition in this system
occurs when $\rho(0)=\frac{1}{2\pi \lambda}$. 
One can determine $\lambda_{c}$ for this theory by simultaneously solving 
\begin{equation}
\begin{split}
\rho(\pi)=0,~\rho(0)=\frac{1}{2\pi\lm_{c}}                 
\end{split}
\end{equation} and \eqref{crisig}. This gives 
\begin{equation}\label{lmccribos}
 \lambda_{c}^{\text{Cri.B.}}=0.403033,~~T_{c}=0.581067\sqrt{N}.
\end{equation}
 For $\lambda>\lambda_{c}^{\text{Cri.B.}}$ the phase transition temperature depends 
on $\lambda$.  

We have numerically solved the relevant equations and plotted the 
phase transition temperature in Fig.\ref{GWWcribos} and Fig.\ref{lmTcribos1}. 
Note that 
as $\lm$ is increased  beyond $\lambda_{c}^{\text{Cri.B.}},$ the phase 
transition temperature from no gap to upper gap phase decreases
in Fig.\ref{lmTcribos1}.\\

\subsubsection{Low temperature expansion}
The self energy equation \eqref{selfboscri} may easily be solved in 
a power series expansion in $\rho_n$. At ${\cal O}(\rho_n^0)$, 
$\sigma$ becomes $\sigma=0$ 
(for all $\lambda$); at higher orders we find 
\begin{equation}\label{selfroexp-susy2}
\sigma=  \sum_{n=1}^{\infty}\frac{1}{n}\left(\rho_{n}+\rho_{-n}\right)+{\cal O}(\rho_n^2).
\end{equation}
Plugging this solution into \eqref{Vrocb} we find 
\begin{equation}\label{Vroexp-cb}
\begin{split}
 V(\rho) &= -\frac{N V_{2}T^2}{2\pi} \sum_{n=1}^{\infty}\frac{1}{n^3}\left(\rho_{n}+\rho_{-n}\right)+\frac{ N V_{2}T^2}{12\pi} \left(\sum_{n=1}^{\infty}\frac{1}{n}\left(\rho_{n}+\rho_{-n}\right)\right)^3+{\cal O}\left(\rho_n^4\right).
\end{split}
\end{equation}
As observed in supersymmetric case, the ${\cal O}(\rho_n^1)$ term is independent of $\lm$ and the ${\cal O}(\rho_n^2)$ term vanishes. Now plugging \eqref{Vroexp-cb} in \eqref{extrmizerho} and completing 
square and using the fact that (which follow from  \eqref{ropn})
\begin{equation}\label{ropnstexcb}\begin{split}
 \rho_{-n}=\rho_{n} &= \frac{V_{2} T^2}{2 N\pi}\frac{1}{n^2}+{\cal O}\left(\frac{T^6}{N^3}\right),
\end{split}\end{equation} we obtain
\begin{equation}\label{ftcribos}
 f\left(\frac{T^2}{N}\right)
=-\frac{V_2}{\pi}\left(\frac{T^2}{N}\right)^{2}\zeta(5)+\frac{1}{12}\left(\frac{V_{2}}{\pi}\right)^{4} \left(\zeta(3)\right)^3 \left(\frac{T^2}{N}\right)^{4}
+{\cal O}\left(\left(\frac{T^2}{N}\right)^{5}\right).
\end{equation}
Note that  $f\left(\frac{T^2}{N}\right)$ is independent of $\lm$. This predicts a simple dependence on $\lm$ for dual regular fermionic theory namely
\begin{equation}\label{predictionrf}
 f_{f}\left(\frac{T^2}{N}\right) = \left(\frac{1-\lm}{\lm}\right)^2 f_{cb}\left(\frac{T^2}{N\frac{1-\lm}{\lm}}\right).
\end{equation}
Also note that at low temperature, upon setting $V_2=4\pi,$ 
${\cal O}\left(\frac{T^4}{N^2}\right)$ term in $f$ (see \eqref{ftcribos} and \eqref{sphereparti-f})
gives $$\ln Z=4 T^4 \zeta(5),$$ which is exactly same as 
that for the gas of multitrace operators of free fermions on the sphere 
(recall the critical boson theory has the same single trace spectrum 
as the free fermion theory). For details see  \eqref{freelogzf}.

In rest of the section, we simply present final results. All the steps 
leading to these results are completely analogous to those employed in 
this subsection and the previous one (subsections \ref{rosusysub} and  
\ref{rocribossub}).

\subsection{Regular fermion theory}
For the regular fermionic theory we obtain
\begin{align}\label{Vrorf}
V(\rho)=V_{2}T^{2}v[\rho]&=-\frac{V_2 N T^2}{6\pi \lambda}(1-\lambda)\tilde c^3  -\frac{V_2 N T^2}{2\pi}\sum_{n=1}^{\infty}(-1)^{n+1}\frac{1}{n^3}(1+n\tilde c)(\rho_{n}+\rho_{-n})e^{-n\tilde c},\nonumber\\
\rho_{-n}=\rho_{n}&= \frac{V_2  T^2}{2\pi N }\frac{(-1)^{n+1}}{n^2}e^{-n\tilde c}(1+n\tilde c),\nonumber\\
\rho(\alpha)&=\frac{1}{2\pi}-\frac{V_2 T^2}{2\pi^2 N}\sum_{m=1}^{\infty}(-1)^{m}\cos m\alpha \frac{1+ m~\tilde c}{m^2}e^{-m \tilde c},\nonumber\\
(1-\lambda)\tilde c&=\frac{V_2 T^2 \lambda}{\pi N}\sum_{m=1}^{\infty}\frac{1+ m~\tilde c}{m^3}e^{-2 m \tilde c}.
\end{align}
One can check that under duality last two equations of \eqref{Vrorf} maps to \eqref{rhocri},\eqref{crisig} respectively.
The critical value of $\lambda$ and temperature is given by 
\begin{equation}
\lambda_c^{\text{Reg.F.}} =0.596967,~~ T=0.477444\sqrt{N}.
\end{equation} The critical value $\lambda_c^{\text{Reg.F.}}$ can be checked 
to equal $1-\lambda_c^{\text{Cri.B.}}$ 
(see \eqref{lmccribos} for $\lambda_c^{\text{Cri.B.}}$). 
Upon raising the temperature, the no gap solution first transitions to the 
lower gap solution for  $\lambda < \lambda_{c}^{\text{Reg.F.}}$ but to the 
upper gap solution for  $\lambda > \lambda_{c}^{\text{Reg.F.}}$.
\begin{figure}[tbp]
  \begin{center}
  \subfigure[]{\includegraphics[scale=.6]{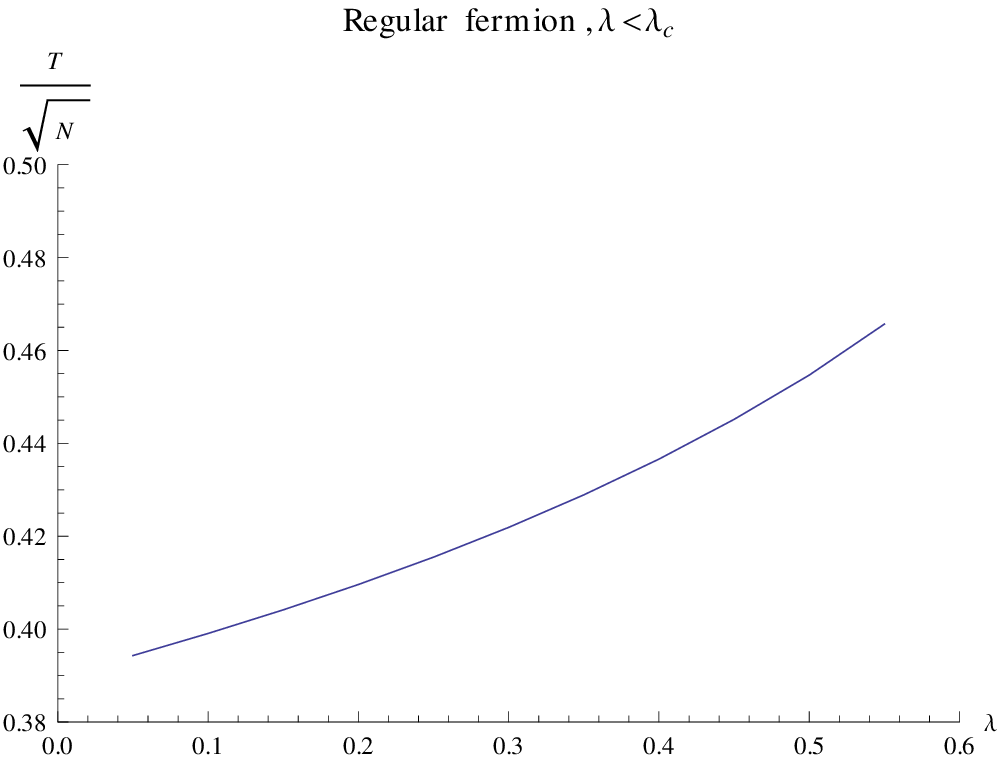}
\label{GWWregfer}
  }
  \qquad\qquad
  \subfigure[]{\includegraphics[scale=.6]{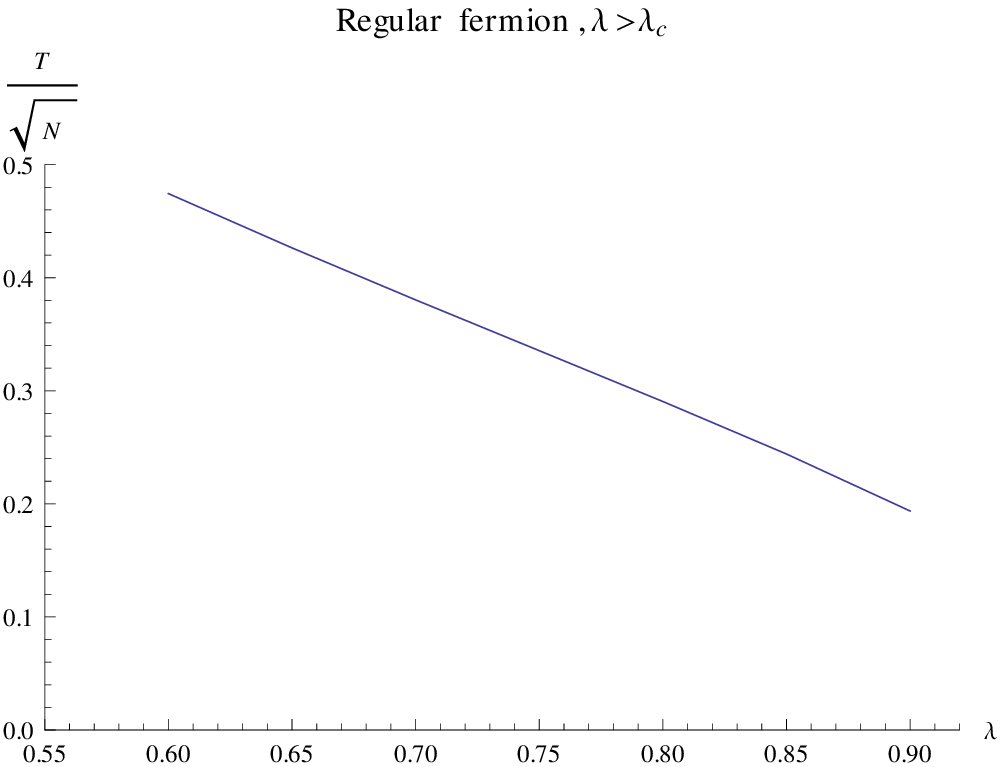}
\label{lmTregfer1}
  }
  \end{center}
  \vspace{-0.5cm}
  \caption{We plot phase transition temperature $\frac{T}{\sqrt{N}}$ as a function of $\lm$ for regular fermionic theory for $\lambda<\lambda_{c}^{\text{Reg.F.}}(=0.596967)$ in Fig.(a) and for $\lambda >\lambda_{c}^{\text{Reg.F.}}(=0.596967)$ in Fig.(b).
In Fig.(a), $T$ denotes the phase transition temperature from no gap phase to lower gap phase. In
Fig.(b), $T$ denotes the phase transition temperature from no gap phase to upper gap phase.}
  \end{figure}
Fig.\ref{GWWregfer} demonstrates that the phase transition temperature
from the no gap phase to the lower gap phase increases as  
we increase $\lm$ keeping $\lm<\lambda_{c}^{\text{Reg.F.}}$. On the 
other hand the phase transition temperature from the no gap phase to 
the upper gap phase decreases upon increasing $\lambda$ when 
$\lambda>\lambda_{c}^{\text{Reg.F.}}$. 

For $\lambda > \lambda_{c}^{\text{Reg.F.}}$, the dependence of the phase 
transition temperature $T$ on $\lambda$ can be predicted 
very simply from  duality. Duality maps the regular fermionic theory at  
$\lambda > \lambda_{c}^{\text{Reg.F.}}$ to the critical bosonic theory 
at $\lm<\lm_{c}^{\text{Cri.B.}}$. 
In this phase, the phase transition temperature, 
$\frac{T}{\sqrt{N}}$ of the critical bosonic theory, is independent 
of $\lm,$ as shown in Fig.\ref{GWWcribos}. This predicts that phase 
transition temperature of the regular fermionic theory is given by 
\begin{equation}
 \frac{T}{\sqrt{N}}=0.581067\sqrt{\frac{1-\lm}{\lm}}.
\end{equation}
One can check that Fig.\ref{lmTregfer1} indeed satisfies this 
simple relationship. 

\subsubsection{Low temperature expansion}
The self energy equation \eqref{tc} may easily be solved in 
a power series expansion in $\rho_n$. At ${\cal O}(\rho_n^0),$ 
$\tilde{c}$ becomes 
$\tilde c=0$ 
(for all $\lambda$); at higher orders we find 
\begin{equation}\label{selfroexp-rf}
 (1-\lm) \tilde c= \lm \sum_{n=1}^{\infty}(-1)^{n+1}\frac{1}{n}\left(\rho_{n}+\rho_{-n}\right)+{\cal O}(\rho_n^2).
\end{equation}
Plugging this solution into first equation of \eqref{Vrorf} we find 
\begin{equation}\label{Vroexp-rf}
\begin{split}
 V(\rho) &= -\frac{N V_{2}T^2}{2\pi} \sum_{n=1}^{\infty}(-1)^{n+1}\frac{1}{n^3}\left(\rho_{n}+\rho_{-n}\right)\\
&+\frac{ N V_{2}T^2 \lm^2}{12\pi\left(1-\lm \right)^2} \left(\sum_{n=1}^{\infty}(-1)^{n+1}\frac{1}{n}\left(\rho_{n}+\rho_{-n}\right)\right)^3+{\cal O}(\rho_n^4).
\end{split}
\end{equation}
As the previous case the ${\cal O}(\rho_n^1)$ term is independent of $\lm$ and the ${\cal O}(\rho_n^2)$ term vanishes.
Now plugging \eqref{Vroexp-rf} in \eqref{extrmizerho} and completing 
square and using the fact that (which follow from second equation in \eqref{Vrorf})
\begin{equation}\label{ropnstexrf}\begin{split}
 \rho_{-n}=\rho_{n} &= \frac{V_{2} T^2}{2 N\pi}(-1)^{n+1}\frac{1}{n^2}+
{\cal O}\left(\frac{T^6}{N^3}\right),
\end{split}\end{equation} we obtain
\begin{equation}\label{ftrf}
 f\left(\frac{T^2}{N}\right)=
-\frac{V_2}{\pi}\left(\frac{T^2}{N}\right)^{2}\zeta(5)+\frac{1}{12}\left(\frac{V_{2}}{\pi}\right)^{4} \left(\zeta(3)\right)^3 \frac{\lm^2}{\left(1-\lm\right)^2}\left(\frac{T^2}{N}\right)^{4}
+{\cal O}\left(\left(\frac{T^2}{N}\right)^{5}\right).
\end{equation}
Note that, ${\cal O}\left(\frac{T^4}{N^2}\right)$ term is independent of $\lm$ and precisely matches with first term in \eqref{ftcribos}. 
It is easy to verify that, \eqref{ftrf} is consistent with \eqref{predictionrf}. Also note that at low temperature, setting $V_2=4\pi,$ we obtain 
$\ln Z=4 T^4 \zeta(5),$ which is exactly same as 
that for the gas of multitrace operators of free fermions on the sphere. For details see \eqref{freelogzf}.

\subsection{Regular boson}
\begin{figure}[tbp]
  \begin{center}
  \subfigure[]{\includegraphics[scale=.6]{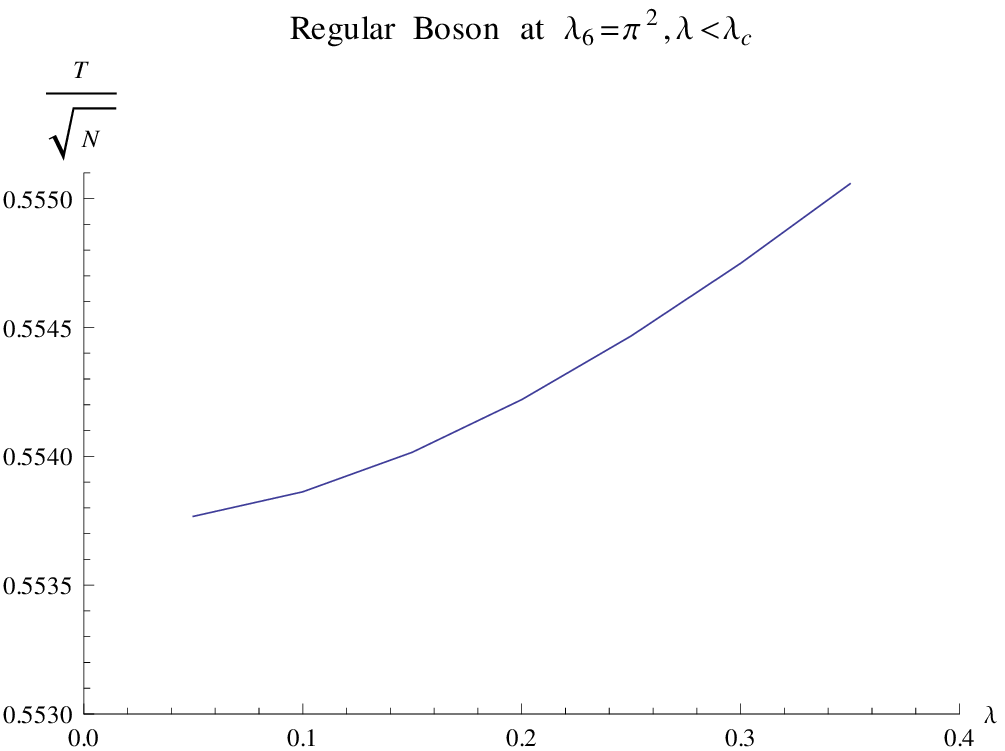}
\label{GWWregbos}
  }
  \qquad\qquad
  \subfigure[]{\includegraphics[scale=.6]{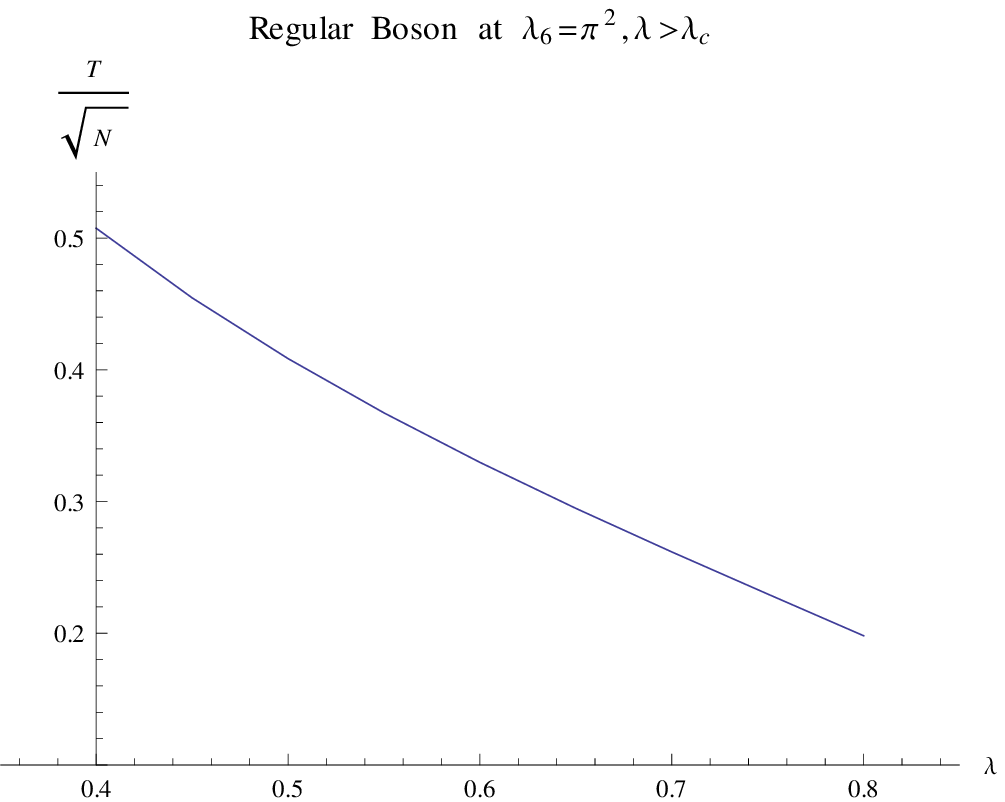}
\label{lmTregbos1}
  }
  \end{center}
  \vspace{-0.5cm}
  \caption{We plot the phase transition temperature 
$\frac{T}{\sqrt{N}}$ as a function of $\lm$ for regular bosonic theory for $\lambda<\lambda_{c}^{\text{Reg.B.}}(=0.360884)$ in Fig.(a) and 
for $\lambda >\lambda_{c}^{\text{Reg.B.}}(=0.360884)$ in Fig.(b).
In Fig.(a), $T$ denotes the phase transition temperature from no gap phase to lower gap phase. In
Fig.(b), $T$ denotes the phase transition temperature from no gap phase to upper gap phase.}
  \end{figure}

For regular bosonic theory we obtain
\begin{align}\label{rhoregu}
V(\rho)=V_{2}T^{2}v[\rho]&=-\frac{V_2 N T^2}{6\pi}\sigma^3 
\left(1+\frac{2}{\sqrt{\frac{\lambda_6}{8\pi^2}+\lambda^2}}\right)  
-\frac{V_2 N T^2}{2\pi}\sum_{n=1}^{\infty}\frac{1}{n^3}(1+n\sigma)(\rho_{n}+\rho_{-n})e^{-n\sigma},\nonumber\\
\rho_{-n}=\rho_{n}&= \frac{V_{2} T^2}{2 N\pi}\frac{1}{n^2}e^{-n\sigma}(1+n\sigma),\nonumber\\
\rho(\alpha)&=\frac{1}{2\pi}+\frac{T^2 V_2}{2 N\pi^2}\sum_{n=1}^{\infty}\frac{1}{n^2}\cos(n\alpha)e^{-n\sigma}(1+n\sigma),\nonumber\\
\sigma \left(1+\frac{2}{\sqrt{\frac{\lambda_6}{8\pi^2}+\lambda^2}}
\right) &=\frac{ V_2 T^2}{N \pi}\sum_{n=1}^{\infty}\frac{1}{n^3}e^{-2n\sigma}(1+n\sigma)
=  \frac{ V_2 T^2}{N \pi}(\sigma \text{Li}_{2}(e^{-2\sigma})+\text{Li}_{3}(e^{-2\sigma})).
\end{align}
For this model, the critical value of $\lambda$, 
which is denoted by $\lambda_{c}^{\text{Reg.B.}}$, 
depends on $\lambda_6.$ When $\lambda_6\rightarrow \infty,$ results are identical to critical bosonic theory. A few other values of $\lambda_c^{\text{Reg.B.}}$ are
\begin{equation}\label{lcril6}\begin{split}
\lambda_{c}^{\text{Reg.B.}}= 0.360884,~T_{c}=0.555125\sqrt{N},&~~\lambda_{6}=\pi^2.\\  
 \lambda_{c}^{\text{Reg.B.}}= 0.373102,~T_{c}=0.559636\sqrt{N},&~~\lambda_{6}=8\pi^2.\\  
\lambda_{c}^{\text{Reg.B.}}= 0.382887,~T_{c}=0.564807\sqrt{N},&~~\lambda_{6}=32\pi^2.\\  
                 \end{split}
\end{equation}
As in previous subsections, for $\lambda < \lambda_{c}^{\text{Reg.B.}}$ the 
no gap solution first transits to the lower gap solution 
(see Fig.\ref{GWWregbos}); the phase 
transition temperature increases as a function of $\lambda$. On the other 
hand for $\lambda > \lambda_{c}^{\text{Reg.B.}}$ the no gap solution 
first transits into the upper gap solution (upon raising the temperature); 
the phase transition temperature decreases as $\lambda$ is increased 
(see  Fig.\ref{lmTregbos1}).

\subsubsection{Low temperature expansion}
The self energy equation \eqref{rboself} may easily be solved in 
a power series expansion in $\rho_n$. At ${\cal O}(\rho_n^0),$ $\sigma$ becomes
$\sigma=0$ 
(for all $\lambda$); at higher orders we find 
\begin{equation}\label{selfroexp-rb}
 \left(1+\frac{2}{\sqrt{\frac{\lambda_6}{8\pi^2}+\lambda^2}}\right) 
\sigma=  \sum_{n=1}^{\infty}\frac{1}{n}\left(\rho_{n}+\rho_{-n}\right)+{\cal O}(\rho_n^2).
\end{equation}
Plugging this solution into first equation of \eqref{rhoregu} we find 
\begin{equation}\label{Vroexp-rb}
\begin{split}
 V(\rho) &= -\frac{N V_{2}T^2}{2\pi} \sum_{n=1}^{\infty}\frac{1}{n^3}\left(\rho_{n}+\rho_{-n}\right)\\
&+\frac{ N V_{2}T^2}{12\pi} \left(1+\frac{2}{\sqrt{\frac{\lambda_6}{8\pi^2}+\lambda^2}}\right)^{-2}\left(\sum_{n=1}^{\infty}\frac{1}{n}\left(\rho_{n}+\rho_{-n}\right)\right)^3+{\cal O}(\rho_n^4).
\end{split}
\end{equation}
As observed in the previous models the ${\cal O}(\rho_n^1)$ term is independent of $\lm$ and the ${\cal O}(\rho_n^2)$ term vanishes. Now plugging \eqref{Vroexp-rb} into \eqref{extrmizerho} and completing 
square and using the fact that (which follow from second equation in \eqref{rhoregu})
\begin{equation}\label{ropnstexrb}\begin{split}
 \rho_{-n}=\rho_{n} &= \frac{V_{2} T^2}{2 N\pi}\frac{1}{n^2}
+{\cal O}\left(\frac{T^6}{N^3}\right),
\end{split}\end{equation} we obtain
\begin{equation}\label{ftrbos}
 f\left(\frac{T^2}{N}\right)
=-\frac{V_2}{\pi}\left(\frac{T^2}{N}\right)^{2}\zeta(5)+\frac{1}{12}\left(\frac{V_{2}}{\pi}\right)^{4} \left(\zeta(3)\right)^3 \left(1+\frac{2}{\sqrt{\frac{\lambda_6}{8\pi^2}+\lambda^2}}\right)^{-2} \left(\frac{T^2}{N}\right)^{4}
+{\cal O}\left(\left(\frac{T^2}{N}\right)^{5}\right).
\end{equation}
Note that the ${\cal O}\left(\frac{T^4}{N^2}\right)$ term is independent of 
$\lm$ and in fact agrees perfectly with the gas of multitraces of the 
free boson, \eqref{freelogzb}.

\subsection{Critical fermion}
\begin{figure}[tbp]
  \begin{center}
  \subfigure[]{\includegraphics[scale=.6]{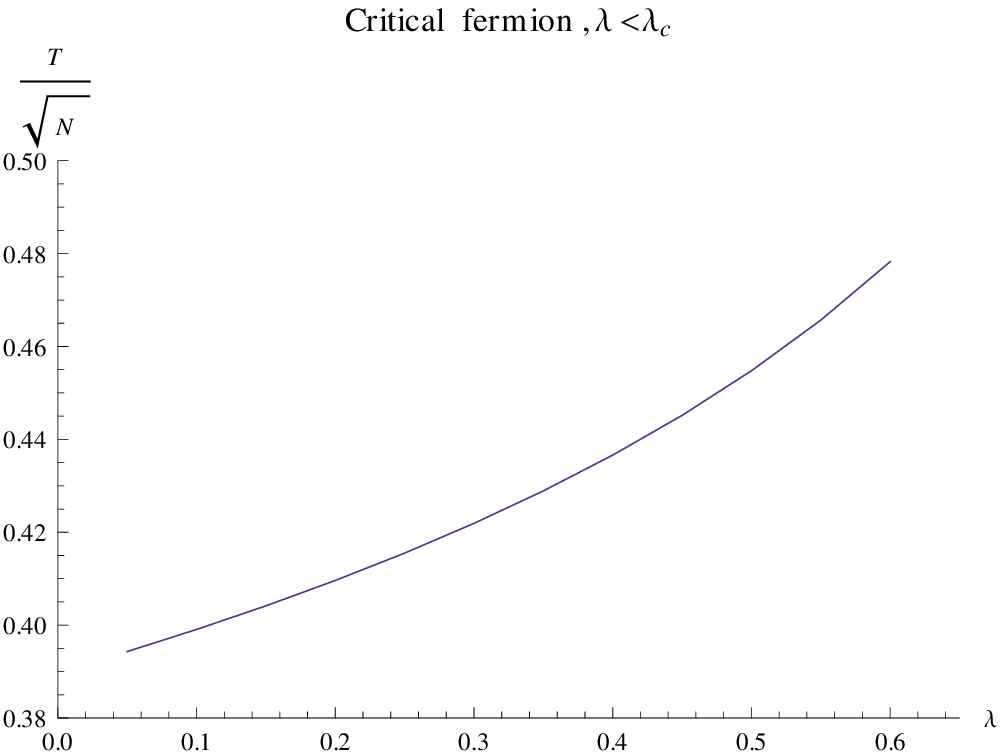}
\label{GWWcrifer}
  }
  \qquad\qquad
  \subfigure[]{\includegraphics[scale=.6]{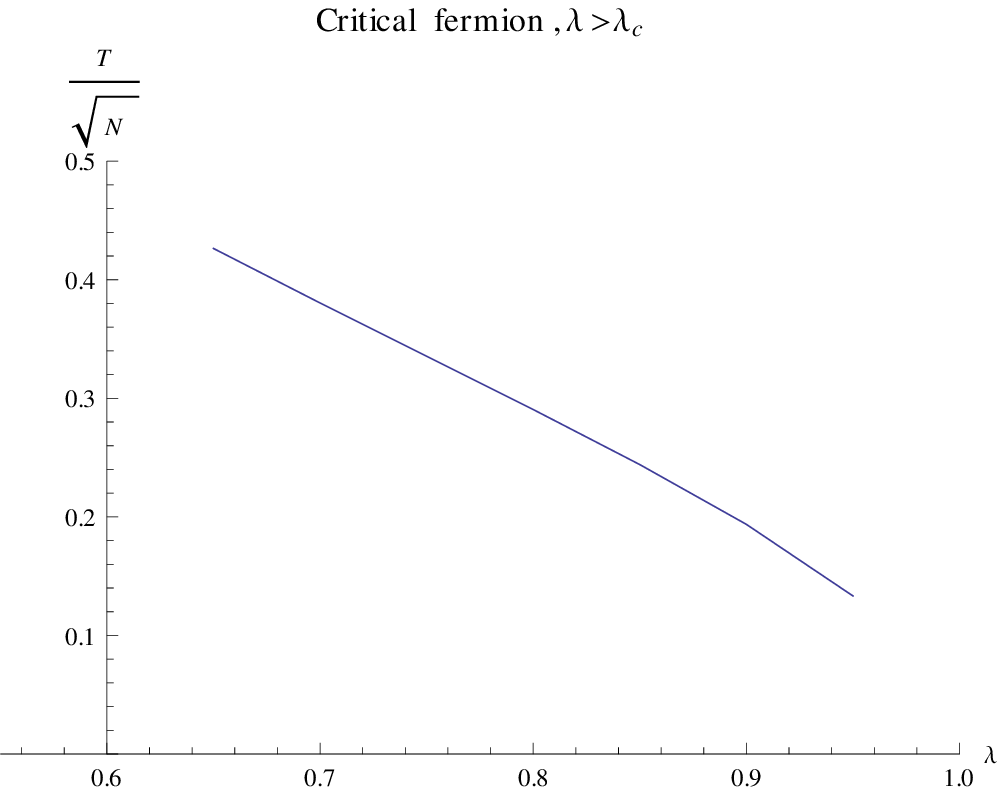}
\label{lmTcrifer1}
  }
  \end{center}
  \vspace{-0.5cm}
  \caption{We plot the phase transition temperature 
$\frac{T}{\sqrt{N}}$ as a function of $\lm$ for critical fermionic for $\lambda<\lambda_{c}^{\text{Cri.F.}}(=0.639116)$ in Fig.(a) and for $\lambda >\lambda_{c}^{\text{Cri.F.}}(=0.639116)$ in Fig.(b).
In Fig.(a), $T$ denotes the phase transition temperature from no gap phase to lower gap phase. In
Fig.(b), $T$ denotes the phase transition temperature from no gap phase to upper gap phase.}
  \end{figure}

For the critical fermionic theory we obtain
\begin{align}\label{criferro}
V(\rho)=V_{2}T^{2}v[\rho]&=-\frac{V_2 N T^2}{6\pi \lambda}(1-\lambda+\hat g)\tilde c^3  -\frac{V_2 N T^2}{2\pi}\sum_{n=1}^{\infty}(-1)^{n+1}\frac{1}{n^3}(1+n\tilde c)(\rho_{n}+\rho_{-n})e^{-n\tilde c},\nonumber\\
\rho_{-n}=\rho_{n}&= \frac{V_2  T^2}{2\pi N }\frac{(-1)^{n+1}}{n^2}e^{-n\tilde c}(1+n\tilde c),\nonumber\\
 \rho(\alpha)&=\frac{1}{2\pi}-\frac{V_2 T^2}{2\pi^2 N}\sum_{m=1}^{\infty}(-1)^{m}\cos m\alpha \frac{1+ m~\tilde c}{m^2}e^{-m \tilde c},\nonumber\\
(1-\lambda+\hat{g})\tilde c&=\frac{V_2 T^2 \lambda}{\pi N}\sum_{m=1}^{\infty}\frac{1+ m~\tilde c}{m^3}e^{-2 m \tilde c},\nonumber\\
\hat{g}(\lm)&=\frac{1}{1-2\pi \lambda\lambda_6^{f}}.
\end{align}
One can check that under duality third and fourth equation of \eqref{criferro} maps to third and forth equation of 
\eqref{rhoregu} with the identification $\hat g(\lm_{f}) = \frac{2 \lambda_{b}}{\sqrt{\lambda_{b}^2+\frac{\lambda_6^{b}}{8\pi^2}}}.$
One can check that, given the $\lambda_b,\lambda_6^{b}$ in \eqref{lcril6}, the $\lambda_{c}^{\text{Cri.F.}}$ for the critical fermionic theory 
satisfies $\lambda_{c}^{\text{Cri.F.}}=1-\lambda_c^{\text{Reg.B.}}.$ For example  for 
$\lambda_{c}^{\text{Reg.B.}}= 0.3731,\lambda^{b}_{6}=8\pi^2$ one gets
$\hat g = 0.69913$ and we obtain $\lambda_{c}^{\text{Cri.F.}} = 0.6269.$
For $\lambda < \lambda_{c}^{\text{Cri.F.}}$ the no gap solution 
first transits to the lower gap solution upon increasing the 
temperature; the phase transition temperature increases 
as $\lambda$ is increased (see Fig.\ref{GWWcrifer}). On the other 
hand for $\lambda<\lambda_{c}^{\text{Cri.F.}}$ the no gap solution 
first transits to the upper gap solution upon raising the 
temperature; the phase transition temperature decreases 
as $\lambda$ is increased (see Fig.\ref{lmTcrifer1}).

\subsubsection{Low temperature expansion}
The self energy equation \eqref{cfers1} may easily be solved in 
a power series expansion in $\rho_n$. At ${\cal O}(\rho_n^0),$ 
$\tilde{c}$ becomes $\tilde c=0$ 
(for all $\lambda$); at higher orders we find 
\begin{equation}\label{selfroexp-cf}
 (1-\lm+\hat g) \tilde c= \lm \sum_{n=1}^{\infty}(-1)^{n+1}\frac{1}{n}\left(\rho_{n}+\rho_{-n}\right)+{\cal O}(\rho_n^2).
\end{equation}
Plugging this solution into first equation of \eqref{criferro} we find 
\begin{equation}\label{Vroexp-cf}
\begin{split}
 V(\rho) &= -\frac{N V_{2}T^2}{2\pi} \sum_{n=1}^{\infty}(-1)^{n+1}\frac{1}{n^3}\left(\rho_{n}+\rho_{-n}\right)\\
&+\frac{ N V_{2}T^2 \lm^2}{12\pi\left(1-\lm+\hat g\right)^2} \left(\sum_{n=1}^{\infty}(-1)^{n+1}\frac{1}{n}\left(\rho_{n}+\rho_{-n}\right)\right)^3+{\cal O}(\rho_n^4).
\end{split}
\end{equation}
As before the ${\cal O}(\rho_n^1)$ term is independent of $\lm$ and the ${\cal O}(\rho_n^2)$ term vanishes. Now plugging \eqref{Vroexp-cf} in \eqref{extrmizerho} and completing 
square and using the fact that (which follow from second equation in \eqref{criferro})
\begin{equation}\label{ropnstexcf}\begin{split}
 \rho_{-n}=\rho_{n} &= \frac{V_{2} T^2}{2 N\pi}(-1)^{n+1}\frac{1}{n^2}
+{\cal O}\left(\frac{T^6}{N^3}\right),
\end{split}\end{equation} we obtain
\begin{equation}\label{ftcf}
 f\left(\frac{T^2}{N}\right)
=-\frac{V_2}{\pi}\left(\frac{T^2}{N}\right)^{2}\zeta(5)+\frac{1}{12}\left(\frac{V_{2}}{\pi}\right)^{4} \left(\zeta(3)\right)^3 \frac{\lm^2}{\left(1-\lm+\hat g\right)^2}\left(\frac{T^2}{N}\right)^{4}
+{\cal O}\left(\left(\frac{T^2}{N}\right)^{5}\right).
\end{equation}
Note that the ${\cal O}\left(\frac{T^4}{N^2}\right)$ term 
is independent of 
$\lm$ and agrees precisely with the high temperature limit of the 
gas of multitrace operators of the free boson theory \eqref{freelogzb}.  

One can check that the  
${\cal O}\left((\frac{T^2}{N})^4\right)$ term 
in \eqref{ftcf} is related to the term of the same order 
in \eqref{ftrbos} by duality.

\section{Exact solution of the Large N capped GWW model}\label{tm}

In the previous section we have solved for the eigenvalue distribution 
$\rho(\alpha)$, and thereby determined the free energy of systems we 
have studied, in the low temperature no-gap phase. It is much more difficult 
to solve the saddle point equations in the intermediate temperature 
lower or upper gap phases, and the high temperature two gap phase. 
In this section we accomplish this task; not for the realistic 
potentials $v[\rho]$ described above, but instead for a toy model 
potential  
$$V(U)=-\frac{N \zeta}{2} \left( \Tr U + \Tr U^\dagger \right).$$
In this case $V(\alpha)= -N \zeta \cos \alpha$.

Our toy model shares the following key qualitative feature with the 
function $v[\rho]$ that comes from real matter CS theories; the potential 
$V[U]$ always has a minimum at $U=I$, and the depth of the potential 
increases as a function of $\zeta$. We hope that our solution of one and 
two gap phases for our toy model, presented in this section, will aid 
in the solution of the saddle point equations in the same phases 
for the potentials $v[\rho]$ that arise in real matter CS theories; however 
we leave the investigation of this issue to the future. 

\subsection{No gap solution}

The `no gap' solution of these equations is identical to the same solution 
in the uncapped GWW model and famously takes the form 
\begin{equation}\label{ngs}
\rho(\alpha)= \frac{1+\zeta \cos \alpha}{2\pi}.
\end{equation}
For all $\lambda<1$, the no gap solution is the correct saddle point for 
small enough $\zeta$. The no gap solution ceases to be a saddle point of 
our model if either $\rho(\alpha)$ goes negative somewhere or if 
$\rho(\alpha)$ exceeds $\frac{1}{2 \pi \lambda}$ somewhere. Which of these 
occurs first (upon monotonically increasing $\lambda$) depends on the 
value of $\lambda$. 

$\rho(\alpha)$ takes its maximum/minimum 
value at $\alpha=0/\pi$; these values are given by $\frac{1\pm \zeta}{2 \pi}$. 
It follows that if $\lambda < \frac{1}{2}$ the no gap solution goes negative 
before it hits the upper bound. In that case the intermediate $\zeta$ phase 
has a single lower gap. On the other hand if $\lambda> \frac{1}{2}$ 
then the intermediate $\zeta$ phase has a single upper gap.

\subsection{Single lower gap solution}

This is the intermediate $\zeta$ phase that occurs if 
$\lambda <\frac{1}{2}$. This phase kicks in for 
$\zeta \geq 1$. The eigenvalue distribution in this phase is same as that 
for the  standard (uncapped) GWW solution 
(see Appendix \ref{gwwv} for a review) and is given by
\begin{equation}\label{evdl} \begin{split}
\rho(\alpha)& = \frac{\zeta \cos \left( \frac{\alpha}{2} \right)}{\pi} 
\sqrt{\frac{1}{\zeta}-\sin^2\frac{\alpha}{2}} ~~~{\rm for}~~ 
\sin^2\frac{\alpha}{2} < \frac{1}{\zeta},\\
\rho(\alpha)& = 0 ~~~ {\rm for}~~ 
\sin^2\frac{\alpha}{2}> \frac{1}{\zeta}.
\end{split}
\end{equation}
Note that we always have a lower gap (a region 
where the eigenvalue distribution vanishes) as $\zeta>1$. 
The eigenvalue distribution is maximum at $\alpha=0$, and the maximum
value is given by $\frac{\sqrt{\zeta}}{\pi}$. It exceed the maximum permissible
value at $\zeta=\frac{1}{4 \lambda^2} >1$ (recall $2 \lambda<1$). At this 
point the system undergoes a second phase transition to the one lower
gap and one upper gap phase.

\subsection{Single upper gap solution}

This is the intermediate $\zeta$ phase when $\lambda> \frac{1}{2}$. This phase 
kicks in at $\zeta=\frac{1}{\lambda}-1 <1$. In Appendix \ref{gwwm} we have 
solved for the eigenvalue distribution in this phase using the general 
analysis of Appendix \ref{cumm}. In this subsubsection we give a much 
simpler derivation of the eigenvalue distribution in this simple phase; 
of course our final result agrees with that of Appendix \ref{gwwm}.

In the case of phases with only one  upper gap the general analysis 
of capped matrix models presented in Appendix \ref{cumm} greatly simplifies if we make 
the following natural choice for the trial eigenvalue distribution 
function $\rho_0(\alpha)$; 
$$\rho_0=\frac{1}{2 \pi \lambda}.$$ 
As the integral of the cotangent against a constant vanishes, with this 
choice we find that $U(\alpha)=V'(\alpha)$ in \eqref{vvec}. It follows that 
the integral equation in \eqref{vvec} reduces precisely to the Gross-Witten-Wadia
integral equation. The boundary conditions on $\psi$ are the converse of 
the GWW boundary conditions; $\psi(\alpha)=0$ for $|\alpha|<a$. We can 
turn these boundary conditions into the GWW boundary conditions by working
with a new variable $\theta= \alpha-\pi$. We have $\psi(\theta)= 0$ for 
$|\theta|>\pi-a$. When rewritten in terms of $\theta$ the GWW force function 
$U(\alpha)=N \zeta \sin \alpha$ turns into $-N \zeta \sin \theta$. 

It follows that $-\psi(\theta)$ is exactly 
the GWW eigenvalue distribution for 
the potential $V(\alpha)=N \zeta \cos \theta$ with one minor difference; 
the eigenvalue distribution density is normalized so that 
$$\int (-\psi) d \alpha =\frac{1}{\lambda}-1,$$
(the RHS is unity in the standard generalized GWW problem). This change 
is easily accounted for (see Appendix \ref{gww} below) and yields 
$$-\psi(\theta)= \frac{\zeta \cos \left( \frac{\theta}{2} \right)}{\pi} 
\sqrt{\frac{\frac{1}{\lambda}-1}{\zeta}-\sin^2\frac{\theta}{2} }. $$
Substituting $\theta=\alpha-\pi$ above we have
\begin{equation}\label{evdd} \begin{split}
\rho(\alpha)& = \frac{1}{2 \pi \lambda}- 
\zeta\frac{ \vline\, \sin \left( \frac{\alpha}{2} \right) \,\vline}{\pi}  
\sqrt{\frac{\frac{1}{\lambda}-1}{\zeta}-\cos^2\frac{\alpha}{2} }
~~~{\rm for }~~\cos^2 \frac{\alpha}{2} <\frac{\frac{1}{\lambda}-1}{\zeta},\\
\rho(\alpha)&=\frac{1}{2 \pi \lambda} ~~~{\rm for }~~\cos^2 \frac{\alpha}{2} 
>\frac{\frac{1}{\lambda}-1}{\zeta}.\\
\end{split}
\end{equation}
This solution has an upper gap region as we have assumed that 
$\zeta>\frac{1}{\lambda}-1$. The minimum value of the eigenvalue 
density occurs at $\alpha=\pi$ and is given by 
$$\frac{1}{2 \pi}\left( \frac{1}{\lambda}-2 \sqrt{\zeta}
\sqrt{\frac{1}{\lambda}-1} \right).$$
The minimum goes to zero for
\begin{equation}\label{sog} 
\zeta>\frac{1}{4 \lambda \left(1-\lambda \right)},
\end{equation}
at which point the system transits to the one upper gap and one lower 
gap phase.

\subsection{One lower gap and one upper gap solution}

In Appendix \ref{tcgww} we have demonstrated that our toy model also 
admits a solution with one lower gap and one upper gap. The upper 
gap extends on the arc on the unit circle from $e^{-ia}$ to $e^{ia}$ and 
the lower gap extends on the arc on the unit circle from $e^{ib}$ to 
$e^{-ib}$ (see Fig.\ref{2cut.eps} in the Appendix)
where $a$ and $b$ are positive and are determined by the equations
\begin{equation}\label{detabm}
\begin{split}
&\frac{1}{4\pi \lambda} \int_{-a}^a d \alpha  \frac{1}
{\sqrt{\sin^2\frac{a}{2}-\sin^2 \frac{\alpha}{2}}
\sqrt{\sin^2\frac{b}{2}-\sin^2 \frac{\alpha}{2}}} =\zeta,\\
&\frac{1}{4\pi \lambda} \int_{-a}^a d \alpha \frac{\cos \alpha}
{\sqrt{\sin^2\frac{a}{2}-\sin^2 \frac{\alpha}{2}}
\sqrt{\sin^2 \frac{b}{2}-\sin^2 \frac{\alpha}{2}}}=
1+\frac{\zeta}{2} \left( \cos a+\cos b \right) .
\end{split}
\end{equation}
In the complement of these two gaps the eigenvalue distribution is given by 
\begin{equation}
\begin{split}
4\pi\rho(\alpha)&= \frac{|\sin\alpha|}{\pi\lambda}\sqrt{\left(\sin^2\frac{\alpha}{2}-\sin^2\frac{a}{2}\right)\left(\sin^2\frac{b}{2}-\sin^2\frac{\alpha}{2}\right)} ~I_1{\rm ~~~~~where~~~~} \\
I_1 &= \int_{-a}^a \frac{d\theta}{(\cos\theta- \cos\alpha)\sqrt{\left(\sin^2\frac{a}{2}-\sin^2\frac{\theta}{2}\right)\left(\sin^2\frac{b}{2}-\sin^2\frac{\theta}{2}\right)}}. \\ 
\end{split}
\label{realrhom}
\end{equation}
By expanding \eqref{realrho} in Appendix 
\ref{tcgww}
near $\alpha =b$ we find 
(see \eqref{ro2cut} in Appendix \ref{ronear_a_b}) 
$\rho(\alpha) = B \sqrt{b-\alpha} +{\cal O}(b-\alpha)^\frac{3}{2}$ 
where $B$ is a known positive constant. Similarly, near $\alpha=a$
$\rho(\alpha) = \frac{ 1}{2 \pi \lambda} + D \sqrt{\alpha-a} +{\cal O}(\alpha-a)^\frac{3}{2}$ where $D$ is another known negative constant
(see \eqref{ro_near_a} in Appendix \ref{ronear_a_b}). 
In other words the eigenvalue density in the `cuts' continuously matches onto the eigenvalue 
densities in the upper and lower gap regions.

Over what range of $\zeta$ does the double cut solution exist? 
In the double cut solution, $a$ and $b$ are functions of $\zeta$ 
and $\lambda$. For some formal purposes it proves useful to invert 
this relationship and regard $\zeta$ and $\lambda$ as functions of 
$a$ and $b$. As analyzed in the Appendix 
\ref{Behave-a-b},
$\zeta$ is a decreasing function of $b$ at fixed $a$ while $\lambda$ 
is an increasing function of $b$ at fixed $a$.
By numerical evaluations, we can also guess that 
both $\zeta$ and $\lambda$ are increasing functions of $a$ at fixed $b$. 
From these facts, we can obtain lower bounds
on the range of $\zeta$ for which the double cut solution exists by 
setting $a=0$ and $b=\pi$. 

Let us first set $b=\pi$ in which case \eqref{detabm} reduce to 
\begin{equation}\begin{split}
\zeta &= \frac{1}{4\pi \lambda} \int_{-a}^a d \alpha  \frac{1}
        {\sqrt{(\sin^2\frac{a}{2}-\sin^2 \frac{\alpha}{2})(1-\sin^2 \frac{\alpha}{2})}} 
        = \frac{1}{2\lambda \cos \frac{a}{2}}, \\
1+\frac{\zeta}{2} \left( \cos a-1 \right) &= \frac{1}{4\pi \lambda} \int_{-a}^a d \alpha \frac{\cos \alpha}
      {\sqrt{(\sin^2\frac{a}{2}-\sin^2 \frac{\alpha}{2})(1-\sin^2 \frac{\alpha}{2})}} 
      = \frac{1}{\lambda}\left( 1- \frac{1}{2 \cos \frac{a}{2}} \right).\\
\end{split}
\label{bpieq}
\end{equation}
\eqref{bpieq} are easily be solved to get determine $a$ and $\zeta$ in 
terms of $\lambda$; we find 
\begin{equation}\begin{split}
\zeta &= \frac{1}{4\lambda(1-\lambda)}, \\
\cos \frac{a}{2} &= 2(1-\lambda). \\
\end{split}
\label{bpisol}
\end{equation}
Note that \eqref{bpisol} matches exactly with the limit of single upper 
gap solution at the edge of its validity (see \eqref{sog}).

Let us next consider the limit $a\rightarrow 0$. In this limit the integration
range in \eqref{detabm} goes to zero, but the integrand simultaneously 
diverges in such a way that the integral is finite. The equations reduce to 
\begin{equation}\begin{split}
\zeta &=  \frac{1}{2 \lambda \sin\frac{b}{2}}, \\
1+\frac{\zeta}{2}(1+\cos b) &=\frac{1}{2 \lambda \sin\frac{b}{2}}.
\end{split}
\label{a0sol}
\end{equation}
This implies that 
\begin{equation}
\zeta=\frac{1}{4 \lambda^2}, ~~~
\sin \frac{b}{2} = 2 \lambda,
\end{equation}
which again matches with the limit of single lower gap solution at the edge 
of its validity. 

It follows that the double gap solution studied in this section exists 
if and only if $\zeta \geq \frac{1}{4 \lambda^2}$ and 
$\zeta \geq \frac{1}{4 \lambda(1-\lambda)}$
(also see Appendix \ref{Behave-a-b}). When $\lambda \leq \frac{1}{2}$
the first inequality is more stringent, and the double gap solution 
exists only when $\zeta \geq \frac{1}{4 \lambda^2}$, i.e. only when 
the GWW single lower gap solution ceases to remain a valid solution. 
On the other hand when $\lambda \geq \frac{1}{2}$ the second inequality 
is more stringent, and the double cut solution exists only for 
$\zeta \geq \frac{1}{4 \lambda^2}$, i.e. only when the single upper gap 
solution ceases to be valid. 

The discussion above makes it plausible that when $\lambda>\frac{1}{2}$ 
the double gap solution of this subsection becomes identical to the single upper gap solution 
when $\zeta= \frac{1}{4\lambda(1-\lambda)}$. It also suggests that when $\lambda<\frac{1}{2}$, 
the solution of this subsection becomes identical with the single lower gap solution at 
$\zeta=\frac{1}{4 \lambda^2}$. In Appendix \ref{spa} and Appendix \ref{spb}
 we have demonstrated that this is indeed the case
by explicitly computing the double gap eigenvalue distribution in these limits, and comparing 
with the respective single gap eigenvalue distributions.

To end this subsection we consider the limit of large $\zeta$ at fixed 
$\lambda$.
In Appendix \ref{lz} we have demonstrated that in this limit 
\begin{equation}\label{lzz}\begin{split}
a&=\pi\lambda-4\sin(\pi\lambda) e^{-\pi\zeta\lambda\sin(\pi\lambda)}+ O[\zeta~e^{-2\pi\zeta\lambda\sin(\pi\lambda)}],\\
b&=\pi\lambda+4\sin(\pi\lambda) e^{-\pi\zeta\lambda\sin(\pi\lambda)}+O[\zeta~e^{-2\pi\zeta\lambda\sin(\pi\lambda)}],\\
\rho(\alpha)&=\frac{1}{\pi^2\lambda} ~\cos^{-1}\sqrt{\frac{\alpha-a}{b-a}} .\\
\end{split}
\end{equation}
Note in particular that the eigenvalue distribution tends exponentially to 
the distribution \eqref{evd} in the limit of large $\zeta.$

\subsection{Level Rank Duality}\label{lrd}

Recall that  
$${\tilde V}(\Tr U^n)= V\left((-1)^{n+1} \Tr U^n \right).$$
For the GWW toy model of this section
\begin{equation} \label{dp}
{\tilde V}(U)=V(U)=-N\frac{\zeta}{2} \left( \Tr U + \Tr U^\dagger \right)
=-(k-N)\frac{\lambda\zeta}{2 (1-\lambda)} 
\left( \Tr U + \Tr U^\dagger \right).
\end{equation} 
In other words the GWW model is self dual under level rank duality together
with the appropriate rescaling of $\zeta$. Earlier in this section we have 
explicitly determined the 
saddle point eigenvalue distribution for the matrix model with the 
potential $V(U)$. This eigenvalue distribution is a function of $\lambda$, 
$\zeta$ and of course $\alpha$ (the phase on the unit circle). Let us 
denote our saddle point eigenvalue distribution by  
$\rho(\lambda, \zeta , \alpha)$. Level rank duality predicts that the 
exact solution to this model, obtained in the previous section, obeys the 
identities
\begin{equation}\label{lrdp}
\rho \left( (1-\lambda), \frac{\zeta \lambda}{1-\lambda}, \alpha \right)
=\frac{\lambda}{1-\lambda} \left(\frac{1}{2 \pi \lambda} -
\rho \left( \lambda, \zeta, \alpha+\pi \right) \right).
\end{equation}
In Appendix \ref{gwwlrd} we have verified in great detail that 
\eqref{lrdp} is indeed correct

\subsection{Summary} \label{summary}

\begin{figure}
  \begin{center}
\subfigure[]{\includegraphics[scale=.4]{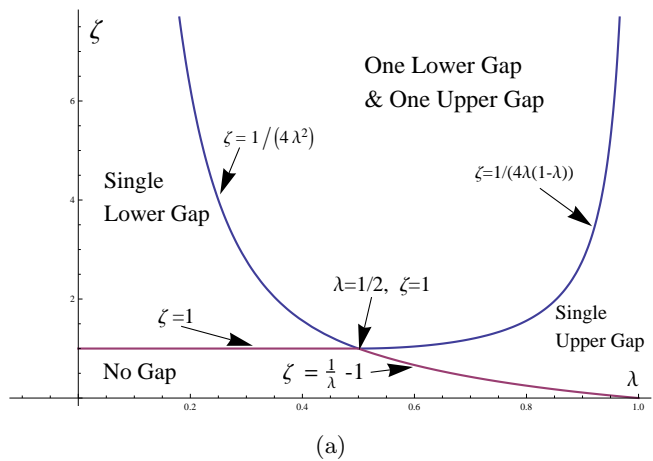}
\label{Phase}}
\caption{The phase diagram for Gross-Witten-Wadia potential.  }
\end{center}
\end{figure}

We end this section with a summary of the behaviour of the capped 
Gross-Witten-Wadia model. Fig.\ref{Phase} summarizes different phases of the GWW model.

\subsubsection{$\lambda<\frac{1}{2}$} 
\begin{figure}[tbp]
  \begin{center}
  \subfigure[]{\includegraphics[scale=.5]{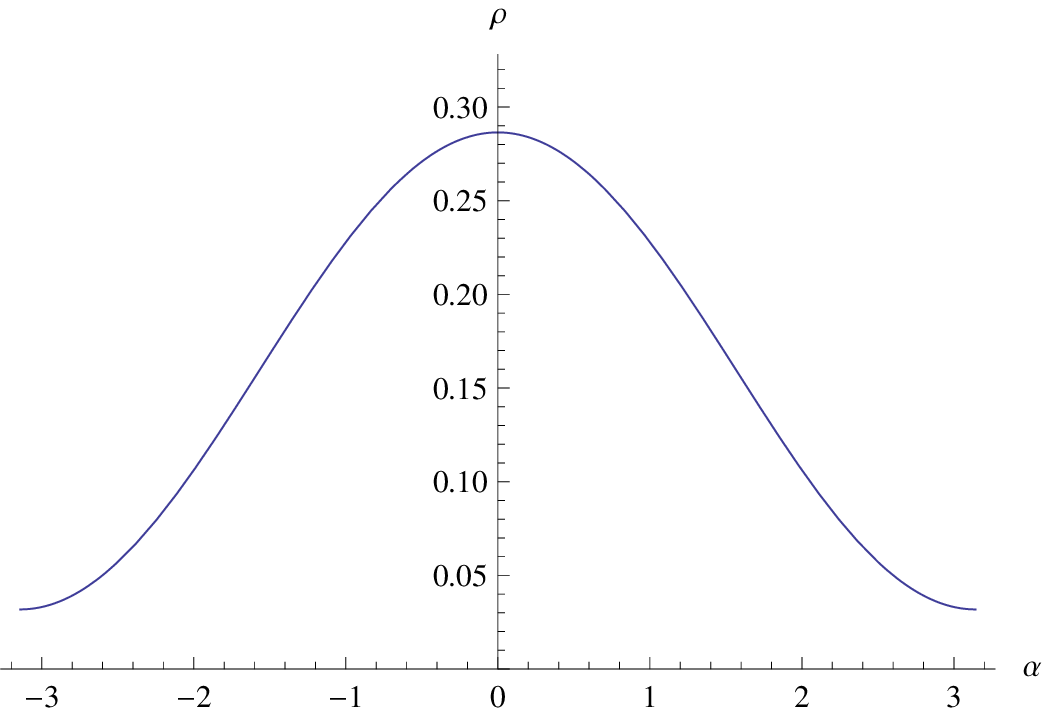}
\label{nocut}
  }
  \qquad\qquad
  \subfigure[]{\includegraphics[scale=.5]{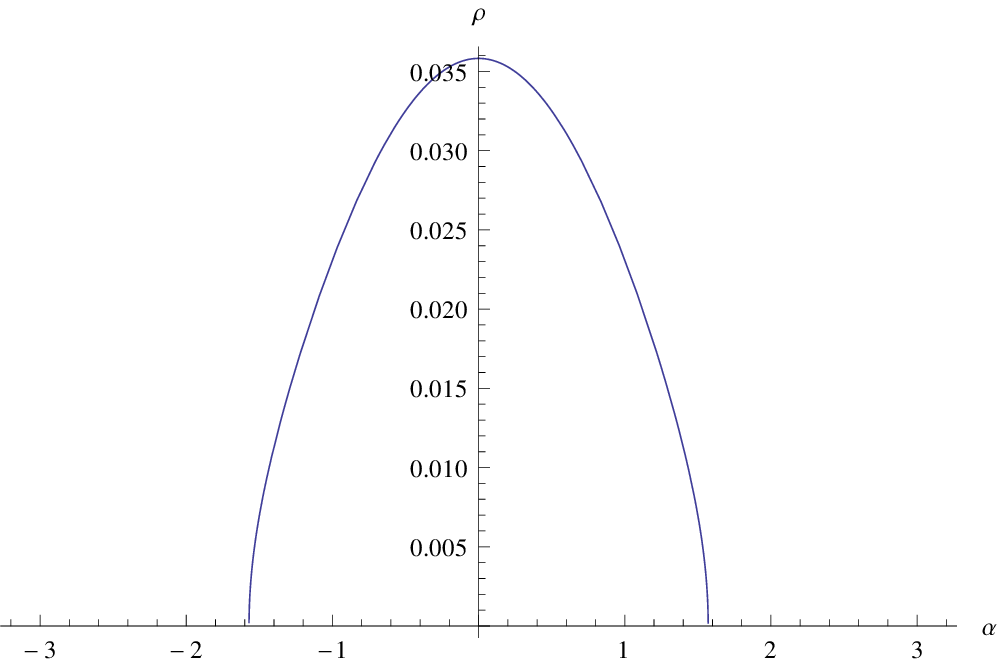}
\label{gwwfig}
  }
\\
  \subfigure[]{\includegraphics[scale=.5]{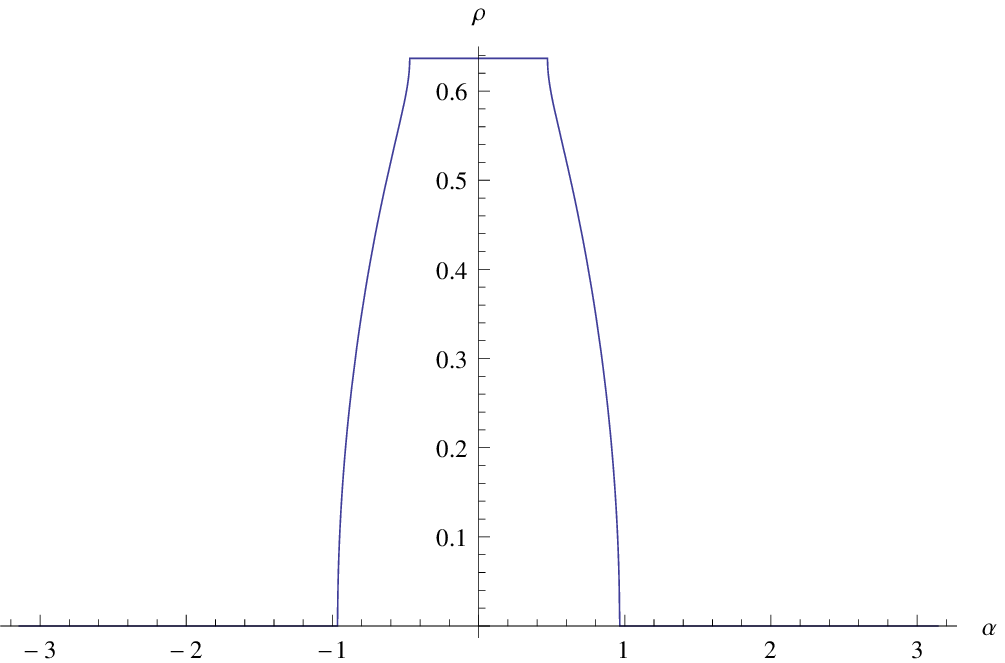}
\label{twocut}
  }
  \qquad\qquad
  \subfigure[]{\includegraphics[scale=.5]{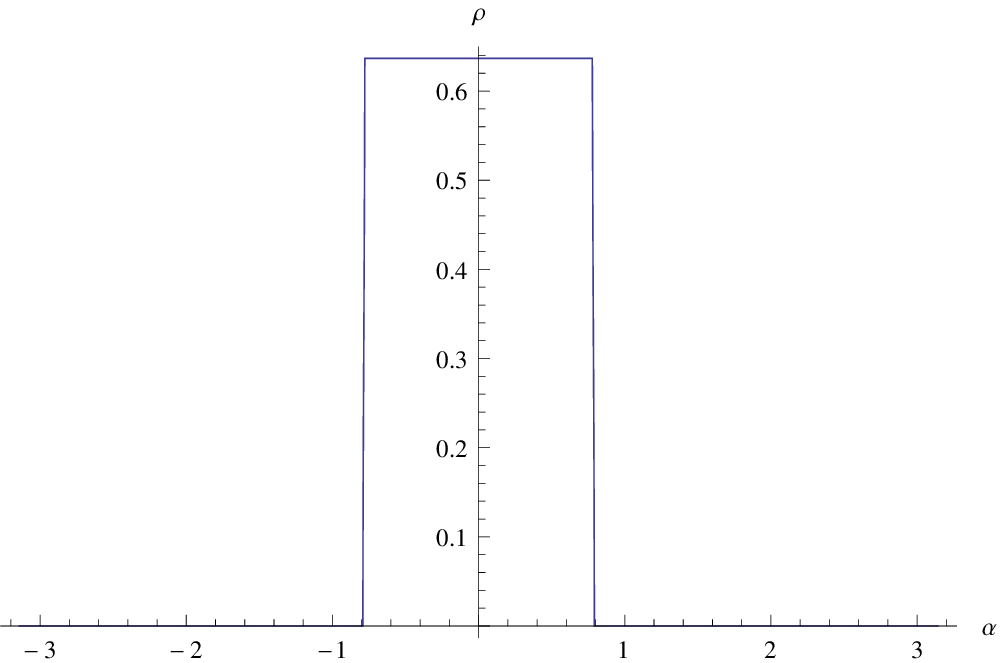}
\label{twocut2}
  }
  \end{center}
  \vspace{-0.5cm}
  \caption{The eigenvalue distribution of the capped GWW model at 
$\lambda=0.25$ for 
increasing values of $\zeta$:~~Fig.\ref{nocut} is for $\zeta=\frac{1}{2},$ Fig.\ref{gwwfig} is for $\zeta=2$,
Fig.\ref{twocut}is for $\zeta=4.6$ and Fig.\ref{twocut2}is for 
$\zeta=11.03.$ In this case as we increase $\zeta$,
eigenvalue density distribution first develops a lower gap as Fig.\ref{gwwfig}. Further increasing $\zeta,$
eigenvalue density distribution develops a upper gap as well Fig.\ref{twocut}.}
  \end{figure}

For $\zeta<1$, the eigenvalue distribution is wavy (i.e. has no gaps) 
and is given by \eqref{ngs} (see  Fig.\ref{nocut} for a plot at 
$\lambda=\frac{1}{4}$ and $\zeta=\frac{1}{2}$). For $1>\zeta>
\frac{1}{4\lambda^2}$, the eigenvalue distribution has a lower gap centered 
about $\pi$ and is given by \eqref{evdl} (see Fig.\ref{gwwfig} for a plot 
at $\lambda=\frac{1}{4}$ and $\zeta=2$). For $\zeta>\frac{1}{4\lambda^2}$, 
the eigenvalue distribution is given by 
\eqref{realrhom} which has a lower gap centered about $\pi$ and an 
upper gap centered around zero (see Fig. \ref{twocut} for a plot at  
$\lambda=\frac{1}{4}$ and $\zeta=4.6$). At large values of $\zeta,$ the 
eigenvalue distribution tends to \eqref{evd} (see \eqref{lzz} for more 
details, and see Fig.\ref{twocut2} for a plot at $\lambda=\frac{1}{4}$ 
and $\zeta=11.03$).

\subsubsection{$\lambda>\frac{1}{2}$} 
\begin{figure}[tbp]
  \begin{center}
  \subfigure[]{\includegraphics[scale=.5]{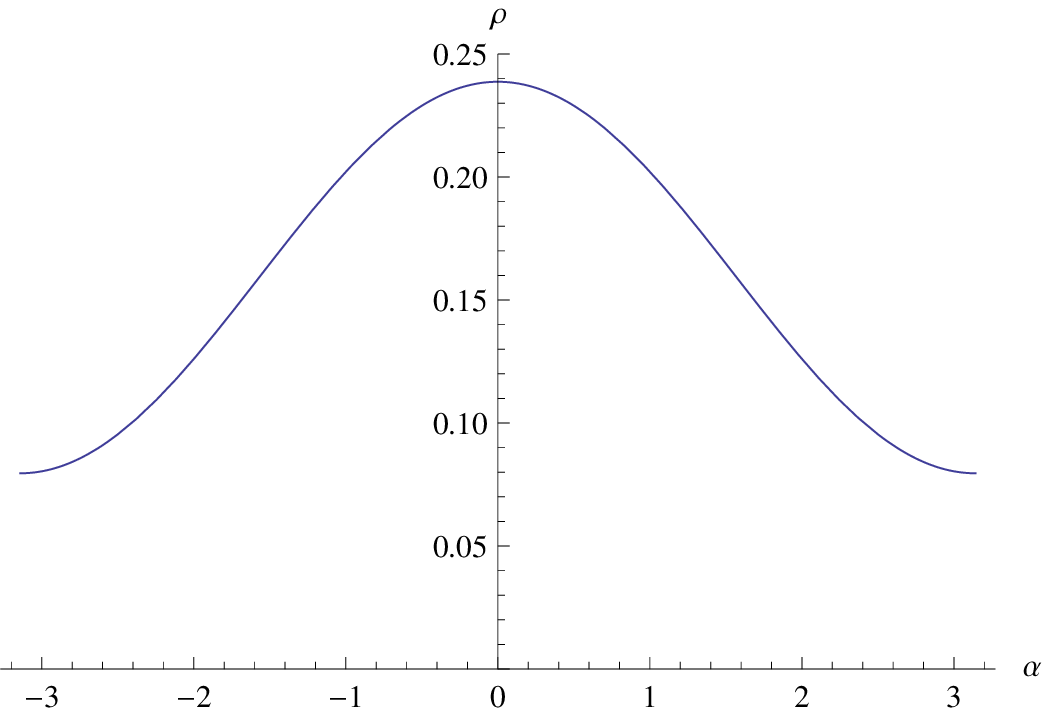}
\label{lbig1}
  }
  \qquad\qquad
  \subfigure[]{\includegraphics[scale=.5]{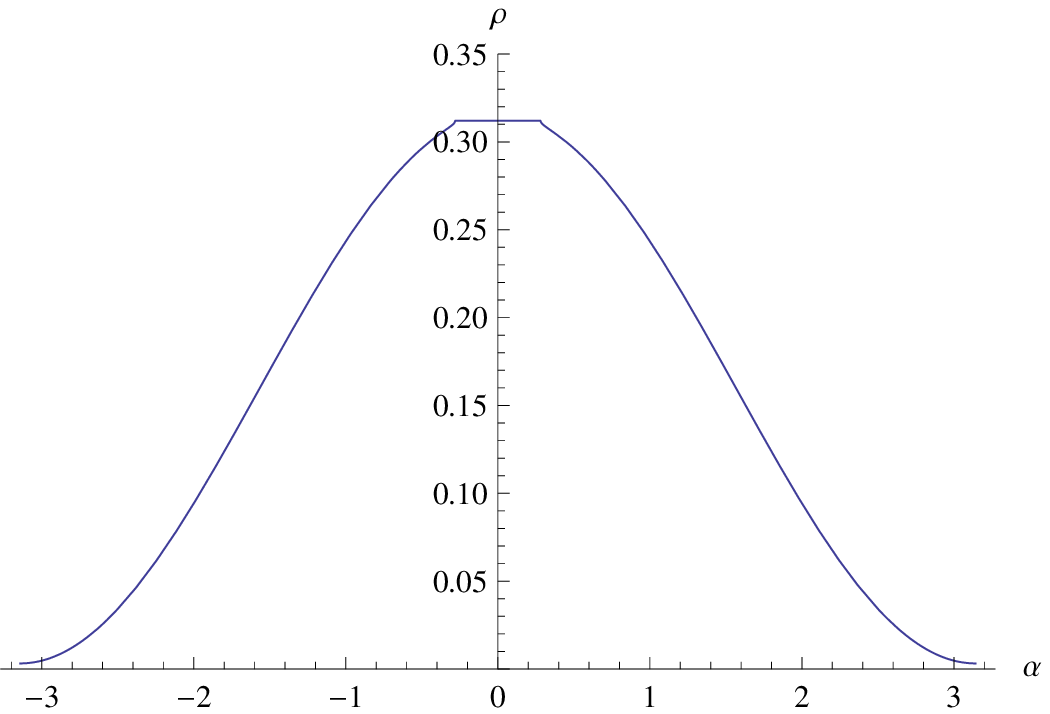}
\label{lbig2}
  }
\\
  \subfigure[]{\includegraphics[scale=.5]{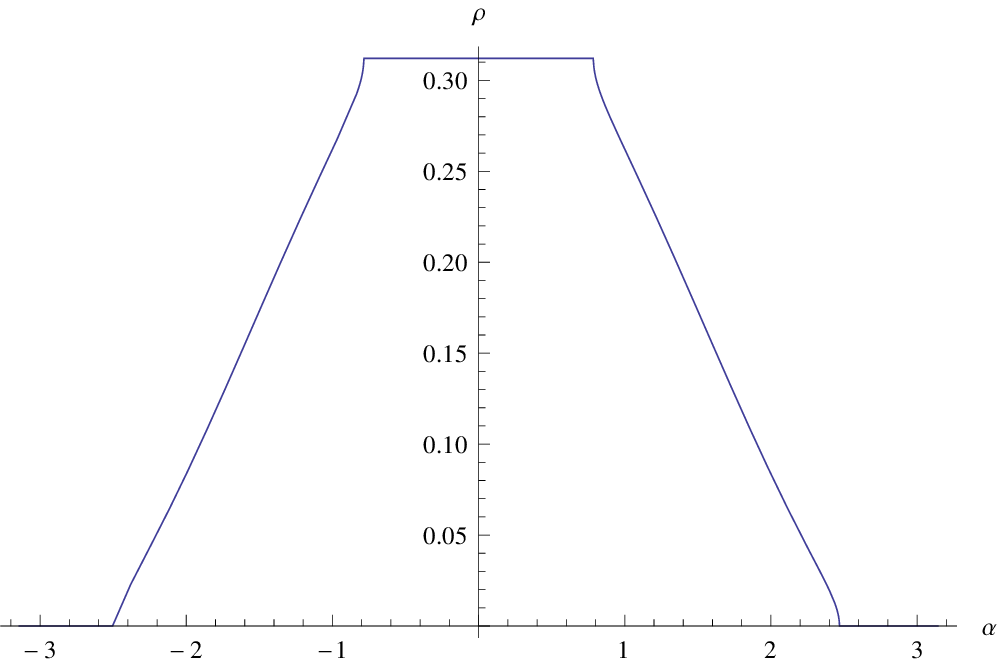}
\label{lbig3}
  }
  \qquad\qquad
  \subfigure[]{\includegraphics[scale=.5]{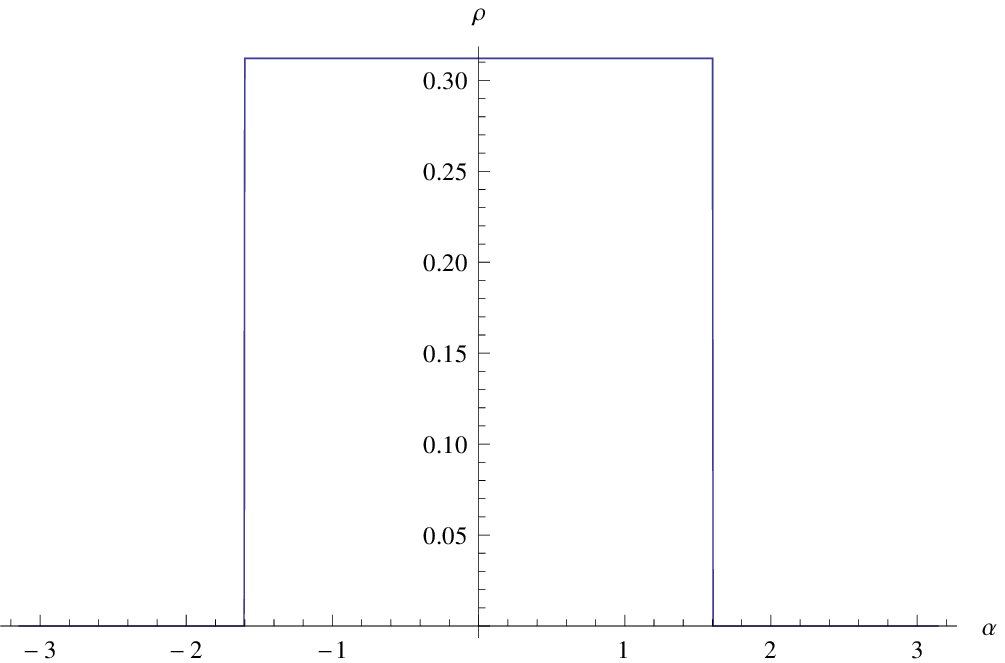}
\label{lbig4}
  }
  \end{center}
  \vspace{-0.5cm}
  \caption{The eigenvalue distribution of the capped GWW model at 
$\lambda=0.51$ for 
increasing values of $\zeta$:~~Fig.\ref{lbig1}is for $\zeta=\frac{1}{2}$, Fig.\ref{lbig2}is for $\zeta=0.98$,
Fig.\ref{lbig3}is for $\zeta=1.13$ and 
 Fig.\ref{lbig4}is for $\zeta=5.06$. In this case as we increase $\zeta$,
eigenvalue density distribution first develops a upper gap as Fig.\ref{lbig2}. Further increasing $\zeta,$
eigenvalue density distribution develops a lower gap as well Fig.\ref{lbig3}.}
  \end{figure}

For $\zeta<\frac{1}{\lambda}-1,$ the eigenvalue distribution is wavy 
(i.e. has no gaps) and is given by \eqref{ngs} (see  Fig.\ref{lbig1}  
for a plot at $\lambda=0.51$ and $\zeta=\frac{1}{2}$). 
For $\frac{1}{\lambda}-1>\zeta>\frac{1}{4 \lambda(1-\lambda)},$ the 
eigenvalue distribution has an upper gap (or plateau) centered around $0$ 
and is given by \eqref{evdd} (see Fig. \ref{lbig2} for a plot 
at $\lambda=0.51$ and $\zeta=0.98$). For $\zeta>\frac{1}{4 \lambda(1-\lambda)},$ 
the eigenvalue distribution is given by \eqref{realrhom} which 
has a lower gap centered about $\pi$ and an 
upper gap centered around zero (see Fig. \ref{lbig3} for a plot at  
$\lambda=0.51$ and $\zeta=1.13$). At large values of $\zeta,$ the 
eigenvalue distribution tends to \eqref{evd} (see \eqref{lzz} for more 
details, and see Fig.\ref{lbig4} for a plot at $\lambda=0.51$ 
and $\zeta=5.06$).

\section{Discussion}

In this paper we have presented a wealth of exact results for the 
thermal partition function of several matter Chern-Simons theories 
as a function of temperature and $\lambda$ in the strict large $N$ limit. 
Our results obey some non trivial consistency checks; they are in 
perfect agreement with several conjectured Giveon-Kutasov type dualities 
between pairs of theories. In every case the partition function also 
reduces, at low temperatures, to the $N$ and $\lambda$ independent 
partition function of a gas of multi trace operators, in agreement 
with the nonrenormalization theorem for multi trace operators in these 
theories\cite{Giombi:2011kc,Aharony:2011jz}. 

While the consistency  checks reported in the previous paragraphs give us 
confidence in our results, some aspects of the procedure employed 
in this paper require fuller discussion. To start with, it would be 
useful to have a more thorough justification for using 
\eqref{purecsi} as the starting point of our analysis. 
Subleading corrections (in $\frac{1}{N}$) to \eqref{purecsi} could modify 
\eqref{csmint} in several ways; 
\footnote{We thank O. Aharony and J. Maldacena for discussions on this point.}
for instance they could spread the discrete delta functions 
in \eqref{csmint} into a distribution function for eigenvalues that 
is peaked about $\alpha_i=\frac{2 \pi n}{k}$ with a width $\delta$ that 
goes to zero in the large $N$ limit. With such a spread out 
eigenvalue distribution function two eigenvalues can presumably 
occupy the same slot. \footnote{The energy price from the Vandermonde 
in the measure, -$\ln \delta$, is presumably negligible 
compared to the energy gain from two eigenvalues occupying the same 
slot.} Hamiltonian methods indicate that the contribution of configurations 
with two nearly equal eigenvalues is very small. Consequently the path 
integral must include a mechanism that suppresses contributions of 
configurations with nearly equal eigenvalues; 
it is important to understand in detail how this 
works. Relatedly, it would be useful to 
thoroughly understand the reason for the exclusion of configurations
with equal eigenvalues in the path integral \eqref{csmintg} 
on $\Sigma_g \times S^1$ where $\Sigma_g$ is a genus $g$ surface. 
We hope to return to these important issues in the future.

The procedure we have employed in the computations of this paper is 
a mix of two different methods. We used  lightcone 
gauge on $R^2$ to compute the effective potential $v[\rho]$, and 
then proceeded to evaluate the Chern-Simons path integral supplemented
with this effective action in temporal gauge. It would of course be 
more satisfactory to rederive all our conclusions from computations that 
utilize temporal gauge all the way through. This would also 
allow the evaluation of the partition 
function of Chern-Simons theory on arbitrary genus $g$ surfaces at finite 
values of the temperature\cite{Banerjee:2012gh,Banerjee:2012aj}. The main obstruction to this procedure are a set of spurious 
divergences that appear in computations in matter Chern-Simons theories 
in temporal gauge~\cite{Giombi:2011kc}; 
it may prove possible to understand these divergences 
well enough to deal with them. We hope to return to this point in the future.

Putting aside all these concerns, the work presented in our paper 
raises several interesting questions. 
One of the most straightforward of these relates to the possible existence 
of a wider class of dualities than those verified in this paper. It should be straight forward to generalize 
the computations of $v(U)$ presented in this paper to a wider class 
of matter Chern-Simons theories and perhaps discover new dualities. 

Giveon Kutasov dualities continue to hold away from the strict vectors 
like limit; for instance they continue to hold at large $N_c$ and $N_f$ 
with $\frac{N_f}{N_c}$ held fixed. It would be interesting to investigate
whether the same is true of nonsupersymmetric dualities, perhaps by 
performing a computation of the $S^2$ partition function at first order 
in $\frac{N_f}{N_c}$.

Relatedly, in this paper we have focused on theories with a single Chern-Simons 
gauge group and matter in the fundamental representation. The story 
is even richer in Chern-Simons theories, like ABJ theory, with two gauge groups - $U(N)$ and 
$U(M)$ and bifundamental matter. At least in the limit in 
which $\frac{M}{N}$ is small the eigenvalues of $U(N)$ should develop upper and 
lower gaps (see \cite{Chang:2012kt} for an analysis of the partition 
function of the free theory). It would be very interesting to analyse 
how the situation evolves as a function of $\frac{M}{N}$, as the 
corresponding Vasiliev theory turns into a vanilla two derivative theory of 
gravity 
\cite{Giombi:2011kc,Chang:2012kt} (IIA theory on $AdS_4 \times CP^3$ with 2-form flux) at large 
$\lm$ when $\frac{M}{N}=1$. 

It would be interesting, and should be possible to generalize the computations
presented in this paper to the computation of the partition function of 
the relevant theories on $S^3$. The results of such a computation must 
agree with the results obtained through localization for supersymmetric 
theories \cite{Kapustin:2009kz}, but will, of course, be more general than those results. 

In section \ref{est}, we have computed eigenvalue density distribution in the no gap phase for several Chern-Simons matter theories.
Following section \ref{tm}, it would be very interesting to 
find out eigenvalue density distribution in the single lower gap, single upper gap and one lower-one upper gap phases of these theories. 

An interesting aspect of the $S^2$ partition functions determined in this 
paper is the presence of phases with upper gaps. Such phases are absent
in the partition function of Yang Mills theories \eqref{mint}. As we have seen 
below, the presence of upper gap phases is a direct consequence of the 
discretization of the eigenvalues of the holonomy matrix, which, in turn, 
is a direct consequence of the summation over flux sectors in the path 
integral. What is interpretation of this new phase in terms of states 
on $S^2$ ? As the existence of this new phase relies on the summation over 
fluxes, it is natural to guess that the new phase is one in which states 
with fluxes on the $S^2$ (states dual to monopole operators) proliferate in 
the system. Note that monopole operators in Chern-Simons theories all have dimensions of order $k$ which is very large in the large $N$ limit. Such operators 
have essentially no impact on the partition function at temperatures of order
unity. However such operators could proliferate at $ VT^2=\zeta N$ 
as studied in this paper; our results indicate this happens when an upper 
gap is formed, i.e. at $\zeta \sim \frac{1}{\lambda^2}$ at small $\lambda$. 
It would be interesting check whether or not this is true, and to flesh it out.
\footnote{We thank S. Shenker for discussions on this topic.}

It has recently been conjectured that level 
$k$ Chern-Simons matter theories  with matter in the fundamental 
/ bifundamental representations, admit a dual description governed by 
parity violating Vasiliev's higher spin equations 
\cite{Vasiliev:1990en,Vasiliev:2003ev}
(See \cite{Giombi:2012ms} for more detail and references) 
at finite values of the 
't~Hooft coupling $\lambda=\frac{N}{k}$ \cite{Giombi:2011kc,Chang:2012kt}. 
The thermal system studied in this paper is thus dual to the Euclidean 
Vasiliev system in global $AdS$ space 
compactified on a circle on circumference $2 \pi R=\frac{1}{T}$. The field 
theory analysis performed in this paper implies that this Vasiliev 
system admits a classical limit in large $N$ limit with $V_{2}T^2=\zeta N$ 
and $\zeta$ held fixed. {\it The saddle points obtained in this paper should
map to classical solutions of this bulk description; however the equations 
of motion governing this classical description are  not necessarily
Vasiliev's equations.} This is because quantum corrections to Vasiliev's 
equations proportional $\frac{V_{2}T^2}{N}$ are not suppressed compared to 
classical effects in the combined high temperature and large $N$ limit 
studied in this paper; such terms could modify Vasiliev's classical 
equations.  It would be fascinating to determine the effective
3 dimensional bulk equations dual to the large $N$ limit of this paper, 
and to study the classical solutions dual to the saddle points 
described in this paper.

Finally, one of the most interesting aspects of the study of Chern-Simons 
matter theories is that we have tantalizing hints that 
 miraculous but well known properties of supersymmetric theories - 
level rank type dualities - are also true of nonsupersymmetric theories. 
There is a more interesting, more  miraculous and better known 
property of the supersymmetric 
ABJ theory, namely that it reduces to a theory of gravity at $M=N$ 
and $\lambda \to \infty$. A really interesting 
question is whether similar result holds for a non supersymmetric theory, for 
instance the theory of minimally coupled bifundamental fermions. 
The study of this question sounds like a fascinating programme for the future.

\acknowledgments

We would like especially  to thank O. Aharony, S.Giombi, J. Maldacena 
and X. Yin for several very useful discussions resulting in the work 
reported in this paper. We would also like to thank Luis Alvarez-Gaume, 
I. Biswas, R. Gopakumar, S. Kim, G. Mandal, D. Prasad, and S. Trivedi for 
useful discussions and O. Aharony, R. Gopakumar, S. Kim, S. Shenker and X. Yin  
for comments on preliminary versions of this manuscript.  The 
work of SM is supported by a Swarnajayanti Fellowship. The 
work of SW is supported by a J.C.Bose Fellowship. 
SM would like to acknowledge the hospitality of NISER Bhubhaneshwar 
when this work was in progress. SW would like to acknowledge the hospitality of CERN Geneva 
when this work was in progress.
We would all also like to    acknowledge our debt to the people of India for 
their generous and steady support to research in the basic sciences.

\appendix

\section{Interpretation of the Fadeev-Popov Determinant}\label{fp}

In this Appendix we present a heuristic
demonstration that this FP determinant simply converts the path integral over 
the two dimensional hermitian zero mode field $\alpha(x)$ into a path 
integral over the two dimensional unitary field $U(x)$. The operator in 
the determinant has no $x$ derivative, so the determinant is a product of 
determinants, one for every $x$ (this is the part of the evaluation 
that is heuristic - strictly we need to regulate to make sense of this 
statement). Specializing for a moment to the group $U(N)$, the determinant
is given by  
$$\prod_{n}  \left( \frac{2 \pi i n}{\beta} +{i(\alpha_m-\alpha_l) \over \beta} \right).$$
The product is taken over all (positive and negative) integer values 
of $n$ when $m \neq l$ but only over all nonzero integer values of 
$n$ when $m=l$. Actually all terms in the product are independent of 
$\alpha$ (and so can be absorbed into the normalization of the path 
integral) when $m=l$. The nontrivial contribution is from $m\neq l$. 

After removing an $\alpha$ independent factor we have 
$$\prod_{x} \prod_{m, l} (\alpha_m-\alpha_l)
  \prod_{n}\left(1+ \frac{\alpha_m-\alpha_l}
{2 \pi  n}\right)$$
where we dropped overall constant regulated suitably.  
The last product vanishes at $\alpha_m-\alpha_l = 2 \pi n$ for every 
integer $n$. It is thus plausible (and true) that it is proportional to 
$$\sin\left(\frac{\alpha_m-\alpha_l}{2} \right).$$
Consequently, the Fadeev-Popov Determinant is heuristically given by  
$$Det_S'(\partial_3 + [A_3, ])= \prod_x \prod_{m\neq l} 
2 \sin \left(\frac{\alpha_m - \alpha_l }{2} \right)= \prod_x \prod_{m < l}\left| e^{i\alpha_{m}}-e^{i\alpha_{l}}\right|^2, $$
where we put the factor $2$ so that (see \S\ref{lrdln})
$$\int DA_3Det_S'(\partial_3+[A_3, ])=\int DU(x),$$
where $U(x)=e^{ \beta A_3}$ and $DU$ is the path integral over $U(x)$ 
with the left-right invariant Haar measure over $U(x)$. The Haar measure is consistent with level rank duality.

\section{Solution to the saddle point equations of `capped' unitary matrix models}
\label{gww}

As we have explained in the main text, Chern-Simons theories coupled to 
fundamental matter fields in the 't~Hooft limit are governed by the saddle point equations of a large $N$ unitary matrix integral with one additional restriction; the 
density of eigenvalues is bounded from above at 
$\frac{1}{2\pi \lambda}$. In this appendix we present a general method 
to solve these `capped' large $N$ unitary matrix integrals. In subsection \ref{rgww} we 
first review the general method to solve ordinary (uncapped) unitary matrix models following the 
method of \cite{Wadia:2012fr}. In subsection \ref{cumm} we then proceed to use 
a very similar method obtain an exact solution to the saddle point of capped Unitary matrix models.
The formulas presented in this appendix apply, with great generality, to any single
trace capped unitary matrix integral; however the final results are a little implicit.
In appendix \ref{gwwc} below we show how this formal solution can be converted into 
explicit formulas in the context of a simple example.

\subsection{Review of standard Unitary matrix integrals at large $N$} \label{rgww}

As we have explained in the main text, the large $N$ eigenvalue density 
function of the standard (i.e. uncapped) unitary matrix integral 
$$\int D U e^{-V(U)} $$
obeys the integral equation
\begin{equation}\label{vvecn} \begin{split}
&N {\cal P} \int d \alpha ~\psi(\alpha) 
 \cot \left( \frac{\alpha_0-\alpha}{2} \right)
=U(\alpha_0)\\
&\int d \alpha~ \psi(\alpha)=A
\end{split}
\end{equation}
with $U(\alpha)=V'(\alpha)$. 
In this subsection we review the standard solution of the integral equation 
\eqref{vvecn}. In the next subsection we will use this solution to obtain
the solution to the capped unitary matrix integral.

\subsubsection{General solution of \eqref{vvecn}}
In order to obtain the general solution of \eqref{vvecn} we 
find it useful to use the following complex variables (see for eg. \cite{Wadia:2012fr})
$$z=e^{i \alpha}, ~~~z_0=e^{i \alpha_0}, ~~~A_i=e^{ia_i}, ~~~B_i=e^{ib_i}.$$
Note that 
$$d \alpha=\frac{dz}{iz}, 
~~~\cot \frac{\alpha_0-\alpha}{2}=i \frac{z_0+z}{z_0-z} .$$
\eqref{vvecn} may be rewritten in these variables as 
\begin{equation}\label{fovv} \begin{split}
&N {\cal P} \int \frac{d z}{z}   \frac{z_0+z}{z_0-z}
\psi(z) =U(z_0),\\
&  \int \frac{dz}{iz} \psi(z)=A
\end{split}
\end{equation}
where the integrals in \eqref{fovv} run clockwise over the unit circle 
in the complex plane. 

Let us suppose that the solution to this equation, $\psi(z)$, has 
support on $n$ connected arcs on the unit circle in the complex plane. 
We denote the beginning and endpoints of these arcs by $A_i=e^{i a_i}$ and 
$B_i=e^{i b_i}$ 
($i =1 \ldots n$). Our convention is that the points 
$A_1$, $B_1$, $A_2$, $B_2$ $\ldots$ $A_n$, $B_n$ sequentially 
follow each other counterclockwise on the unit circle. We refer to the 
$n$ arcs $(A_i,B_i)$ as `cuts'. The arcs $(B_i, A_{i+1})$ (as also 
$(B_n, A_1)$ ) are referred to as gaps. By definition, $\psi$ has support
only along the cuts on the unit circle.

We use the (as yet unknown) function $\psi(z)$ to define an analytic function 
$\Phi(u)$ on the complex plane
\begin{equation}\label{gwws1}
\Phi(u)=\sum_{i} \int_{A_i}^{B_i} \frac{dz}{iz} \frac{u+z}{u-z}\psi(z) 
\end{equation}
where the integral is taken counterclockwise along the $n$ cuts, 
$(A_i, B_i)$ of the unit circle in the complex plane.
It is easy to see that the analytic function $\Phi(u)$ is discontinuous 
along these $n$ cuts. Let $\Phi(z)^+$ denote the limit of 
$\Phi(u)$ as it approaches a cut from $|u|>1$, and let $\Phi^-(z)$ 
denote the limit of $\Phi(u)$ as it approaches a cut from $|u|<1$.
\footnote{We use analogous notation for other analytic functions below. 
Completely explicitly, let $z$ denote a complex number 
of unit norm. The symbol $F^+(z)$ will denote the limit of 
$F(u)$ as $u \to z$ from above (i.e. from $|u| >1$), while $F^-(z)$ is 
the limit of the same function  as $u \to z$ from below (i.e. from $|u|<1$). 
Note that along a cut $F^-(z)=-F^+(z)$.}
Then
\begin{equation}\label{disc}
\Phi^+(z)-\Phi^-(z)= 4 \pi \psi(z).
\end{equation}
\footnote{\eqref{disc} may be derived as follows. Let $D(u)$ represent 
any function that is analytic in a region that encloses all $n$ cuts. 
The integral 
\begin{equation}\label{ef}
\int du \Phi(u) D(u)
\end{equation}
 over a contour that encloses all $n$ 
cuts is, by definition given by $\int dz D(z)(\Phi^+(z)-\Phi^-(z))$ 
where
the integration runs anticlockwise along the $n$ cuts. 
On the other hand, by substituting \eqref{gwws1} into \eqref{ef}, interchanging
the order of integration, and evaluating the contour integral over $u$ 
by use of Cauchy's theorem, we find the alternate expression 
$\int dz 4 \pi \psi(z) D(z)$. The equality of these two expressions for arbitrary
$D(z)$ implies \eqref{disc}.}
On the other hand  
\begin{equation}\label{princ}
\Phi^+(z)+\Phi^-(z)= 2  
\sum_i 
{\cal P}
\int_{A_i}^{B_i} \frac{d\omega}{i\omega} 
\frac{z+\omega}{z-\omega}\rho(\omega)= \frac{2 U(z)}{i N}
\end{equation}
(the first equality is the definition of the principal value, while 
the second equality follows using \eqref{fovv}).
Moreover it follows immediately from \eqref{fovv} and \eqref{gwws1} that  
\begin{equation}\label{asymp}
\lim_{u \to \infty} \Phi(u)= A.
\end{equation}

As the RHS of \eqref{princ} is a known function, we are posed with the 
problem of determining $\Phi(u)$ given its principal value along a cut. 
The problem of determining an analytic function from its discontinuity 
along a cut is, of course, standard. A simple variable change allows 
us to turn the principal value of $\Phi(u)$ into the discontinuity of 
a new function, as we now explain. 

Consider the function 
\begin{equation}\label{hz}
h(u)=\sqrt{(A_1-u)(B_1-u)(A_2-u)(B_2-u) \ldots (A_n-u)(B_n-u)}.
\end{equation} 
We define $h(u)$ to have cuts precisely on the $n$ arcs on the unit 
circle that extend from $(A_i, B_i)$. 
This definition fixes the function $h(u)$ up to an overall sign. This sign 
will cancel out in our solution for $\Phi$ below, and so is uninteresting. 
For future use
we note that when $u=e^{i \alpha}$ 
\begin{equation}\label{hsous}
h^2(u)=\prod_{m=1}^n 4 e^{i \frac{a_m+b_m}{2}} e^{i \alpha}
 \left(\sin^2 \left( \frac{a_m-b_m}{4} \right)
- \sin^2\left( \frac{\alpha}{2}-\frac{a_m+b_m}{4}
\right) \right). {}\; 
\end{equation} 
\footnote{This may be seen by using 
\begin{equation}\begin{split} 
(e^{ia_1}-e^{i\alpha})(e^{ib_1} -e^{i \alpha})
&=2 e^{i \left(\frac{a_1+b_1}{2} +\alpha \right)} 
\left( \cos\left( \alpha-\frac{a_1+b_1}{2}
\right) - \cos \left( \frac{a_1-b_1}{2} \right) \right) \\
&
=4  e^{i \left( \frac{a_1+b_1}{2} +\alpha \right)} \left(\sin^2 \left( \frac{a_1-b_1}{4} \right)
- \sin^2\left( \frac{\alpha}{2}-\frac{a_1+b_1}{4}
\right) \right).
\end{split}
\end{equation} }
We use the function $h(z)$ to define a new function, $H(z)$, via the equation
$$\Phi(z)=h(z) H(z).$$
Using the fact that $h^+(z)=-h^-(z)$ along the cut, \eqref{princ} turns
into 
\begin{equation}\label{princd}
H^+(z)-H^-(z)= \frac{2 U(z)}{i N h^+(z)}.
\end{equation}
Note that at large $v$, $H(v) ={\cal O}(\frac{1}{v^n})$ so that 
\begin{equation}\label{ci}
\int_{C_\infty} dv \frac{H(v)}{2 \pi i (v-u)}=0
\end{equation}
where the the contour $C_{\infty}$ 
runs counterclockwise over a very large circle at 
infinity. Assuming that the function $H(u)$ has singularities only along 
the cuts $(A_i, B_i)$ and that the function $U(u)$ is analytic on the cuts, 
the use of Cauchy's theorem on \eqref{ci} yields
\begin{equation}\label{Huc}
H(u)=-\int_{C_{cuts}} dv \frac{H(v)}{2 \pi i (v-u)}=
\frac{1}{\pi} \int_{L_{arcs}} dz \frac{U(z)}{ N h^+(z)(z-u)}
=\frac{1}{2\pi} \int_{C_{cuts}} dv \frac{U(v)}{ N h(v)(v-u)}.
\end{equation}
In the equation above, $v$ is a variable on the complex plane while $z$ 
is a variable on the unit circle of the complex plane. The contour 
$C_{cuts}$ encloses each of the $n$ cuts $(A_i, B_i)$ but does not enclose 
the point $u$. The integration region $L_{arcs}$ runs counterclockwise 
along $n$ cuts on the unit circle, integration region for each cut is from
$A_i$ to $B_i$.
The second equality in \eqref{Huc} is obtained 
by using \eqref{princd}. The third equality 
uses $h^+(z)=-h^-(z)$ together with the assumption that 
$U(z)$ has no singularities on the $n$ cuts.
 
In practical situations, \eqref{Huc} may itself be evaluated using 
contour techniques. Suppose, for instance, that $U(z)$ is a meromorphic 
function whose 
poles, located at the points $z_1 \ldots z_r$. Then we find 
\begin{equation}\label{Hucp}
H(u)=\frac{1}{2\pi} \int_{C'} dz \frac{U(z)}{ N h(z)(z-u)}
- \frac{iU(u)}{N h(u)}-\sum_{m=1}^r i {\rm Res}_{z=z_k} \frac{U(z)}{Nh(z)(z-u)}
\end{equation}
where the contour $C'$ runs counterclockwise over an infinitely large circle. In simple examples the last contour integral either 
vanishes or evaluates to something simple.

\subsubsection{The GWW problem}\label{gwwv}

As an example let us consider the GWW matrix model\cite{Gross:1980he,Wadia:1980cp,Wadia:2012fr}. Let the potential be 
given by 
$$V(U)=-N\frac{\zeta}{2} \left( \Tr U + \Tr U^\dagger \right).$$
In terms of eigenvalues, we have $V(\alpha)= -N \zeta \cos \alpha$ so that 
$U(\alpha)=V'(\alpha)=N \zeta \sin \alpha$. In the complex coordinate employed
above this corresponds to 
$$U(u)= \frac{N \zeta}{2i} \left( u-\frac{1}{u} \right)$$
where $u = e^{i\alpha}$.

\begin{figure}
  \begin{center}
  \subfigure[]{\includegraphics[scale=.33]{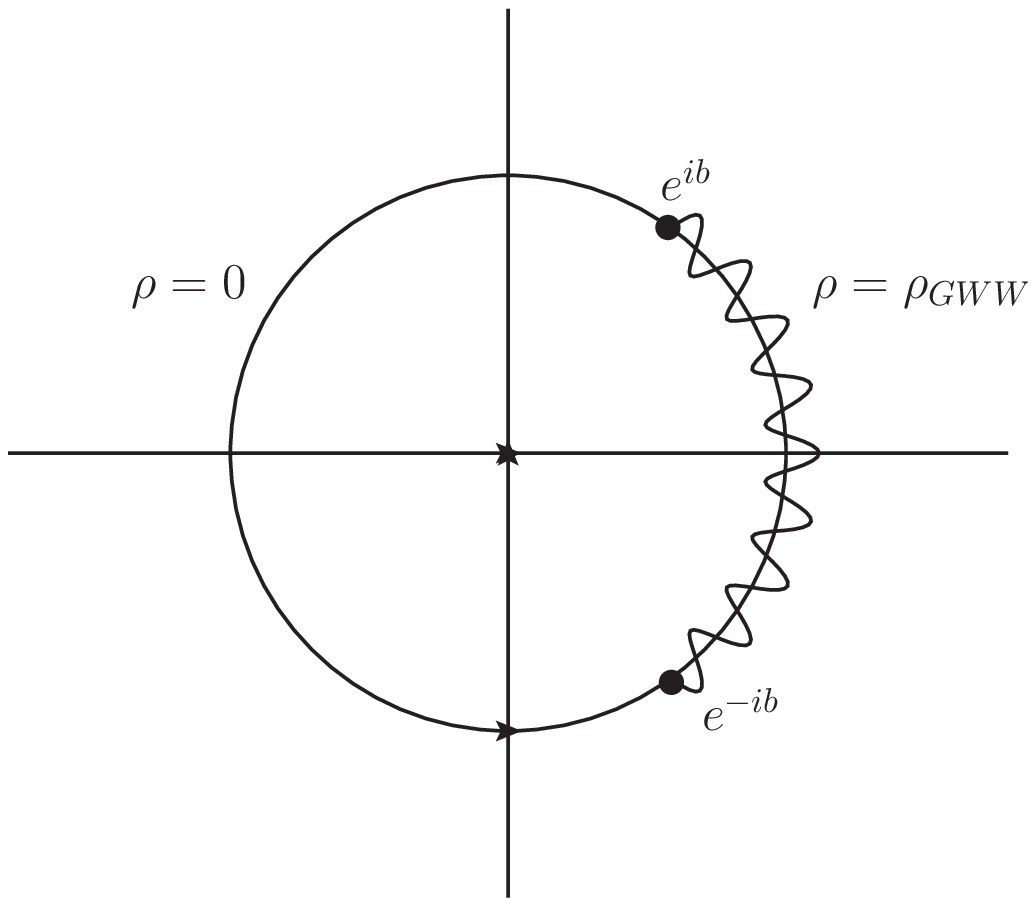}
\label{1-cutGWW}
  }
  \qquad\qquad
  \subfigure[]{\includegraphics[scale=.33]{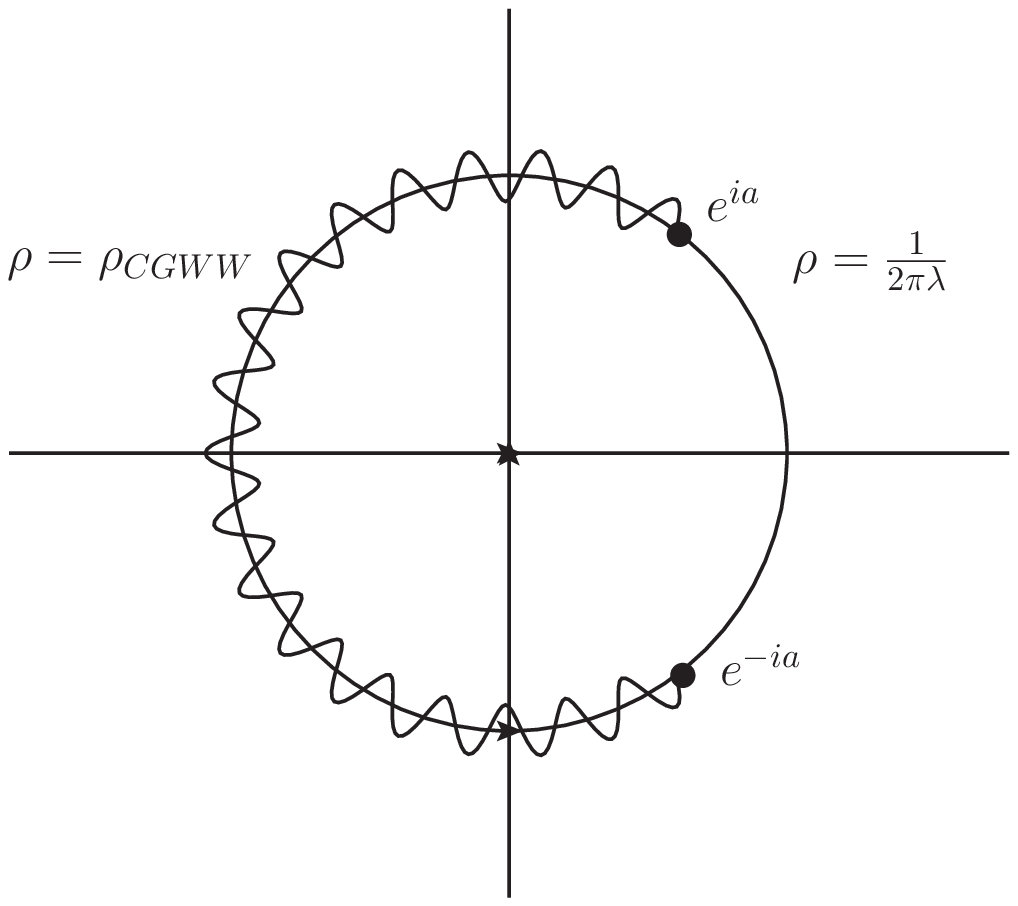}
\label{capped-GWW}
  }
  \qquad\qquad
  \subfigure[]{\includegraphics[scale=.33]{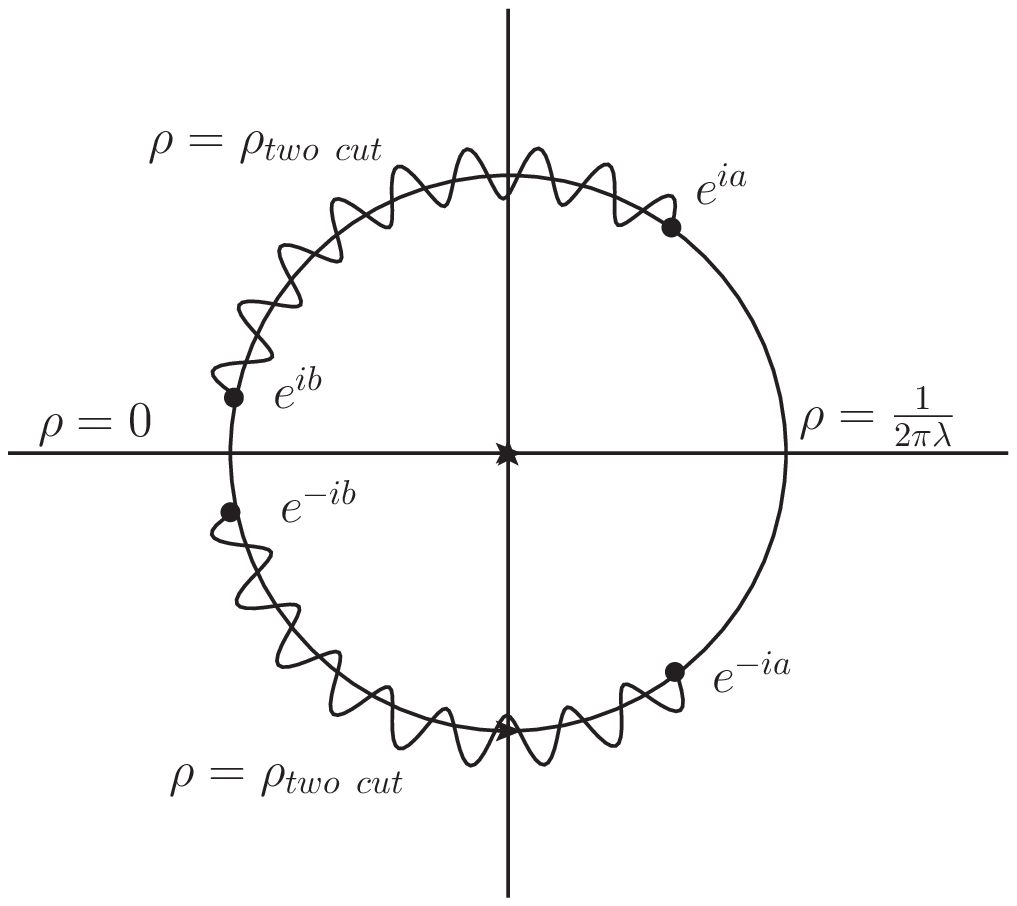}
\label{2cut.eps}
  }
\caption{Fig.\ref{1-cutGWW}, \ref{capped-GWW}, \ref{2cut.eps} show one lower cut solution, one upper cut one, two cut one, respectively.  }
  \end{center}
\end{figure}

We search for a single cut solution with $A_1=e^{-i b}$ and $B_1=
e^{i b}$ (See figure \ref{1-cutGWW}). 
The solution for $H(u)$ is given by 
\eqref{Huc} with the appropriate function $h(u)$. We pause to 
carefully define the function $h(u)$ and enumerate some of its 
properties. Let $u=|u|e^{i \theta}$ with $\theta \in (-\pi, \pi]$, then
we will define the $h(u)$ as
\begin{equation}
h(u) = \left((u-e^{ib })
(u - e^{-ib})
\right)^{\frac{1}{2}}.
\end{equation}
The single cut of this function is taken to lie in an arc on the 
unit circle extending from  $A_1$ to 
$B_1$. The function $h$ is discontinuous on this arc with   
\begin{equation}
h^{\pm}(z) = \pm 2 e^{\frac{i\theta}{2}}
\sqrt{
\sin^2 \frac{b}{2} - 
\sin^2 \frac{\theta}{2}  }
\label{Behaviour in the cut}
\end{equation}
where $z = e^{i\theta}$.
Away from this cut (i.e. on the gap) on the unit circle
\begin{eqnarray}
&h(z) = 2i e^{\frac{i\theta}{2}}
\sqrt{\sin^2 \frac{\theta}{2} - 
\sin^2 \frac{b}{2}  } \qquad &\theta \ge 0, \nonumber \\
&h(z) = -2i e^{\frac{i\theta}{2}}
\sqrt{\sin^2 \frac{\theta}{2} - 
\sin^2 \frac{b}{2}  } \qquad &\theta \le 0. 
\end{eqnarray}
(recall that $\sin^2 \frac{\theta}{2} \geq \sin^2 \frac{b}{2}$
everywhere on the gap). On the real axis $h(x)$ is everywhere real and  
is positive for $x>1$, but is negative for $x<1$.  In particular $h(0)=-1$.%
\footnote{In more detail along the real axis
\begin{eqnarray}
&h(x) = 
\sqrt{\left(
x - \cos b
\right)^2 + 
\sin^2 b  } > 0 \qquad &x > 1, \nonumber \\
&h(x) = -\sqrt{\left(
x - \cos b
\right)^2 + 
\sin^2 b  } < 0 \qquad &x < 1.
\end{eqnarray}
}
At infinity 
\begin{equation}
h(z) \to z \quad \text{at} \quad |z| \to \infty .
\end{equation}

$H(u)$ is now easily obtained from \eqref{Hucp} as follows. 
At large $z$, $\frac{U(z)}{ N h(z)(z-u)} = \frac{ \zeta}{2 iz}$ so that the contour 
integral at infinity in \eqref{Hucp} evaluates to $\frac{\zeta}{2}$. 
The only singularity in $U(z)$ is a pole at zero; using  
$h(0)=-1$ it follows that the residue contribution of this pole to 
\eqref{Hucp} is $\frac{\zeta}{2u}$. Putting these contributions together we 
conclude that 
\begin{equation}\label{gwwaU0}
H(u)=\frac{\zeta}{2}  \left[ \left( 1+ \frac{1}{u} 
\right) -\frac{u - \frac{1}{u}}{h(u)} \right],
\end{equation}
in other words, 
\begin{equation}\label{gwwaU}
\Phi(u)=\frac{\zeta}{2}  \left[ \left( 1+ \frac{1}{u} 
\right)h(u) -\left( u - \frac{1}{u} \right) \right].
\end{equation}
Using \eqref{disc} and \eqref{Behaviour in the cut} we conclude that 
\begin{equation}\label{gwws2}
\psi(\alpha)=\frac{\zeta}{\pi}\cos \frac{\alpha}{2}
\sqrt{\sin^2\frac{b}{2}-\sin^2\frac{\alpha}{2}}.
\end{equation}

It remains to determine $b$. At large $u$,
$$h(u)=u - \cos b  + {\cal O} \left( \frac{1}{u} \right).$$
It follows from \eqref{gwwaU} that at large $u$ 
$$\Phi(u)=\frac{\zeta}{2}(1-\cos b) 
+{\cal O}\left(\frac{1}{u} \right) = 
\zeta \sin^2\frac{b}{2}+{\cal O}\left(\frac{1}{u} \right).$$
Comparing with \eqref{fovv}, we conclude 
\begin{equation}\label{gs}
\sin^2\frac{b}{2}=\frac{A}{\zeta}.
\end{equation}

In summary the eigenvalue distribution is given by 
\begin{equation}\label{gwws3}
\psi(\alpha)=\frac{\zeta}{\pi}\cos \frac{\alpha}{2}
\sqrt{\frac{A}{\zeta}-\sin^2\frac{\alpha}{2}}.
\end{equation}
We recover the famous GWW \cite{Gross:1980he,Wadia:1980cp,Wadia:2012fr} result upon setting $A=1$ in this result.

\subsection{The capped unitary matrix model}\label{cumm}

In this subsection we turn to the study of the capped unitary matrix 
model. As we have explained in the main text, the saddle point equations 
for this model are given by $\rho_0 + \psi$ where $\rho_0$ is any continuous function that obeys the boundary conditions
 and $\psi$ obeys the integral equation
\eqref{vvec}. \eqref{vvec} is a special case of \eqref{vvecn} with 
$U(\alpha)$ and $A$ specified in the last two of \eqref{vvec}. Consequently 
the analysis of the previous subsection applies to this case. In particular 
the eigenvalue density $\psi$ is given by \eqref{disc} for the analytic 
function $\Phi(u)=h(u)H(u)$ where $H(u)$ is given by 
\begin{equation}\label{Huca}
H(u)=\frac{1}{2\pi} \int_{C_{cuts}} dv \frac{U(v)}{ N h(v)(v-u)}
\end{equation}
(see \eqref{Huc}) with 
\begin{equation}\label{ua}
U(z)=V'(z)
-N {\cal P} \int \frac{dw}{w} \rho_0(w)\frac{z+w}{z-w}, ~~~
A=1-\int \frac{dw}{iw} \rho_0(w)
\end{equation}
where both integrals run over the full unit circle (more precisely, 
the integral can be taken to exclude all lower gaps 
because $\rho_0$ vanishes there). 

The contribution to $H(u)$ \eqref{Huca} from the second term in $U(z)$ 
in \eqref{ua} (the term proportional to 
${\cal P} \int \frac{dw}{w} \rho_0(w)\frac{z+w}{z-w}$) may be evaluated 
as follows.  We interchange the order of integration in \eqref{Huca}
to do the $v$ integral first, followed by the $w$ integral. The $v$ 
integral may conveniently be performed by deforming the integration 
contour to infinity; in that process we pick up the contribution from the 
poles at $v=w$. The integral at infinity vanishes because the integrand
falls of at least as fast as $\frac{1}{z^2}$. The contribution of the 
poles at $v=w$ 
$$2 i {\cal P} \int \frac{dw \rho_0(w)}{h(w)(w-u)}$$
where the integral runs over the all cuts and all upper gaps. However 
it is easily seen that the contribution of the 
cuts to this integral vanishes \footnote{The reason for this is that 
$h(w)$ changes sign across a cut. Consequently the principal value - 
the average of an integral over the upper and lower contour - vanishes
over the cut.}. In other words the integral in this expression 
receives contributions only from the upper gap, where 
$\rho_0(w)=\frac{1}{2 \pi \lambda}$. Putting it all together we conclude
that 
\begin{equation}\label{hg} \begin{split}
H(u)&=H_V(u)+H_{\rho_0}(u)\\
H_V(u)&=\frac{1}{2\pi} \int_{C_{cuts}} dv \frac{V'(v)}{ N h(v)(v-u)}\\
H_{\rho_0}(u)&=+i 
\frac{{\cal P} \int_{L_{unit ~circle}} \frac{dw}{w} \rho_0(w)\frac{u+w}{u-w}}{h(u)}
+\frac{i}{\pi \lambda} \int_{L_{ugs}} \frac{dw }{h(w)(w-u)}\\
\end{split}
\end{equation}
where $C_{cuts}$ is any contour that encloses all $n$ cuts but excludes the 
point $u$, $L_{unit~circle}$ runs counterclockwise 
over the unit circle and $L_{ug}$ is the sum over arcs that run 
counterclockwise all upper gap regions on the unit circle.
The analytic function $\Phi(z)$ is given by 
\begin{equation}\label{hgp} \begin{split}
\Phi(u)&=\Phi_V(u)+\Phi_{\rho_0}(u)\\
\Phi_V(u)&=h(u)H_V(u)
= h(u)\frac{1}{2\pi} \int_{C_{cuts}} dv \frac{V'(v)}{ N h(v)(v-u)}\\
\Phi_{\rho_0}(u)&=h(u) H_{\rho_0}(u)= i {\cal P} \int_{C_{unit~circle}} 
\frac{dw}{w} \rho_0(w)\frac{u+w}{u-w}
+ \frac{i h(u)}{\pi \lambda} \int_{L_{ugs}} \frac{dw }{h(w)(w-u)} .
\\
\end{split}
\end{equation}
The discontinuity in $\Phi_{\rho_0}$ is easily evaluated; using  
\eqref{disc} we find that across the cuts,
\footnote{Each of the integrals in the third of \eqref{hg} also has a 
discontinuity across upper gap regions, but these discontinuities 
cancel between the two different integrals in \eqref{hgp}. As a consequence
$\Phi_{\rho_0}$ is analytic across all upper gaps. It is also, of course, 
analytic across all lower gaps.}
\begin{equation}\label{disct}
\Phi_{\rho_0}^+(z) -\Phi^-_{\rho_0}(z) \equiv
4 \pi \psi_{\rho_0}(z) = \frac{2i h^+(z)}{\pi \lambda}
\int_{L_{ugs}}\frac{dw}{h(w)(w-z)} -4\pi \rho_0(z).
\end{equation}
Let us define 
\begin{equation}\label{psiv}
\Phi_{V}^+(z) -\Phi^-_{V}(z) \equiv 4 \pi \psi_{V}(z) .
\end{equation}
It follows that the eigenvalue density, $\rho(z)$, of the capped unitary 
matrix model is given in the cuts by  
\begin{equation}\label{rhofin}
\rho(z)=\psi_V(z) + \psi_{\rho_0}(z)+ \rho_0(z)=\psi_V(z) + 
\frac{2 i h^+(z)}{\pi \lambda}
\int_{L_{ugs}}\frac{dw}{h(w)(w-z)}.
\end{equation}
Of course $\rho(z)$ vanishes in lower gap regions and equals 
$\frac{1}{2 \pi \lambda}$ in upper gap regions. 

At large $u$, 
\begin{equation}\label{lup}
\Phi_{\rho_0}(u)= \frac{i h(u)}{\pi \lambda} 
\int_{L_{ugs}} \frac{dw }{h(w)(w-u)}
-\int \frac{dw}{iw} \rho_o(w) +{\cal O}(\frac{1}{u}).
\end{equation}
It follows that at large $u$  
\begin{equation}\label{bcpv}
\Phi_V(u)=1- \frac{i h(u)}{\pi \lambda} 
\int_{L_{ugs}} \frac{dw }{h(w)(w-u) } +{\cal O}(\frac{1}{u}).
\end{equation}

In summary, the saddle point eigenvalue distribution of the capped 
unitary matrix model is given by \eqref{rhofin} with $\psi_V$ defined 
in \eqref{psiv} in terms of the function $\Phi(u)_V$ defined in 
\eqref{hgp}. The contour of integration $L_{ugs}$ runs over all upper 
gap arcs on the unit circle. The locations of these arcs is determined by 
the normalization condition \eqref{bcpv}. 
Note that the arbitrary function $\rho_0$ that we utilized in intermediate
steps in our computation has disappeared from the final answer as it 
should.

Two comments about our solution are in order. First, even though this 
is not immediately apparent, it may be verified that the RHS of 
the eigenvalue distribution function \eqref{rhofin} is real. Second, 
while the eigenvalue distribution \eqref{rhofin} is not analytic at 
branch points (boundaries of the cuts) it is continuous. 
$$\frac{2 i h^+(z)}{\pi \lambda}
\int_{L_{ugs}}\frac{dw}{h(w)(w-z)}$$
vanishes at the boundary of every lower gap (as $h(z)$ vanishes at that 
boundary). Near the boundary of an upper gap, on the other hand, the 
integral $\int_{L_{ugs}}\frac{dw}{h(w)(w-z)}$ diverges, and it is possible 
to demonstrate that 
$$\frac{2 i h^+(z)}{\pi \lambda}\int_{L_{ugs}}\frac{dw}{h(w)(w-z)}=
\frac{1}{2 \pi \lambda}$$
in this limit. As this is also the value of the eigenvalue density 
function in the upper gap region, it follows that $\rho(z)$ is everywhere 
continuous. 

\section{Solution of the Capped GWW model}\label{gwwc}

In this appendix we use the results of Appendix \ref{gww} to find an explicit solution to the 
capped GWW matrix model. As reviewed in the introduction to this paper, the capped GWW model has four phases. 
The no gap and lower gap phases of this model are identical to that of the original 
GWW model, reviewed in Appendix \ref{gww}. The new phases of this model are the 
upper gap phase and the two gap phase. In the rest of this appendix we proceed to 
determine the solution in these phases applying the general formalism of subsection 
\ref{cumm} above.

\subsection{One upper gap solution of the capped GWW model}\label{gwwm}

In this subsection we apply the analysis presented above to obtain 
the solution with a single upper gap (but no lower gap) of the capped 
GWW model, i.e. the special case
$$V'(u)= \frac{N \zeta}{2i} \left( u-\frac{1}{u} \right).$$
We search for a solution with no lower gap and one upper gap that extends
from counterclockwise from $e^{-ia}$ to $e^{ia}$ ($a$ is positive). 
In other words the cut in our problem is 
the {\it complement} of the cut in subsubsection \ref{gwwv}, see Fig.\ref{capped-GWW}. 

The one upper cut solution may be obtained by applying the general 
method outlined above. The function 
$h(z)= \sqrt{(e^{ia}-z)(e^{-ia}-z)}$ in this section {\it differs} from 
$h(z)$ in subsubsection \ref{gwwv} in the location of its branch cut, 
which is taken to run counterclockwise from $e^{-ia}$ to $e^{ia}$. We 
pause to enumerate some of the properties of the function $h(z)$ 
employed in this section. The function we work with obeys  
\begin{equation}\begin{split} 
-h^-(e^{i\alpha})= 
h^+(e^{i\alpha})=& + 2i e^{i\frac{\alpha}{2}}\sqrt{\sin^2\frac{\alpha}{2} - \sin^2\frac{a}{2}} {\rm~~~if~~~} \pi>\alpha>a, \\
                   & -2i e^{i\frac{\alpha}{2}}\sqrt{\sin^2\frac{\alpha}{2} - \sin^2\frac{a}{2}} {\rm~~~if~~~} -\pi<\alpha<-a ,\\     
                   h(e^{i \alpha}) =& 2e^{i\frac{\alpha}{2}}\sqrt{\sin^2\frac{a}{2}-\sin^2\frac{\alpha}{2}} {\rm ~~~if~~~} a>\alpha>-a ,\\
h(x) =& +\sqrt{(x-\cos a)^2 +\sin^2a} {\rm ~~~for~~~} x>1, \\
      & -\sqrt{(x-\cos a)^2 +\sin^2a} {\rm ~~~for~~~} x<-1 ,\\
h(z\rightarrow \infty)& \rightarrow z ,\\
h(0)& =1,
\end{split}
\label{hprop}
\end{equation}
where $h^-$ and $h^+$ are the values of function along the branch cut just 
inside and just outside the unit circle.

Using these properties, the contour integral in the second of \eqref{hg}
is easily evaluated and we find 
\begin{equation}\label{Honecut}
H_{V}(u)=\frac{\zeta}{2}  \left[(1 -\frac{1}{u}) -
\frac{u - \frac{1}{u}}{h(u)}  \right] 
\end{equation}
(the contribution of the contour at infinity is $\frac{\zeta}{2}$). It follows
that 
\begin{equation}\label{puonecut1}
\Phi_V(u)= \frac{\zeta}{2}  \left[h(u)(1 -\frac{1}{u}) -
(u - \frac{1}{u})  \right] .
\end{equation}
It follows from \eqref{psiv}, \eqref{hprop} and 
\eqref{puonecut1} that 
\begin{equation}\label{puonecut2}
4 \pi \psi_V(e^{i \mu})=
-4\zeta  \sin\frac{\mu}{2} 
\sqrt{\sin^2\frac{\mu}{2}-\sin^2\frac{a}{2}}.
\end{equation}
In order to simplify the second term on the RHS of \eqref{rhofin}, 
simply recall that 
$$h^+(e^{i\mu})=
2ie^{i\frac{\mu}{2}}\sqrt{\sin^2\frac{a}{2}-\sin^2\frac{\mu}{2}}$$
and note that 
$$\frac{-i dw}{h(w)}=\frac{d\theta e^{i\frac{\theta}{2}}}
{2\sqrt{\sin^2\frac{a}{2}-\sin^2\frac{\theta}{2}} }.$$
It follows from \eqref{rhofin} that  
\beal{ 
 4\pi\rho(e^{i\mu})&=-4\zeta \left|\sin\frac{\mu}{2}\right| \left(\sin^2\frac{\mu}{2}-\sin^2\frac{a}{2}\right)^{\frac{1}{2}} -\frac{2 h^+(e^{i\mu})}{\pi\lambda}\int^{a}_{-a}\frac{d\theta \cos\frac{\theta}{2}(1-e^{-i\mu})}{8\left(\sin^2\frac{a}{2}-\sin^2\frac{\theta}{2}\right)^{\frac{1}{2}}\left(\sin^2\frac{\mu}{2}-\sin^2\frac{\theta}{2}\right)}\nn
&=-4\zeta  \left|\sin\frac{\mu}{2}\right| \left(\sin^2\frac{\mu}{2}-\sin^2\frac{a}{2}\right)^{\frac{1}{2}}+\frac{2}{\lambda} \label{rhfoc}
}
where the last equality follows from explicitly evaluating the integral. 

We now turn to the evaluation of $a$. At large $u,$  $\Phi_V(u)= -\frac{\zeta}{2}(\cos a+1)+O[\frac{1}{u}]$.  Now 
 \eqref{bcpv} at large $u$ gives
\begin{equation}
 -\frac{\zeta}{2}(\cos a+1)=1-\frac{1}{\lambda}
\end{equation}which implies
\begin{equation}\label{adef}
 \cos^2\frac{a}{2}=\frac{\frac{1}{\lambda}-1}{\zeta}.
\end{equation}

In summary, the saddle point eigenvalue distribution is given by 
\begin{equation}\begin{split}\label{ocf}
&\rho(e^{i\mu})=-\frac{\zeta}{\pi}  \vline \sin\frac{\mu}{2} \vline
\left(\sin^2\frac{\mu}{2}-\sin^2\frac{a}{2}\right)^{\frac{1}{2}}+\frac{1}{2 \pi\lambda}~~~(|\mu|>a)\\
&\rho(e^{i\mu})=\frac{1}{2 \pi\lambda}~~~(|\mu|<a)
\end{split}
\end{equation}
where $a$ is defined in \eqref{adef}.

\subsection{Two cut solution of the capped GWW model}\label{tcgww}

In this section we search for a solution of the capped GWW model
\footnote{As in the previous subsubsection 
$V'(\alpha)=\zeta N \sin \alpha$.} that 
has one upper gap centered around $\alpha=0$ and one lower gap centered 
around $\alpha=\pi$. We assume that our cuts 
extend along the unit circle from $e^{ia}$ to $e^{ib}$ and from 
$e^{-ib}$ to $e^{-ia}$ respectively 
($a$ and $b$ lie in $(0, \pi)$ with $b>a$, see also Fig.\ref{2cut.eps}).
In order to obtain the eigenvalue distribution, we use the general 
discussion presented earlier in this subsection, with 
$h(u)$ as
\begin{equation}
h(u) = \left((u-e^{ia})(u-e^{-ia})(u-e^{ib})(u-e^{-ib})
\right)^{\frac{1}{2}}.
\end{equation}
$h(u)$ is take to have cuts along the arcs enumerated above.
We pause to summarize some of the properties of the analytic function 
$h(u)$.

Along the unit circle outside the cuts, the $h$ is analytic as
\beal{
h(e^{i\theta}) =&  4 e^{i\theta} \sqrt{ ( \sin^2{a \over 2}-\sin^2{\theta \over 2}) 
(\sin^2{b \over 2} - \sin^2{\theta \over 2}) }, \, \quad -a<\theta<a, \\
h(e^{i\theta}) = & - 4 e^{i\theta} \sqrt{ (\sin^2{\theta \over 2} - \sin^2{a \over 2}) 
(\sin^2{\theta \over 2} - \sin^2{b \over 2}) }, \, \quad -\pi<\theta<-b, \, b<\theta<\pi .
}
Along the cuts, $h$ becomes discontinuous with 
\beal{
h^{\pm}(e^{i\theta}) =& \pm 4 i e^{i\theta} \sqrt{ (\sin^2{\theta \over 2} - \sin^2{a \over 2}) 
(\sin^2{b \over 2} - \sin^2{\theta \over 2}) }, \, \quad a<\theta<b, \\
h^{\pm}(e^{i\theta}) = & \mp 4 i e^{i\theta} \sqrt{ (\sin^2{\theta \over 2} - \sin^2{a \over 2}) 
(\sin^2{b \over 2} - \sin^2{\theta \over 2}) }, \, \quad -b<\theta<-a, 
}
where $h = h^{+}$ on the outside of the circle and 
$h = h^{-}$ on the inside. Moreover $h$ is real and positive along the 
real axis; explicitly for real $x$
 \beal{
h(x) =&  \sqrt{ (1-2x \cos a + x^2)  (1-2x \cos b + x^2) }, \, \quad -\infty<x<\infty  .
}
Note in particular that $h(0)=1$. At large $u$ we have  
\beal{
h(u) \sim +u^2 - u (\cos a + \cos b), \quad u \sim \infty.
}

Proceeding as in the previous subsubsection we find 
$$H_V(u)=\frac{\zeta}{2}  \left[ -\frac{1}{u} -
\frac{u - \frac{1}{u}}{h(u)} \right]$$
and 
$$\Phi_V(u)= \frac{h(u)\zeta}{2}  \left[ -\frac{1}{u} -
\frac{u - \frac{1}{u}}{h(u)} \right],$$
which follows that 
$$4\pi\psi_V(u)=- h^+(u) \frac{\zeta}{u} $$
The full eigenvalue density can now be computed using \eqref{rhofin}. 
The second term on the RHS of \eqref{rhofin} can be simplified 
by moving back to angle variables.
Substituting 
$w=e^{i\theta},~~ u= e^{i\alpha}$,  
converting the integral over $w$ to integral over $\theta$ and using 
\eqref{detab} 
below we find the following 
explicitly real expression for the density distribution
\begin{equation}
\begin{split}
4\pi\rho(\alpha)&= \frac{\vline\,\sin\alpha\,\vline}{\pi\lambda}\sqrt{\left(\sin^2\frac{\alpha}{2}-\sin^2\frac{a}{2}\right)\left(\sin^2\frac{b}{2}-\sin^2\frac{\alpha}{2}\right)} ~I_1{\rm ~~~~~where~~~~} \\
I_1 &= \int_{-a}^a \frac{d\theta}{(\cos\theta- \cos\alpha)\sqrt{\left(\sin^2\frac{a}{2}-\sin^2\frac{\theta}{2}\right)\left(\sin^2\frac{b}{2}-\sin^2\frac{\theta}{2}\right)}}. \\ 
\end{split}
\label{realrho}
\end{equation}

It remains to evaluate $a$ and $b$. Setting the coefficients of 
terms of ${\cal O}(u)$ and ${\cal O}(1)$ in \eqref{bcpv} to zero 
(and substituting $w=e^{i \alpha}$ to simplify the integrals)
\footnote{Under this substitution 
$h(w)=4 e^{i \alpha}\sqrt{\sin^2\frac{a}{2}-\sin^2 \frac{\alpha}{2}}
\sqrt{\sin^2\frac{b}{2}-\sin^2 \frac{\alpha}{2}}$ where the 
square root is positive.}
we obtain two equations
\begin{equation}\label{detab}
\begin{split}
&\frac{1}{4\pi \lambda} \int_{-a}^a d \alpha  \frac{1}
{\sqrt{\sin^2\frac{a}{2}-\sin^2 \frac{\alpha}{2}}
\sqrt{\sin^2\frac{b}{2}-\sin^2 \frac{\alpha}{2}}} =\zeta\\
&\frac{1}{4\pi \lambda} \int_{-a}^a d \alpha \frac{\cos \alpha}
{\sqrt{\sin^2\frac{a}{2}-\sin^2 \frac{\alpha}{2}}
\sqrt{\sin^2 \frac{b}{2}-\sin^2 \frac{\alpha}{2}}}=
1+\frac{\zeta}{2} \left( \cos a+\cos b \right) 
\end{split}
\end{equation}
which determine $a$ and $b$. \eqref{realrho} and \eqref{detab}   
provide a complete solution of the two cut model. 

\subsection{Special Limits of the capped GWW model}

\subsubsection{$\rho(\alpha)$ as $b \to \pi$} \label{spb}

Let us first consider the eigenvalue density in $b\rightarrow \pi$ limit. 
Setting $b=\pi$ in \eqref{realrho} and we obtain
\begin{equation}\begin{split}
4\pi\rho(\alpha)&= \frac{\vline\,\sin\alpha\,\vline \cos\frac{\alpha}{2}}{\pi\lambda}\sqrt{\left(\sin^2\frac{\alpha}{2}-\sin^2\frac{a}{2}\right)} 
          \int_{-a}^a \frac{d\theta}{(\cos\theta- \cos\alpha)\cos\frac{\theta}{2}\sqrt{\left(\sin^2\frac{a}{2}-\sin^2\frac{\theta}{2}\right)}}. \\
\end{split}
\end{equation}

Using  
$$ \int_{-a}^a \frac{d\theta}{(\cos\theta- \cos\alpha)\cos\frac{\theta}{2}\sqrt{\left(\sin^2\frac{a}{2}-\sin^2\frac{\theta}{2}\right)}}
= \frac{\pi}{\cos^2\frac{\alpha}{2}}\left( \frac{1}{\sin\frac{\alpha}{2}\sqrt{\sin^2\frac{\alpha}{2}-\sin^2\frac{a}{2}}}- \frac{1}{\cos\frac{a}{2}} \right)$$
together with \eqref{bpieq} we find 
\begin{equation}\label{bpilim}
\rho(\alpha)= \frac{1}{2\pi\lambda}- \frac{\zeta}{\pi} 
\vline\,\sin\frac{\alpha}{2}\,\vline\sqrt{\sin^2\frac{\alpha}{2}-\sin^2\frac{a}{2}}
\end{equation}
(where $\zeta$ and $a$ are given in \eqref{bpieq}) in perfect agreement 
with \eqref{evdd}.

\subsubsection{$\rho(\alpha)$ as $a \to 0$} \label{spa}

Now let us examine the $a\rightarrow 0$ limit of \eqref{realrhom}. While the 
limits of integration coincide in this limit, and the integrand 
diverges in a precisely compensating manner, leading to a finite result.
The integral is easily evaluated in this limit; combining the result 
 with the first equation in \eqref{detabm} we find
\begin{equation}
\rho(\alpha)= \frac{\zeta}{\pi}\cos\frac{\alpha}{2}\sqrt{\sin^2\frac{b}{2}-\sin^2\frac{\alpha}{2}}
\end{equation}
(where $\zeta$ and $\lambda$ satisfy \eqref{a0sol}), in perfect agreement 
with \eqref{evdl}.

\subsubsection{The large $\zeta$ limit} \label{lz} 
In this section we elaborate on the evaluation of the integrals appearing in Eq.\eqref{realrho}, \eqref{detab}.
To start with, let us consider
\begin{equation}
 I=2\int_{0}^{a}d\theta \frac{1}{\sqrt{\left(\sin^2(\frac{a}{2})-\sin^2(\frac{\theta}{2})\right)\left(\sin^2(\frac{b}{2})-\sin^2(\frac{\theta}{2})\right)}}.
\end{equation}
Since $b$ is very near $a$ in the large $\zeta$ limit,  
the integral receives most of the contribution 
from the integration region near $a$, and it diverges. 
To separate out the divergent piece, let us divide the integral into two parts
$I = I_1 + I_2$,
\begin{equation}
\begin{split}\label{I}
  I_1 &= 2\int_{a-\gamma}^{a}d\theta \frac{1}{\sqrt{\left(\sin^2(\frac{a}{2})-\sin^2(\frac{\theta}{2})\right)\left(\sin^2(\frac{b}{2})-\sin^2(\frac{\theta}{2})\right)}},\\
I_{2}&=  2\int_{0}^{a-\gamma}d\theta \frac{1}{\sqrt{\left(\sin^2(\frac{a}{2})-\sin^2(\frac{\theta}{2})\right)\left(\sin^2(\frac{b}{2})-\sin^2(\frac{\theta}{2})\right)}},           
 \end{split}
\end{equation}
where $I_2$ is finite whereas $I_1$ contains a divergent piece. 
Our aim is to compute $a,b$ in terms of $\zeta.$ 
Now we take $b=a+\frac{\gamma}{M},$ where $M$ is large. We make a change of variable $\theta=a-\frac{\gamma}{M}y,$ 
and expanding $I_1$ and $I_2$ in $O\left(\theta ^2\right)$ and $O\left(\frac{1}{M}\right)^2$ respectively we obtain
\beal{
 I_1&= 8 \csc (a) \sinh ^{-1}\left(\sqrt{M}\right) +\frac{\gamma}{M}  
\left(2\sqrt{M (M+1)} \cot a \csc a -4 \cot a
   \csc a \sinh ^{-1}\sqrt{M}\right) \nn &~+ O\left(\gamma ^2\right) \nn
I_2&= 4 \csc a \left(\log \sin \left(a-\frac{\gamma }{2}\right)-\log \sin \left(\frac{\gamma
   }{2}\right)\right) \nn& 
+\frac{\gamma}{M}  \csc a \left(2\frac{\sin (a-\gamma )}{\cos a-\cos (a-\gamma
   )}+2\cot a \left(\log \sin \left(\frac{\gamma }{2}\right)-\log \sin \left(a-\frac{\gamma
   }{2}\right)\right)\right) \nn &+O\left(\frac{1}{M}\right)^2.
\label{I1I2}
}
When re-expressed as a function of $\epsilon (=\frac{\gamma}{M})$ and $M,$ $I_1+I_2$ must (of course) be  independent of $M$. Within the perturbative
expansion presented above this works as follows. Terms of order $O(\frac{1}{M^2})$ in the expansion of $I_2$, when re-expressed
as a function of $M$ and $\epsilon$, yields expressions of order $\frac{1}{M^2}$, $\frac{\epsilon}{M}$ and $\epsilon^2$ (along with 
logarithmic corrections). As we have not computed these second order corrections, it follows that the leading $M$ dependence of 
the $O(1)$ part of our answer must be at order $\frac{1}{M^2}$, and that the leading part $M$ dependence of the $O(\epsilon)$
part of this integral occurs at $O(\frac{1}{M})$; we have explicitly checked that this is the case. These sub leading $M$ 
dependences in the answer are an artifact of our perturbative expansion, and would vanish if we carried out our expansion to higher 
orders. The correct answer for the integral upto $O(\epsilon)$ is simply given by taking $M$ to infinity in the coefficients 
of the $O(1)$ and $O(\epsilon)$ parts of our integral.%
\footnote{Note that taking $M$ to infinity in the coefficients 
of higher order terms - e.g. the coefficient of $\epsilon^4$ - would yield a divergent result at this order in perturbation theory. 
This is because we have not yet included the third order terms in $I_2$ that would cancel this divergence. As a consequence, 
the correct procedure to estimate the integral expanded to first order is to truncate $I_1+I_2$ to first order in $\epsilon$ and then to take 
$M$ to $\infty$ in the final answer.}
We finally arrive at
\begin{equation}\begin{split}
 I&=I_1+I_2\\
&=2\left(-\epsilon \frac{\cos(a)}{\sin^{2}(a)}\left(-1+\log\left(\frac{8\sin(a)}{\epsilon}\right)\right)+2 \frac{1}{\sin(a)} \log\left(\frac{8\sin(a)}{\epsilon}\right)\right)+O[\epsilon^2].
\end{split}
\end{equation}
In a similar way one perform second integral appearing in Eq.(\ref{detab}) we obtain
\begin{equation}\label{I3}
 \begin{split}
I_3&=2\int_{0}^{a}d\theta \frac{\cos(\theta)}{\sqrt{\left(\sin^2(\frac{a}{2})-\sin^2(\frac{\theta}{2})\right)\left(\sin^2(\frac{b}{2})-\sin^2(\frac{\theta}{2})\right)}}\\
&=2\left(2~a+2~\cot(a) \log\left(\frac{8\sin(a)}{\epsilon}\right)-\epsilon\frac{1}{\sin^{2}(a)}\left(-1+\log\left(\frac{8\sin(a)}{\epsilon}\right)\right)\right)+O[\epsilon^2].
\end{split}
\end{equation}
Now  using $I$,$I_3$  in  \eqref{detab} we get two equations for $a$ and $\epsilon.$ Solving these equations we obtain
\begin{equation}\label{epsia}
\begin{split}
\epsilon&=8\sin(\pi \lambda) e^{-\pi\zeta\lambda\sin(\pi\lambda)}+O[\zeta~e^{-2\pi\zeta\lambda\sin(\pi\lambda)}] \\
a&=\pi\lambda-4\sin(\pi \lambda) e^{-\pi\zeta\lambda\sin(\pi\lambda)}+ O[\zeta~e^{-2\pi\zeta\lambda\sin(\pi\lambda)}]\\
   &=\pi\lambda-\frac{1}{2}\epsilon+ O[\zeta~\epsilon^2]            .
\end{split}
\end{equation}
Now in a similar way, 
one can compute the leading term in the eigenvalue density 
using Eq.(\ref{realrho}). In order to proceed we make following change of 
variables,
\begin{equation}
 \theta=a-\epsilon~y,~~b=a+\epsilon,~~\alpha=a+\epsilon \alpha_1.
\end{equation}
Note that $\alpha_1$ takes value between $0$ to $1.$ The value of $\rho$ takes the form
\begin{equation}
4\pi\rho=\frac{4}{\pi\lambda} ~\cos^{-1}\sqrt{\alpha_1}.
\end{equation}
As a check of our result, we show $\int_{-\pi}^{\pi}\rho d\alpha =1.$ The integral of $\rho$ can be written as
\begin{equation}
\begin{split}
\int_{-\pi}^{\pi}\rho d\alpha&=2\int_{a}^{b}\rho d\alpha +\int_{-a}^{a}d\alpha \rho_{0}\\
& =2\int_{0}^{1}\epsilon~\rho~ d\alpha_{1} +\int_{-(\pi\lambda-\frac{1}{2}\epsilon)}^{\pi\lambda-\frac{1}{2}\epsilon}d\alpha \frac{1}{2\pi\lambda}\\ 
&=\frac{\epsilon}{2\pi\lambda}+(1-\frac{\epsilon}{2\pi\lambda})\\
&=1.             
\end{split}
\end{equation}
\subsubsection{Eigenvalue density near the end points of cuts}
\label{ronear_a_b}

In this subsubsection we study the behaviour of the two cut eigenvalue density 
function $\rho(\alpha)$ given in \eqref{realrho} as $\alpha$ tends to the edges $a$ and $b$ 
of the distribution function.  We work at fixed $\zeta$ and $\lambda$ (hence 
fixed $a$ and fixed $b$).

 Let us first consider the limit $\alpha \to b$. Clearly, to leading order  
 \begin{equation}\label{ro2cut}\begin{split}
 4\pi\rho(\alpha)&= B \sqrt{b-\alpha}+ {\cal O}((b-\alpha)^\frac{3}{2}) \\
 B&=\frac{| \sin\frac{b}{2} |}{\pi\lambda}\sqrt{(\sin^2\frac{b}{2}-\sin^2\frac{a}{2}) \sin\frac{b}{2}\cos\frac{b}{2}}\\
          &\int^a_{-a} \frac{d\theta}{2(\sin^2\frac{b}{2}-\sin^2\frac{\theta}{2})\sqrt{(\sin^2\frac{a}{2}-\sin^2\frac{\theta}{2}) 
 (\sin^2\frac{b}{2}-\sin^2\frac{\theta}{2})}}.\\
 \end{split}
          \end{equation}
It is not difficult to check that the coefficient of $\sqrt{b-\alpha}$ agrees with the coefficient of the same term in the 
Gross-Witten-Wadia solution when $a$ is taken to zero, and that it vanishes when $b \to \pi$. 

We now study the limit $\alpha \to a$. The estimation of the eigenvalue density function in this limit is more subtle; 
we follow the method described in subsection\ref{lz}. Let $\alpha-a=\epsilon$. The expression for $\rho(\alpha)$ 
in \eqref{realrho} may be recast as 
$$\rho(\alpha)=I_1+I_2$$
where
 \begin{equation} \begin{split}
 \label{Iro}
  I_1 &= \frac{\sin(a)\sqrt{(\cos(a)-\cos(b))\sin(a)}}{\pi \lambda} \epsilon^{\frac{1}{2}} ~ \\
 &\int_{a-\gamma}^{a}d\theta \frac{1}{(\cos(\theta)-\cos(\alpha))\sqrt{\left(\sin^2(\frac{a}{2})-
 \sin^2(\frac{\theta}{2})\right)\left(\sin^2(\frac{b}{2})-\sin^2(\frac{\theta}{2})\right)}} ~~,\\
 I_{2}&=  \frac{\sin(a)\sqrt{(\cos(a)-\cos(b))\sin(a)}}{\pi \lambda} \epsilon^{\frac{1}{2}} ~ \\
 &\int_{0}^{a-\gamma}d\theta \frac{1}{(\cos(x)-\cos(\alpha))\sqrt{\left(\sin^2(\frac{a}{2})
 -\sin^2(\frac{\theta}{2})\right)\left(\sin^2(\frac{b}{2})-\sin^2(\frac{\theta}{2})\right)}} .           
  \end{split}
 \end{equation}
As in the previous section we have divided the integral in \eqref{realrho} (which can be taken to run over the range $(0,a)$ 
into an integral from $(0, a-\gamma)$ and an integral from $(a-\gamma,  a)$). We take $\gamma=M\epsilon$ where $M$ is a large 
but number that is held fixed in the limit $\alpha \to a$ i.e. $\epsilon \to 0$.

Inside the integral in $I_1$, it is legitimate to Taylor expand $\alpha$ and the integration variable $\theta$ about $a$. 
The contributions
of successive terms in this Taylor expansion are suppressed compared to leading terms by increasing powers of $\epsilon$. It is not 
difficult to verify that 
   \begin{equation}\label{io} \begin{split}
   I_1&=\left( \frac{2}{ \lambda} +  h_1(M) \right)+ \epsilon\,\, h_2(M) + {\cal O}(\epsilon^\frac{3}{2})\\
   h_1(M)&=-\frac{4}{\pi\lambda\sqrt{M}} + O(\frac{1}{M^{\frac{3}{2}}})\\
    h_2(M)&=\frac{\sqrt{M} }{2\pi\lambda (\cos a-\cos b)\sin a}\left(1+5\cos (2a)-6\cos a \cos b\right).
   \end{split}
    \end{equation}
The expression for $I_2$ may be processed as follows. Of course $I_2=(I_2-I_3)+I_3$,  
where we choose $I_3$ as
 \begin{equation}
  I_3= \frac{\sin(a)\sqrt{(\cos(a)-\cos(b))\sin(a)}}{\pi \lambda} \epsilon^{\frac{1}{2}} \int_{0}^{(a-\gamma)}\frac{4}{\sqrt{\cos(a)-\cos(b)}\sqrt{(\cos(a)-\cos(x))^{3}}}dx.
 \end{equation}
The point of this manipulation is as follows. In the expression for $I_2-I_3$ we first subtract the 
integrands before performing the integral.
In order to proceed further we proceed as follows.
\begin{equation}
\begin{split}
 & I_2-I_3\\&=  \sqrt{\epsilon} \frac{\sin(a)\sqrt{(\cos(a)-\cos(b))\sin(a)}}{\pi \lambda}  ~ \\
   &\int_{0}^{a-\gamma}\left(-\frac{4}{\sqrt{\cos(a)-\cos(b)}}+\frac{4}{\sqrt{\cos(x)-\cos(b)}}\right)\frac{1}{\sqrt{(\cos(a)-\cos(x))^{3}}}dx\\
 &= \sqrt{\epsilon} A+  2\epsilon \sqrt{M}\frac{\sin a}{\pi\lambda (\cos a-\cos b)}+O[\epsilon^{\frac{3}{2}}],
\end{split}
\end{equation}
where 
\begin{equation}
 \begin{split}\label{AA}
  &A= \frac{\sin(a)\sqrt{(\cos(a)-\cos(b))\sin(a)}}{\pi \lambda}  ~ \\
   &\int_{0}^{a}\left(-\frac{4}{\sqrt{\cos(a)-\cos(b)}}+\frac{4}{\sqrt{\cos(x)-\cos(b)}}\right)\frac{1}{\sqrt{(\cos(a)-\cos(x))^{3}}}dx.\\
 \end{split}
\end{equation}

The $I_3$ integral can be performed easily and is given by
\begin{equation}
 \begin{split}
I_{3}&=-h_{1}(M)+h_3(M)-\frac{4\sqrt{\epsilon} E\left(a\left|\csc ^2\left(\frac{a}{2}\right)\right.\right)}{\pi\lambda\sqrt{\cot\left(\frac{a}{2}\right)}}+O[\epsilon^{\frac{3}{2}}]\\  
& h_3(M)= -\epsilon \sqrt{M}\frac{3 \cot a}{\pi\lambda } +O[\frac{1}{M^{\frac{3}{2}}}],
\end{split}
\end{equation}
where symbol $E$ refers to elliptic function
 of second kind.
Adding all up, we finally obtain eigenvalue distribution  as $\alpha \to a$ to be
\begin{equation}
\begin{split}
\rho(\alpha)&=\frac{1}{2\pi\lambda}+\frac{1}{4\pi} D \sqrt{\alpha -a}+ {\cal O}(\alpha-a)^\frac{3}{2} \qquad 
\\
&\text{with} \qquad
D= A-\frac{4 E\left(a\left|\csc ^2\left(\frac{a}{2}\right)\right.\right)}{\pi\lambda\sqrt{\cot\left(\frac{a}{2}\right)}}.
\end{split}
\label{ro_near_a}
\end{equation}
We notice once again that near $a,$ the eigenvalue
approaches $\frac{1}{2\pi\lambda}$ as $\epsilon^{\frac{1}{2}}.$
As a check of our result, we take $b=\pi.$ In this case result should reduce to that obtained from one cut solution.
In this limit the $A$ in \eqref{AA} can be easily computed and is given by
\begin{equation}\begin{split}
 A&=\frac{4 E\left(a\left|\csc ^2\left(\frac{a}{2}\right)\right.\right)}{\pi\lambda\sqrt{\cot\left(\frac{a}{2}\right)}}-\frac{(\sin a)^{\frac{3}{2}}}{\sqrt{2}\lambda\cos^{2}\frac{a}{2}}\\
&=\frac{4 E\left(a\left|\csc ^2\left(\frac{a}{2}\right)\right.\right)}{\pi\lambda\sqrt{\cot\left(\frac{a}{2}\right)}}-4 \zeta\sin\frac{a}{2}\sqrt{\sin\frac{a}{2}\cos\frac{a}{2}},
\end{split}
\end{equation}
where in the last line we have used \eqref{bpisol}.
So finally we obtain 
as $\alpha \to a$ 
$$\rho(\alpha)=\frac{1}{2\pi\lambda}-\frac{\zeta}{\pi}\sin\frac{a}{2}\sqrt{\sin\frac{a}{2}\cos\frac{a}{2}}+ {\cal O}(\alpha-a)^\frac{3}{2}$$
 which is in perfect agreement with \eqref{bpilim} as we take $\alpha \to a.$

\subsection{Level rank duality of the solution to the capped GWW model}
\label{gwwlrd}

In this subsection we will directly verify that the 
explicit solution of the capped GWW model obeys \eqref{lrdp}. 
We will find the following notation useful. We denote the `no gap' saddle point 
eigenvalue distribution of the capped GWW model by 
$\rho_{ng}(\lambda, \zeta, \alpha)$. In a similar manner we denote 
the one lower gap, one upper gap and two gap eigenvalue distributions by 
$\rho_{lg}(\lambda, \zeta, \alpha)$, 
$\rho_{ug}(\lambda, \zeta, \alpha)$  and 
$\rho_{tg}(\lambda, \zeta, \alpha)$ respectively.

\paragraph{ No Gap Solutions:}
In the no gap phase 
\begin{equation}\label{ngsn}
\rho_{ng}(\lambda, \zeta, \alpha)= \frac{1+\zeta \cos \alpha}{2\pi}.
\end{equation}
Using \eqref{ngsn} it is easy to directly  verify that 
\begin{equation}\label{lrdpng}
\rho_{ng} \left( (1-\lambda), \frac{\zeta \lambda}{1-\lambda}, \alpha \right)
=\frac{\lambda}{1-\lambda} \left(\frac{1}{2 \pi \lambda} -
\rho_{ng} \left( \lambda, \zeta, \alpha+\pi \right) \right)
\end{equation}
so that level rank duality maps the no gap phase to itself.
\paragraph{One gap solutions:}
In the one lower gap phase (see \eqref{evdl}) 
\begin{equation}\label{evdln} \begin{split}
\rho_{lg}(\lambda, \zeta, \alpha)& 
= \frac{\zeta \cos \left( \frac{\alpha}{2} \right)}{\pi} 
\sqrt{\frac{1}{\zeta}-\sin^2\frac{\alpha}{2}} ~~~{\rm for}~~ 
\sin^2\frac{\alpha}{2} < \frac{1}{\zeta}\\
& = 0 ~~~~~~~~~~~~~~~~~~~~~~~~~~~~~~~ {\rm for}~~ 
\sin^2\frac{\alpha}{2}> \frac{1}{\zeta}.
\end{split}
\end{equation}
On the other hand in the upper gap phase (see \eqref{evdd}) 
\begin{equation}\label{evddn} \begin{split}
\rho_{ug}(\lambda, \zeta,\alpha)& = \frac{1}{2 \pi \lambda}- 
\zeta\frac{ \vline\, \sin \left( \frac{\alpha}{2} \right) \,\vline}{\pi}  
\sqrt{\frac{\frac{1}{\lambda}-1}{\zeta}-\cos^2\frac{\alpha}{2} }
~~~{\rm for }~~\cos^2 \frac{\alpha}{2} <\frac{1-\lambda }{\lambda \zeta}\\
&=\frac{1}{2 \pi \lambda} ~~~~~~~~~~~~~~~~~~~~~~~~~~~~~~~~~~~~~~~~~~
{\rm for }~~\cos^2 \frac{\alpha}{2} 
>\frac{1-\lambda}{\lambda \zeta}.\\
\end{split}
\end{equation}
Using \eqref{evddn} and \eqref{evdln} it is straightforward
to directly verify that 
\begin{equation}\label{lrdppog} \begin{split}
\rho_{lg} \left( (1-\lambda), \frac{\zeta \lambda}{1-\lambda}, \alpha \right)
&=\frac{\lambda}{1-\lambda} \left(\frac{1}{2 \pi \lambda} -
\rho_{ug} \left( \lambda, \zeta, \alpha+\pi \right) \right)\\
\rho_{ug} \left( (1-\lambda), \frac{\zeta \lambda}{1-\lambda}, \alpha \right)
&=\frac{\lambda}{1-\lambda} \left(\frac{1}{2 \pi \lambda} -
\rho_{lg} \left( \lambda, \zeta, \alpha+\pi \right) \right).\\
\end{split}
\end{equation}
It follows that level rank duality exchanges the one lower gap and 
one upper gap solutions. 
\paragraph{ Two gap solution:}
The two gap solution is listed in \eqref{realrhom} with the end points 
$a$ and $b$ given in \eqref{detabm}. We believe it is true that  
\begin{equation}\label{lrdptg}
\rho_{tg} \left( (1-\lambda), \frac{\zeta \lambda}{1-\lambda}, \alpha \right)
=\frac{\lambda}{1-\lambda} \left(\frac{1}{2 \pi \lambda} -
\rho_{tg} \left( \lambda, \zeta, \alpha+\pi \right) \right),
\end{equation}
so that level rank duality maps the two gap solution to itself. 

We have not succeeded in directly verifying \eqref{lrdptg} starting 
with \eqref{realrhom}. However we have generated some numerical 
evidence that \eqref{realrhom} obeys \eqref{lrdptg}, as we now proceed 
to describe. 

$a$ and $b$, the end points of the gaps in the two cut solution, are both
functions of $\lambda$ and $\zeta$. As we have explained in Appendix 
\ref{Behave-a-b} it is formally useful to invert this dependence and 
regard $\lambda$ and $\zeta$ as a function of $a$ and $b$. In particular 
the dependence of $\lambda(a,b)$ on $a$ and $b$ is listed in the first of 
\eqref{detab-2}. Level rank duality interchanges upper and lower gaps 
and so requires that 
\begin{equation}\label{lab}
\lambda(\pi-b,\pi -a)= 1-\lambda(a,b).
\end{equation}
Using the explicit integral representation for $\lambda(a,b),$ 
\eqref{detab-2} we have numerically verified \eqref{lab} to high precision
for several randomly selected values of $a$ and $b$. 

We have also generated numerical evidence for the self duality of the 
two gap eigenvalue distributions themselves in the following manner. 
Our solution for the eigenvalue distribution \eqref{realrhom} takes 
the following structural form 
\begin{equation}\label{evdstr}
2 \pi \lambda \rho_{tg}(\lambda, \zeta, \alpha)
=\int_{-a}^a d\theta \upsilon(a, b, \alpha, \theta),
\end{equation}
where the explicit form of the function $\upsilon$ may be read off from 
\eqref{realrhom}. On the RHS of \eqref{evdstr} $a$ and $b$ are the 
functions of $\lambda$ and $\zeta$ determined by \eqref{detabm}. 
As we have explained above, however, for the purposes of the current section 
it is more convenient to invert this dependence and regard $\lambda$ and 
$\zeta$ as functions of $a$ and $b$. Let us define 
$$H(a, b, \alpha)=2 \pi \lambda(a, b) 
\rho_{tg}(\lambda(a,b) , \zeta (a, b), \alpha).$$
\eqref{evdstr} may be rewritten as 
\begin{equation}\label{evdstrn}
H(a,b, \alpha)
=\int_{-a}^a d\theta \upsilon(a, b, \alpha, \theta).
\end{equation}
Now the prediction \eqref{lrdptg} may be rewritten as 
\begin{equation}\label{hph}
H(a, b, \alpha)+ H(\pi-b, \pi-a, \alpha-\pi)=1.
\end{equation}
Using the explicit integral representation for
\eqref{evdstrn} we have numerically verified \eqref{hph} to high 
precision for several randomly selected values of $a$, $b$ and $\alpha$. 

While the numerical evidence for \eqref{lab} and \eqref{hph}
is impressive, it would certainly be useful to find an analytic proof of 
these equations. We leave this for future work. 
\footnote{
The analytic proof is given in \cite{Takimi:2013zca}
which is a sequel of this paper.}
\paragraph{ Level Rank duality of the phase transition points:}
As we have explained above, for $\lambda<\frac{1}{2}$, 
capped GWW model undergoes two phase transitions at 
$$\zeta^s_l(\lambda)=1,$$ for no gap to lower gap and 
$$\zeta^s_h(\lambda)=\frac{1}{4 \lambda^2},$$ 
for lower gap to two gap respectively. 
On the other hand, for $\lambda>\frac{1}{2}$, the 
two phase transitions occur at 
$$\zeta^b_l(\lambda)=\frac{1-\lambda}{\lambda},$$
for no gap to upper gap and 
$$\zeta^b_h(\lambda)=\frac{1}{4 \lambda(1-\lambda)},$$ 
for upper gap to two gap respectively.
Under level 
rank duality a value of $\lambda<\frac{1}{2}$ maps to a value of $\lambda$
greater than $\frac{1}{2}$. Level rank duality requires that 
\begin{equation}\label{ptlr}
\begin{split}
\lambda \zeta^s_l(\lambda)&=(1-\lambda)\zeta^b_l(1-\lambda),\\
\lambda \zeta^s_h(\lambda)&=(1-\lambda)\zeta^b_h(1-\lambda).\\
\end{split}
\end{equation}
\eqref{ptlr} is very easily verified using the explicit expression 
for $\zeta^s_l$, $\zeta^b_l$, $\zeta^s_h$, and $\zeta^b_h$ presented above.

\subsection{Behavior of $(\lambda,\zeta)$ with respect to $(a,b)$ in the one lower gap and one upper gap solution}\label{Behave-a-b}
\begin{figure}
  \begin{center}
  \subfigure[]{\includegraphics[scale=.5]{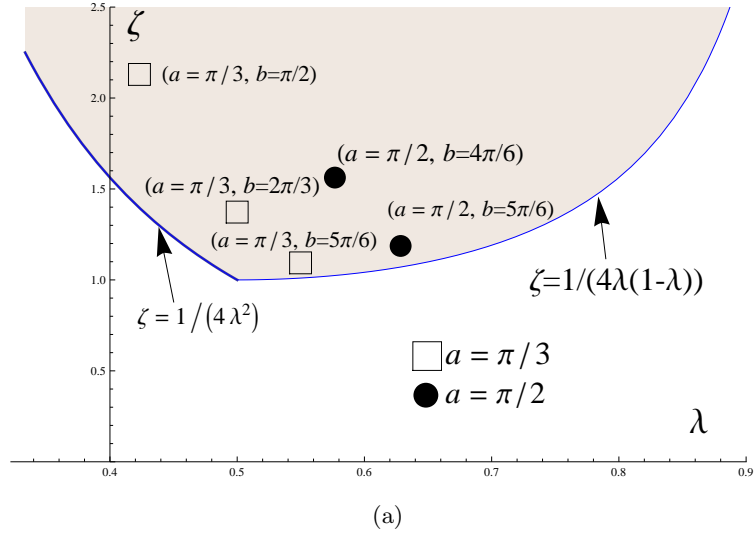}
\label{numplot}
  }
  \end{center}
  \vspace{-0.5cm}
  \caption{Plotting $(\lambda(a,b),\zeta(a,b))$
with the values of $(a,b) = 
(\frac{\pi}{3},\frac{\pi}{2}),
(\frac{\pi}{3},\frac{2\pi}{3}),
(\frac{\pi}{3},\frac{5\pi}{6}),
(\frac{\pi}{2},\frac{2\pi}{3}) \text{and} 
(\frac{\pi}{2},\frac{5\pi}{6})$.
From this graph, we can guess that $\lambda,\zeta$ are increasing functions 
of $a$, and we can also see that
$\lambda$ is surely a increasing function of $b$ and $\zeta$ 
is a decreasing function with respect to $b$. 
The region $\zeta > \frac{1}{4\lambda^2}, \lambda<\frac{1}{2}$ and $\zeta > \frac{1}{4\lambda(1-\lambda)},\lambda 
\ge\frac{1}{2}$ is shaded. 
All points are located on the inside of the shaded region.}
\end{figure}

The set of equations \eqref{detab} is equivalent to the following set
\begin{equation}\label{detab-2}
\begin{split}
&\lambda(a,b) = 
\frac{1}{4\pi } \int_{-a}^a d \alpha 
\left(
G(a,b,\alpha)
+ \frac{1}{G(a,b,\alpha)}
\right), \\
&\lambda(a,b)\zeta(a,b)
= \frac{1}{4\pi } \int_{-a}^a d \alpha  \frac{1}
{\sqrt{\sin^2\frac{a}{2}-\sin^2 \frac{\alpha}{2}}
\sqrt{\sin^2\frac{b}{2}-\sin^2 \frac{\alpha}{2}}}
\end{split}
\end{equation}
where
\begin{equation}
G(a,b,\alpha) \equiv
\frac
{\sqrt{\sin^2\frac{b}{2}-\sin^2 \frac{\alpha}{2}}}
{\sqrt{\sin^2\frac{a}{2}-\sin^2 \frac{\alpha}{2}}}.
\end{equation}
One can confirm it by subtracting $\frac{1}{2}(\cos a + \cos b)$ times 
first line from the second line in \eqref{detab}.
First let us observe the behavior of $\lambda$ with respect to the 
change of $b$ with fixed $a = a_{o}$.
In the case $\pi \ge b > b' \ge a_{o}  \ge 0$,
since $
1 \ge \sin^2 \frac{b}{2}
> \sin^2 \frac{b'}{2}
\ge \sin^2 \frac{a_0}{2} \ge 0,$
it becomes
\begin{equation}
G(a_o,b,\alpha) > 
G(a_o,b',\alpha) \ge 1 .
\end{equation}
Therefore 
\begin{equation}
 \left(G(a_o,b,\alpha)
+\frac{1}{G(a_o,b,\alpha)}\right)
>  \left(G(a_o,b',\alpha)
+\frac{1}{G(a_o,b',\alpha)} \right) \ge 2.
\end{equation}
Hence in the fixed $a_o$, $\lambda(a_o,b)$ is a increasing function
with respect to $b$, i.e,
\begin{equation}
\lambda(a_o,b) \ge \lambda(a_o,b') \quad \text{at} \quad
\pi \ge b \ge b' \ge a_{o}  \ge 0.
\end{equation}
The range of $\lambda$ in the fixed $a = a_o$ is 
\begin{equation}
\frac{a_o}{\pi} \le \lambda(a_o,b) \le 1- \frac{1}{2}\cos \frac{a_o}{2}
\end{equation}
where 
\begin{equation}
\lambda(a_o,a_o) = \frac{a_o}{\pi}, 
\quad
\lambda(a_o,\pi) = 
1- \frac{1}{2}\cos \frac{a_o}{2}.
\end{equation}
From the factor $1/\sqrt{\sin^2\frac{b}{2}-\sin^2 \frac{\alpha}{2}}$
in the second equation of \eqref{detab-2}, 
we can immediately see that 
$\lambda\zeta(a_o,b)$ is the decreasing function with respect to
$b$.
Moreover, since the $\lambda(a_o,b)$ is increasing function, $\zeta(a_o,b)$ 
becomes decreasing function.
The range of $\zeta(a_o,b)$ is
\begin{equation}
\frac{1}{2\left(1-\frac{1}{2}\cos \frac{a_o}{2}\right)\cos \frac{a_o}{2}}
\le \zeta(a_o,b) \le \infty ,
\end{equation}
where
\begin{equation}
\zeta(a_o, \pi)
= \frac{1}{2\left(1-\frac{1}{2}\cos \frac{a_o}{2}\right)\cos \frac{a_o}{2}}
, \quad
\zeta(a_o, a_o) = \infty.
\end{equation}
We can also see that
\begin{equation}
\zeta(a_o,b) \ge \zeta(a_o,\pi) \ge \min \left(\frac{1}{2\left(1-\frac{1}{2}\cos \frac{a_o}{2}\right)\cos \frac{a_o}{2}}\right)= 1.
\end{equation}
Hence $\zeta$ must be larger than 1, and we have exactly shown that
$\zeta$ has a unique minimum at $a =0, b= \pi$.
As a summary, $\lambda$ is a increasing function of $b$ while $\zeta$ is a
 decreasing function of $b$.
We can also guess the behavior of $\lambda$ and $\zeta$ with respect to $a$,
with fixed $b =b_o$ by a numerical calculation.
(See Fig.\ref{numplot}).
From this, we can guess that $\lambda$ is increasing function of 
$a$ and the also $\zeta$ is increasing function of $a$.
\begin{figure}
  \begin{center}
  \subfigure[]{\includegraphics[scale=.5]{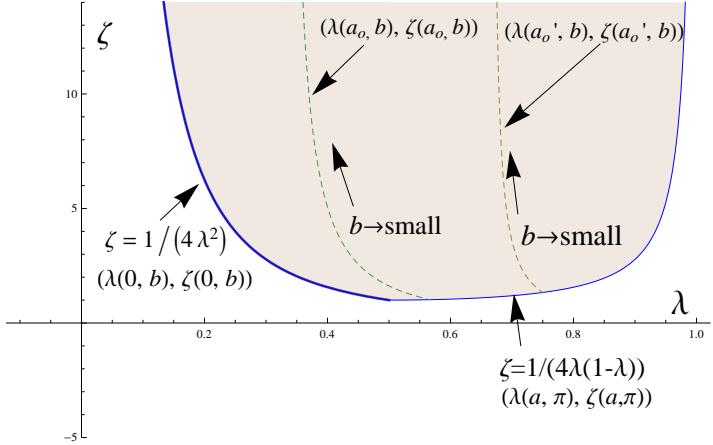}
\label{a-line}
  }
  \caption{
The schematically drawn two lines $(\lambda(a_o,b),\zeta(a_o,b))$ and $(\lambda(a_o',b),\zeta(a_o',b))$ with fixed $a_o$ and $a_o'$ with $a_o < a_o'$.
They are drawn by dashed line. For each line, $\zeta$ becomes a decreasing function of $\lambda$.}
  \qquad\qquad
\end{center}
\end{figure}
With fixed $a$, let us consider the one parameter line
$(\lambda(a_o,b), \zeta(a_o,b))$ with parameter $b$ in the 
$\lambda-\zeta$ plane.
(See Fig.\ref{a-line}).
Since the $\lambda(a_o,b)$ is increasing function with respect to $b$
while $\zeta(a_o,b)$ is the decreasing function, 
$\zeta = \zeta(\lambda)$ behaves as the
decreasing function with respect to $\lambda$.
The $\zeta$ is divergent at
\begin{equation}
\zeta(a_o,b) \to \infty \quad \text{at} \quad
\lambda(a_o,b) \to \frac{a_o}{\pi} \qquad (b \to a_o).
\end{equation}
In the fixed $b = b_o$ case, 
we can also consider the one parameter line 
$(\lambda(a,b_o), \zeta(a,b_o))$ with parameter $a$ in the 
$\lambda-\zeta$ plane.
In the case of fixed $b = b_o = \pi$,
the one parameter line
$(\lambda(a,\pi), \zeta(a,\pi))$ is represented as 
\begin{equation}
\zeta = \frac{1}{4 \lambda (1- \lambda)} \qquad \frac{1}{2} < \lambda 
\le 1.
\end{equation}
On the other hand, 
in the case of fixed $a = a_o =0$, the one parameter line
$(\lambda(0,b), \zeta(0,b))$ is represented as the
\begin{equation}
\zeta = \frac{1}{4 \lambda^2} \qquad 0 < \lambda \le \frac{1}{2}.
\end{equation}
Next let us consider the 
line $(\lambda(a_o,b), \zeta(a_o,b))$ with fixed $a_o \ne 0$.
In this case, due to  
$\zeta(a_o,\pi) = \frac{1}{4\lambda(a_o,\pi)(1-\lambda(a_o,\pi))}$
and $\frac{1}{2} < \lambda(a_o,b) < \lambda(a_o, \pi)$,
we can see that
\begin{eqnarray}
\zeta(a_o,b) > \zeta(a_o,\pi) 
=  \frac{1}{4\lambda(a_o,\pi)(1-\lambda(a_o,\pi))} > 
\frac{1}{4\lambda(a_o,b)(1-\lambda(a_o,b))} >1.
\end{eqnarray}
for $b < \pi$.	
Then the lines 
$(\lambda(a_o,b), \zeta(a_o,b))$
are located on the inside 
of the region $\zeta \ge \frac{1}{4 \lambda(1-\lambda)}$ when
$\lambda(a_o,b) \ge 1/2$.
Also for the case $\lambda(a_o,b) \le 1/2$,
we can show that the lines $(\lambda(a_o,b), \zeta(a_o,b))$
are located on the inside of the region $\zeta \ge \frac{1}{4 \lambda^2}$. 
To show it, we should note that the value of $(a,b)$ is uniquely determined
if we specify the 2 parameter
$(\lambda, \zeta)$.
This means that there are not any intersection between two lines,
$(\lambda(a_o,b),\zeta(a_o,b))$ and $(\lambda(a_o',b),\zeta(a_o',b))$ 
with fixed $a_o \ne a_o'$. In particular,
there is no intersection of line $(\lambda(a_o,b),\zeta(a_o,b))$ and 
the $(\lambda(0,b),\zeta(0,b))$, which is written as $\zeta 
= \frac{1}{4 \lambda^2}$.
For the lines $(\lambda(a_o,b),\zeta(a_o,b))$ with fixed $0 < a_o \le 
\frac{\pi}{2}$, it becomes
\begin{equation}
\zeta(a_o,b) \gvertneqq 1 = \frac{1}{4 \lambda(a_o,b)^2}
\quad \text{at} \quad  \lambda(a_o,b) = \frac{1}{2}.
\end{equation}
So for $\lambda(a_o,b) < \frac{1}{2}$,
the line $(\lambda(a_o,b),\zeta(a_o,b))$ must be located 
inside the $\zeta > \frac{1}{4\lambda^2}$ 
since the line can not have the intersection with the line
$(\lambda(0,b),\zeta(0,b))$.
\footnote{For the line $(\lambda(a_o,b),\zeta(a_o,b))$ with $a_o \ge 
\frac{\pi}{2}$ we do not have to worry about the region with $\lambda 
< \frac{1}{2}$. Because the range of the $\lambda$ of the line is
$1/2 < a_o/\pi < \lambda < 1$.}
This shows that two gap solution exists if and only if
$\zeta \ge \frac{1}{4\lambda^2}$ and $\zeta \ge \frac{1}{4\lambda(1-\lambda)}$.
Our numerical result also shows it.
(See Fig.\ref{numplot})

\section{Level-rank duality of the saddle point equation in the 
multi trace potential}
\label{multitrace}

We consider the multi-trace potential,
\begin{equation}
V = 
\sum_{m} \sum_{n} A_{m,n}(\tr U^n)^{m} +c.c
= 
\sum_{m} \sum_{n} A_{m,n}\left(\sum_{i=1}^{N} e^{in\alpha_i}\right)^{m} +c.c.
\end{equation}
The corresponding potential in the dual theory is obtained by
$\tr U^n \leftrightarrow (-1)^{n+1}\tr U^n$, 
\begin{equation}
\tilde{V} = 
\sum_{m} \sum_{n} A_{m,n}(\tr (-1)^{n+1}U^n)^{m} +c.c
= 
\sum_{m} \sum_{n} A_{m,n}\left(\sum_{i=1}^{k-N} -e^{in(\alpha_i + \pi)}
\right)^{m} +c.c.
\end{equation}
Derivative of the potentials 
with respect to an eigenvalue $\alpha_{l}$ are
\begin{eqnarray}
V'_{l} &\equiv& \frac{d}{d \alpha_l}V =
\sum_{m} \sum_{n} A_{m,n}
\left(\sum_{i=1}^{N} e^{in\alpha_i}\right)^{m-1} 
\left(im n e^{in \alpha_l}\right)
\label{Vprime}
+c.c,
\\
\tilde{V}'_{l} &\equiv& \frac{d}{d \alpha_l}\tilde{V} =
\sum_{m} \sum_{n} A_{m,n}
\left(\sum_{i=1}^{k-N} -e^{in(\alpha_i+ \pi)}\right)^{m-1} 
\left(-im n e^{in (\alpha_l+\pi)}\right)
+c.c.
\label{Vtprime}
\end{eqnarray}
In \eqref{Vprime}, 
please note that only the factor $\left(im n e^{in \alpha_l}\right)$ 
carries the label $l$ dependence of the $V'_{l}$ 
(there is no longer label $l$ dependence in the 
$\left(\sum_{i=1}^{N} e^{in\alpha_i}\right)^{m-1}$ because 
it is the sum over the label).
Similarly, in \eqref{Vtprime}, 
the label $l$ dependence in $\tilde{V}'_{l}$ 
is carried only by the factor
$\left(-im n e^{in (\alpha_l+\pi)}\right)$.
Saddle point equations are obtained for each eigenvalue $\alpha_{l}$
by taking the derivative as, 
\begin{eqnarray}
0 &=& \frac{d}{d \alpha_l}\left(
V - \sum_{i \ne j} \ln \left(
2 \sin \frac{\alpha_{i} - \alpha_{j}}{2}
\right)
\right)
=
V'_{l} - \sum_{i, i\ne l} \cot\left( 
\frac{\alpha_l - \alpha_i}{2}
\right),
\\ 
0 &=& \frac{d}{d \alpha_l}\left(
\tilde{V} - \sum_{i \ne j} \ln \left(
2 \sin \frac{\alpha_{i} - \alpha_{i}}{2}
\right)
\right)
=
\tilde{V}'_{l} - \sum_{i, i\ne l} \cot\left( 
\frac{\alpha_l - \alpha_i}{2}
\right).
\end{eqnarray}

Now let us consider the large $N$ limit in the original theory.
When we take the large $N$ limit, 
we first fix the residual 
Weyl permutation, exchanging the order of the eigenvalues.
After fixing the Weyl permutation, the order of the eigenvalues obey following
\begin{equation}
0 \le \alpha_1 < \alpha_2 < \ldots < \alpha_{N} \le 2 \pi.
\end{equation}
Then there is a one to one relationship between 
the value of the eigenvalue $\alpha_l$ and the label $l$ such that
\begin{equation}
i < l \Leftrightarrow \alpha_i < \alpha_l . 
\label{incresing}
\end{equation}
In the large $N$ limit, these labels can be regarded as the continuum variable
$x = \frac{l}{N}, (l = 1,\ldots N)$.
Moreover due to the \eqref{incresing}, 
$\beta(x) \equiv \alpha_l, (x = \frac{l}{N})$
becomes the increasing continuum function with respect to 
$x$, so 
\begin{equation}
l < i \Leftrightarrow  \frac{l}{N} = x < x' = \frac{i}{N}
\Leftrightarrow \alpha_l = \beta(x) < \beta(x') = \alpha_i. 
\label{increasing-2}
\end{equation}
And from these facts, 
the trace is written by the integration over the 
$\beta(x)$ as,
\begin{equation}
\sum_{i = 1}^{N} \to 
N\int^{1}_{0} dx 
=N\int^{2\pi}_{0} \frac{dx}{d\beta} d\beta
=N\int^{2\pi}_{0} d\beta \rho(\beta) .
\end{equation}
From \eqref{increasing-2}, we would like to emphasize that
{\it value of eigenvalue variable $\beta$ 
represents the label $l$ of the eigenvalue $\alpha_{l} = \beta$ 
in the discretized
representation.}
According to these, 
$V'(\beta_0)$ is just a large $N$ representation of the 
$V'_{l}$ with $\beta_0 \equiv \beta(\frac{l}{N})  = \alpha_l$,
\begin{eqnarray}
V'_{l} &=& 
\sum_{m} \sum_{n} A_{m,n}
\left(\sum_{i=1}^{N} e^{in\alpha_i}\right)^{m-1} 
\left(im n e^{in \alpha_l}\right)
+c.c 
\nonumber \\
\to V'(\beta_0) &=&
\sum_{m} \sum_{n} A_{m,n}
\left(N \int^{2\pi}_0 d \beta \rho(\beta) e^{in\beta}\right)^{m-1} 
\left(im n e^{in \beta_0}\right) +c.c.
\end{eqnarray}
So, 
with keeping in mind that the variable $\beta$ represents the label of 
eigenvalue, 
$V'(\beta_0 + \pi)$ becomes the $V'_{l'}$ with the label $l'$ 
such that $\alpha_{l'} = \alpha_{l} +\pi = \beta_0 + \pi$,
\begin{eqnarray}
V'(\beta_0+\pi) = V'_{l'}&=&
\sum_{m} \sum_{n} A_{m,n}
\left(\sum_{k=1}^{N} e^{in\alpha_k}\right)^{m-1} 
\left(im n e^{in \alpha_{l'}}\right) +c.c
\nonumber \\
&=&
\sum_{m} \sum_{n} A_{m,n}
\left(N \int^{2\pi}_0 d \beta \rho(\beta) e^{in\beta}\right)^{m-1} 
\left(im n e^{in (\beta_0 + \pi)}\right) +c.c.
\nonumber \\
\label{V'}
\end{eqnarray}
This is also the case in the dual theory. 
At the large $k-N$ limit, 
$\tilde{V}'(\beta_0)$ is
\begin{eqnarray}
\tilde{V}'(\beta_0) &=&
\sum_{m} \sum_{n} A_{m,n}
\left((-1)^{n+1}\sum_{i=1}^{k-N} e^{in(\alpha_i +\pi )}\right)^{m-1} 
\left(-im n e^{in \alpha_l}\right) +c.c
\nonumber \\
&=&
\sum_{m} \sum_{n} A_{m,n}
\left(-(k-N) \int^{2\pi}_0 d \beta \tilde{\rho}(\beta) 
e^{in(\beta+\pi)}\right)^{m-1} 
\left(-im n e^{in (\beta_0 + \pi)}\right) +c.c,
\nonumber\\
\label{tildeV'}
\end{eqnarray}
with $\beta_0 = \alpha_l$.

Now let us return to the large $N$ limit of the saddle point equation and 
the discussion of the level-rank duality.
The large $N$ (or large $k-N$) representation of the saddle point equations are
\begin{eqnarray}
V'_{l} &=& \sum_{i, i\ne l} 
\cot\left( 
\frac{\alpha_l - \alpha_i}{2}
\right)
\to V'(\beta_0) = N {\cal P} \int  d\beta \rho(\beta)
\cot\left( 
\frac{\beta_0 - \beta}{2}
\right) ,
\label{saddle-multi}
\\
\tilde{V}'_{l} &=& \sum_{i, i\ne l} 
\cot\left( 
\frac{\alpha_l - \alpha_i}{2}
\right)
\to \tilde{V}'(\beta_0) = (k-N) {\cal P} \int  d\beta \tilde{\rho}(\beta)
\cot\left( 
\frac{\beta_0 - \beta}{2}
\right) .
\label{saddle-multi2}
\end{eqnarray}
As we did 
from \eqref{vtd} to \eqref{vtppp} in the subsection \ref{lrdln}, 
assuming that $\rho(\beta)$ is the solution of the 
\eqref{saddle-multi}, following $\tilde{\rho}(\beta)$ 
\begin{equation}
\tilde{\rho}(\beta) = \frac{\lambda}{1-\lambda}\left(
\frac{1}{2\pi\lambda} - \rho(\beta + \pi)
\right) \label{eigen}
\end{equation}
is a solution of the \eqref{saddle-multi2} for the dual theory 
if 
\begin{equation}
\tilde{V}'(\beta_0)
= -V'(\beta_0+\pi) \label{task}
\end{equation}
is satisfied with 
\eqref{eigen}. 
Then we only have to show that \eqref{task} is satisfied
with \eqref{eigen}. 
We can check the statement \eqref{task} directly.
From \eqref{tildeV'}, by substituting 
\eqref{eigen}, we can check
\begin{eqnarray}
\tilde{V}'(\beta)
&=&
\sum_{m} \sum_{n} A_{m,n}
\left(-(k-N) \int^{2\pi}_0 d \beta \tilde{\rho}(\beta) 
e^{in(\beta+\pi)}\right)^{m-1} 
\left(-im n e^{in (\beta_0 + \pi)}\right) +c.c.
\nonumber \\
&=&
\sum_{m} \sum_{n} A_{m,n}
\left(N \int^{2\pi}_0 d \beta \rho(\beta + \pi) e^{in(\beta+\pi)}\right)^{m-1} 
\left(-im n e^{in (\beta_0 + \pi)}\right) +c.c. 
\nonumber \\
&=&
-\sum_{m} \sum_{n} A_{m,n}
\left(N \int^{\pi}_{-\pi} d \beta \rho(\beta) e^{in\beta}\right)^{m-1} 
\left(im n e^{in (\beta_0 + \pi)}\right) +c.c. \label{tildeV2}
\end{eqnarray}
Comparing \eqref{tildeV2} with the \eqref{V'},
we can see \eqref{task} is satisfied with \eqref{eigen}.

Then this concludes the proof of the level-rank duality of the saddle point equation in the 
multi-trace potential terms.

\section{High temperature limit of the partition function of 
a gas of non renormalized multitrace operators}\label{mtrace}
In this appendix we compute the partition function of a gas of non 
renormalized multitrace operators, in the high temperature limit, 
for the theory of regular bosons, regular fermions and the supersymmetric 
theory with a single chiral multiplet in the fundamental representation.

The single letter partition function for bosonic theory is given by
\begin{equation}
 f_{b}(x)=x^{\frac{1}{2}}\frac{1+x}{(1-x)^2},
\end{equation} where $x=e^{-\frac{1}{T}}.$ In the large temperature limit we obtain
\begin{equation}
 x=1-\frac{1}{T}+{\cal O}\left(\frac{1}{T^2}\right),~~f_{b}(x)=2 T^2+{\cal O}(T^0).
\end{equation}
The multi-trace partition function is given by
\begin{equation}
 \ln Z=\sum_{n=1}^{\infty}\frac{1}{n}f_{b}(x^{n})^2
\end{equation} which in the large temperature limit is given by 
\begin{equation}\label{freelogzb}
 \ln Z=4~T^4\sum_{n=1}^{\infty}\frac{1}{n^5}+{\cal O}(T^2)=4~T^4\zeta(5)+{\cal O}(T^2).
\end{equation} $\ln Z$ in \eqref{freelogzb} gives the value of partition function
 for the gas of multi-trace operators of free bosons on the sphere of volume $V_2=4\pi.$
 
For the fermionic theory, the single letter partition function is given by
\begin{equation}
 f_{f}(x)=\frac{2 x}{(1-x)^2},
\end{equation} where $x=e^{-\frac{1}{T}}.$ In the large temperature limit we obtain
\begin{equation}
 f_{f}(x)=2 T^2+{\cal O}(T^0).
\end{equation}
The multi-trace partition function is given by
\begin{equation}
 \ln Z=\sum_{n=1}^{\infty}\frac{1}{n}f_{f}(x^{n})^2
\end{equation} which in the large temperature limit is given by 
\begin{equation}\label{freelogzf}
 \ln Z=4~T^4\sum_{n=1}^{\infty}\frac{1}{n^5}+{\cal O}(T^2)=4~T^4\zeta(5)+{\cal O}(T^2),
\end{equation} which is same as that of boson obtained in \eqref{freelogzb}. $\ln Z$ in \eqref{freelogzf} gives the value of partition function
 for the gas of multitrace operators of free fermions on the sphere of volume $V_2=4\pi.$

For the ${\cal N}=2$ supersymmetric theory with a chiral multiplet, 
the multitrace partition function is given by
\begin{equation}
 \ln Z=\sum_{n=1}^{\infty}\frac{1}{n}\left(f_{b}(x^{n})+(-1)^{n+1}f_{f}(x^{n})\right)^2
\end{equation} which in the large temperature limit is given by 
\begin{equation}\label{freelogzsusy}
 \ln Z=8~T^4\sum_{n=1}^{\infty}\frac{1}{n^5}+8~T^4\sum_{n=1}^{\infty}(-1)^{n+1}\frac{1}{n^5}=\frac{31}{2}~T^4\zeta(5)+{\cal O}(T^2).
\end{equation} $\ln Z$ in \eqref{freelogzsusy} gives the value of partition function
 for free ${\cal N}=2$ supersymmetric theory with a chiral multiplet on the sphere of volume $V_2=4\pi.$ This exactly
matches with first term in \eqref{ftsusy1}.

\bibliographystyle{JHEP}
\bibliography{ccs}
\end{document}